\documentclass[review]{elsarticle}

\usepackage[colorlinks,citecolor=blue,linktoc=all,linkcolor=cyan]{hyperref}
\usepackage{graphicx}

\usepackage[T1]{fontenc}
\usepackage{dsfont}               
\usepackage{mathrsfs}             
\usepackage{slashed}              
\usepackage{amsmath}
\usepackage{amssymb}
\usepackage{amsbsy}
\usepackage{amsfonts}

\numberwithin{equation}{section}
\numberwithin{table}{section}
\numberwithin{figure}{section}

\journal{Progress in Particle and Nuclear Physics}

\usepackage{titlesec}
\usepackage{sectsty}
\titleformat{\section}{\normalfont\Large\bfseries}{\thesection}{1em}{}
\titleformat{\subsection}{\normalfont\large\bfseries}{\thesubsection}{1em}{}
\titleformat{\subsubsection}{\normalfont\normalsize\bfseries}{\thesubsubsection}{1em}{}

\newcommand{\mev}{\text{MeV}}
\newcommand{\gev}{\text{GeV}}
\newcommand{\tev}{\text{TeV}}

\newcommand{\iab}{\text{ab}^{-1}}
\newcommand{\fb}{\text{fb}}

\begin{document}
	
	\begin{frontmatter}
		
		\title{The FASER experiment at the Large Hadron Collider}

		\author[Jamie.Boyd@cern.ch]{Jamie Boyd}

		\address[mymainaddress]{CERN, CH-1211 Geneva 23, Switzerland}

		\begin{abstract}
			The FASER experiment is located in the Large Hadron Collider (LHC) complex at CERN, 480~m downstream of the ATLAS collision point and aligned with the beam-collision-axis. The experiment was designed to search for light, weakly-interacting new-particles which could be produced in the LHC collisions, and, for the first-time, to study high-energy neutrinos of all flavours originating at a particle collider. This review article presents the status of FASER up to early-2026. This includes details of the FASER detector design, operation, performance and physics results, as well as briefly mentioning upgrades that have been installed since the start of FASER. In addition, future plans for the experiment are detailed. 
		\end{abstract}
		
		
	\end{frontmatter}

		\small{ Preprint submitted to Progress in Particle and Nuclear Physics on 20 April 2026}
	\newpage
	
	\thispagestyle{empty}
	\tableofcontents


	\newpage
	\section{Introduction}\label{sec:intro}
    
The ForwArd Search ExpeRiment (FASER)~\cite{Feng:2017uoz} is an experiment at the CERN Large Hadron Collider (LHC) designed to search for light, weakly-coupled new particles and to study high-energy neutrinos of all flavours. FASER is situated about 480~m from the LHC interaction point 1 (IP1), which is inside the ATLAS experiment, and is closely aligned to the beam collision-axis line of sight (LOS). In this location the detector has world leading sensitivity to several light, weakly-coupled new-physics scenarios, as well as being able to detect and study thousands of collider neutrino interactions for the first time. 

\begin{figure}[hbt!]
    \centering
    \includegraphics[width=0.7\textwidth]{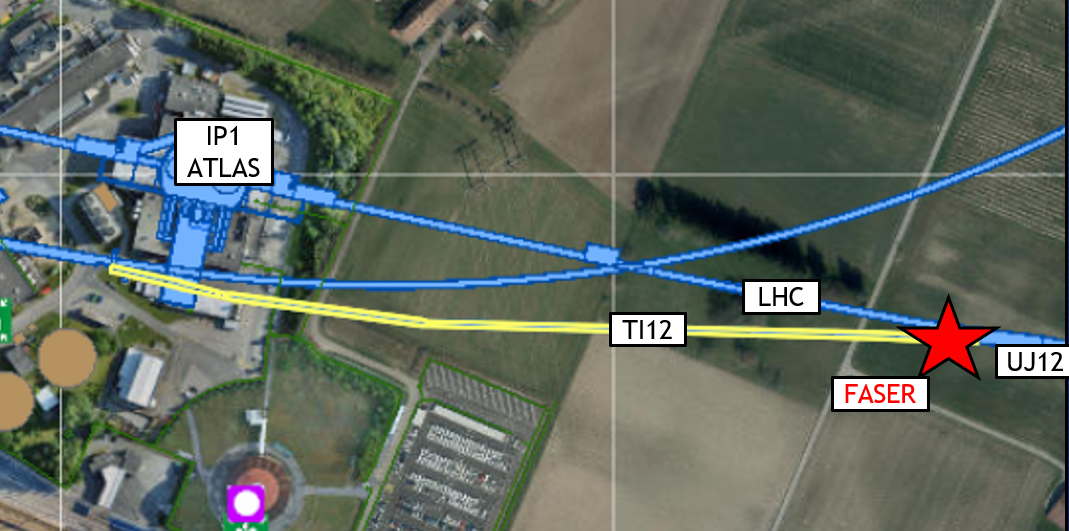}
    \caption{The FASER location: TI12 tunnel, 480~m downstream of the ATLAS interaction point. The detector is located along the beam collision axis line-of-sight. (This figure is taken from Ref.~\cite{FASER:2018bac}.)}
    \label{fig:TI12}
\end{figure}

\begin{figure}[hbt!]
    \centering
    \includegraphics[width=0.7\textwidth]{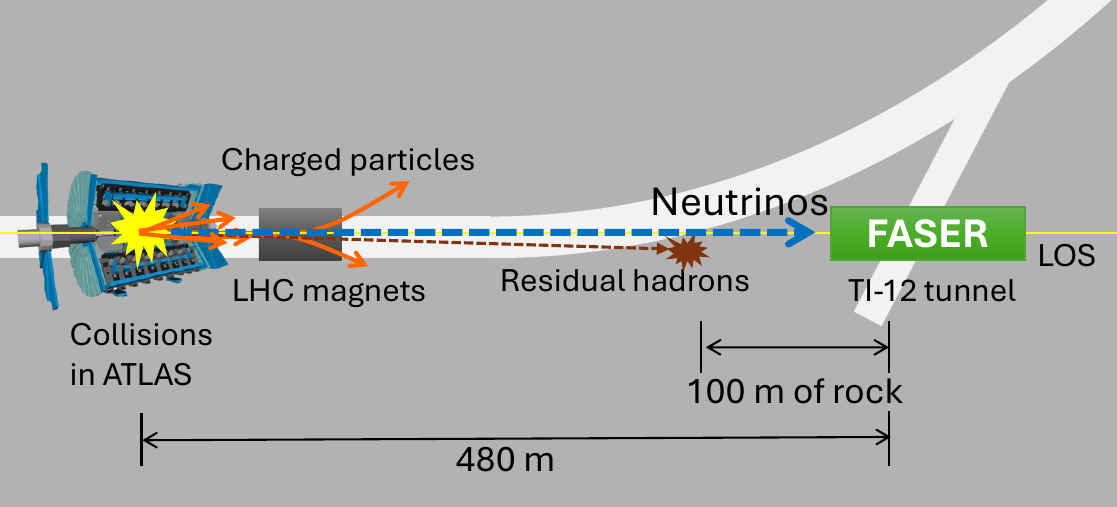}
    \caption{A sketch showing the location of FASER in the LHC complex. (This figure is modified from Ref.~\cite{Ariga:2025qup} and is not to scale.)}
    \label{fig:location-sketch}
\end{figure}

The detector is located in the TI12 tunnel, formerly used as a transfer line between the SPS and LEP accelerators, but unused for the LHC. As can be seen in Fig.~\ref{fig:TI12}, this location allows the detector to be aligned with the LOS, after the LHC machine has bent away. The detector is about 5~m from the LHC beam pipe. 
Fig.~\ref{fig:location-sketch} shows a sketch of the location, highlighting how the detector is well shielded from background particles produced in the collisions, with about 100~m of rock shielding the location from particles traveling along the LOS.

The FASER electronic detector consists of several scintillators to allow incoming charged particles to be vetoed, a magnetized decay volume, a tracking spectrometer, and an electromagnetic calorimeter. It is complemented by a tungsten/emulsion passive neutrino detector called FASER$\nu$.

\subsection{A brief history of FASER}

The idea to search for light, weakly-coupled new particles in the very forward region of the LHC collisions was proposed in Ref.~\cite{Feng:2017uoz} in 2017. The idea was developed by a small team with great help from the CERN Physics Beyond Collider study group~\cite{PBCwebpage}. Initial studies by CERN technical groups assessed the feasibility of realizing such an experiment in the TI12 or TI18~\footnote{TI18 is situated 480~m from IP1 on the other side of IP1, and now houses the SND@LHC experiment~\cite{SNDLHC:2022ihg}.} tunnels. These investigations, focused on the available space on the LOS and the possibility of increasing this with small civil engineering work, estimating the background rates and radiation levels, the available services and technical infrastructure, as well as access and transport considerations. In parallel, the experimental collaboration was formed and explored different options for the detector design. After receiving funding from the Heising-Simons~\cite{HS-foundation} and Simons~\cite{S-foundation} foundations in November 2018 the FASER Collaboration decided to propose to install the detector in LHC Long Shutdown 2 (LS2), at that time scheduled from 2019 - 2020~\footnote{Later the LS2 schedule was revised due to delays to various projects caused by the Covid-19 pandemic.}
. Given the very tight timescale the approach was to design a simple and robust detector, where possible, using spare parts from existing experiments to reduce costs and time. The CERN magnet group agreed to build the three FASER permanent dipole magnets and the ATLAS and LHCb Collaborations agreed to donate to FASER spare tracker and calorimeter modules, respectively. FASER submitted a Letter of Intent~\cite{Ariga:2018zuc} to the LHC Experiments Committee (LHCC) in July 2018, followed by a Technical Proposal~\cite{FASER:2018bac} in November 2018 with final approval by CERN given in March 2019~\cite{approval-RBminutes}.

In order to measure the background rate during LHC running, a small emulsion detector was installed into the FASER location in September 2018, and was exposed to 12~fb$^{-1}$ of 13~TeV $pp$ collision data. It was quickly realized that this could also be used to search for first neutrino events at the LHC. Analysis, of this emulsion led to the first observation of neutrino candidates~\cite{FASER:2021mtu} at a particle collider. Based on the experience with this emulsion detector the FASER$\nu$ subdetector was added to the FASER design, with an Letter of Intent~\cite{FASER:2019dxq} and Technical Proposal~\cite{Abreu:2020ddv} submitted to the LHCC, and with FASER$\nu$ being approved by CERN in December 2019~\cite{approval2-RBminutes}.

During 2020, the TI12 site was prepared for FASER which involved carrying out the needed civil engineering works, and the installation of the required technical infrastructure. This work was successfully carried out, despite complications due to the Covid-19 pandemic which closed access to CERN for several periods in 2020. 

The detector components were designed, produced and tested during 2019 and 2020 and a near-full assembly-test was carried out on the surface in the summer of 2020. The detector was then installed into the TI12 tunnel during March 2021. This was followed by a long period of in situ commissioning using cosmic-rays, and then during a dedicated low-energy beam test in October 2021. First physics collisions occurred in July 2022, and FASER successfully took data during the LHC physics runs in 2022 - 2025. More details on the detector operations are given in Section~\ref{sec:operations}. Several physics results have been released during this period as detailed in Section~\ref{sec:results}.

The FASER Collaboration has grown considerably since the start. Originally, 14 people from 8 institutes in 4 countries signed the letter of intent, this approximately doubled for the technical proposal which was signed by 35 people from 15 institutes and 7 countries. By the time the first physics results were released in March 2023 the collaboration consisted of 85 people from 22 institutes and 9 countries, and currently (April 2026) it is made up of 123 people from 27 institutes in 11 countries.

    \clearpage
    \newpage
	\section{Physics Case}\label{sec:physicscase}
    
The physics objectives of FASER can be divided into two classes, i) to search for light weakly-interacting new-particles, such as those predicted in many dark-sector models, and, ii) to study high-energy neutrinos produced in the LHC collisions.
In this section the physics motivation for these two areas are discussed in more detail.

\subsection{Searches for new particles}

FASER searches for light, weakly-interacting particles which may be produced in the $pp$ collisions at IP1, travel long distances through rock without interacting and then decay to visible particles inside the detector decay volume. The sensitivity reach for FASER for several beyond the Standard Model (BSM) theories predicting long-lived particles is reported in Ref.~\cite{FASER:2018eoc}. Dark photon models are one of the main targets of FASER searches, while other scenarios considered include models with dark Higgs bosons, heavy neutral leptons,  light gauge bosons and axion-like particle models with various coupling configurations. Below the dark photon model is used to demonstrate the FASER search methodology.

\begin{figure}[hbt!]
    \centering
    \includegraphics[trim=0.cm 0.cm 0.cm 0.cm, clip=true,angle=0,width=0.9\textwidth]{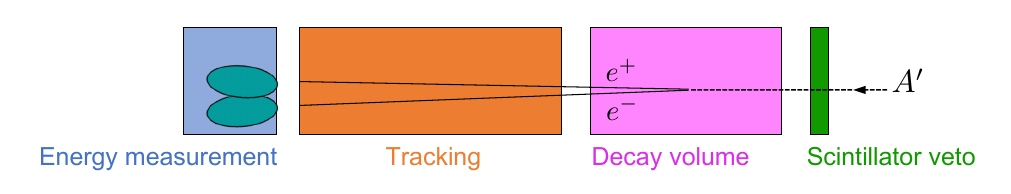}
    \caption{Sketch showing the detector signature of a dark photon ($A'$) decaying to an electron-positron pair inside the decay volume of the FASER experiment. The $A'$ enters the detector from the right. This figure is taken from Ref.~\cite{FASER:2018eoc}. }
    \label{fig:DarkPhotonSketch}
\end{figure}

The dark photon ($A'$) appears in dark-sector models in which the dark sector contains a new spin-1 vector boson with suppressed couplings to Standard Model (SM) particles via its kinetic mixing with the photon, with the coupling labeled $\epsilon$. In such models the $A'$ acts as a mediator between the SM and a hypothetical dark matter particle ($\chi$). The $A'$ phenomenology is defined by $\epsilon$ and the dark photon mass ($m_{A'}$), and there is a large region of viable parameter space yet to be constrained by experiments. If $m_{A'} < 2m_\chi$, the $A'$ decays to SM particles, and FASER is mostly sensitive to the mass range $2m_\mu > m_{A'} > 2m_e$ such that the $A'$ decays to an electron and positron pair. For sufficiently weak couplings, the $A'$ is long-lived and can travel significant distances before decaying. The relevant parameter space for which FASER has sensitivity is $\mathcal{O}(10^{-6}) < \epsilon <\mathcal{O}(10^{-4})$ with $m_{A'} \lesssim 100$~MeV, such that the $A'$ may travel from IP1 to FASER without interacting and decay inside FASER to an $e^+ - e^-$ pair. A sketch of the corresponding detector signature is shown in Figure~\ref{fig:DarkPhotonSketch}. 

\begin{figure}[t]
\centering
\includegraphics[width=0.49\textwidth]{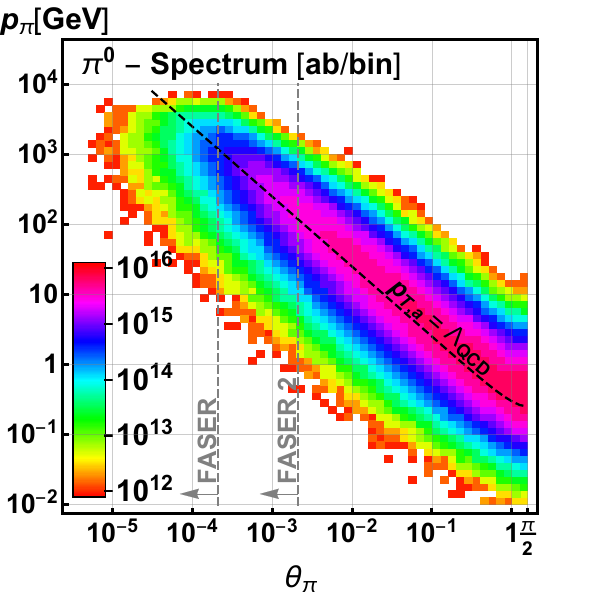}
\caption{Simulated $\pi^0$ production at the LHC, showing the production angle with respect to the beamline versus the pion energy. (Taken from Ref.~\cite{FASER:2018eoc}.)
}
\label{fig:hadron-production-angle}
\end{figure}

The main production process for dark photons relevant for FASER is through light meson decays, in particular the decay of $\pi^0$s. Neutral pions may decay to a dark photon and a SM photon, with a branching fraction proportional to $\epsilon^2$. As shown in Fig.~\ref{fig:hadron-production-angle}, pion production at the LHC is strongly peaked in the very forward direction, such that $\mathcal{O}{(1\%)}$ of the pions produced with energy $E_{\pi^0}> 10$ GeV are within the FASER angular acceptance of $|\theta| < 0.21$~mrad, despite the fact that this covers only $\mathcal{O}{(10^{-8})}$ of the total solid angle.  
The figure also shows that these forward pions have a large boost, as much as $\mathcal{O}(1~\mathrm{TeV})$, along the beam direction in the lab frame, allowing dark photons produced in their decay to reach FASER even with relatively short lifetimes. 

\begin{figure}[hbt!]
    \centering
    \includegraphics[trim=0cm 0.cm 0.cm 0.cm, clip=true, width=0.6\textwidth]{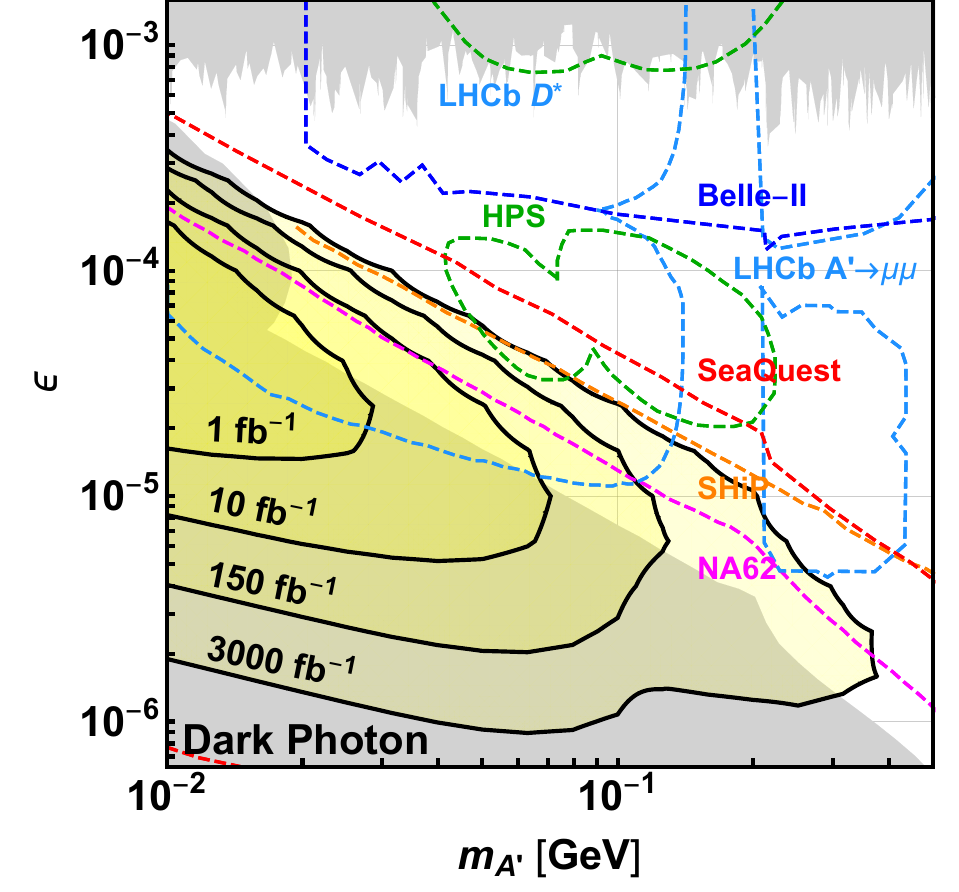}
    \caption{The expected sensitivity of FASER for dark photons as a function of the mass ($m_{A'}$) and coupling ($\epsilon$), for different values of integrated luminosity. Experimental constraints (grey) and projections (coloured lines) are shown at the time this figure was produced in 2018. 
   This plot is taken from Ref.~\cite{FASER:2018eoc}, which also gives the  details of the experimental projections and constraints shown.
    }
    \label{fig:darkphotonSensitiviy}
\end{figure}

During Run~3 of the LHC, with a total of around 350~fb$^{-1}$ of data,  around $\mathcal{O}(10^{14})$  $\pi^0$s are produced within the FASER angular acceptance. Therefore, a significant number of signal events can be detected in FASER, even taking into account the large suppression due to the $\pi^0 \to A' \gamma$ branching fraction ($\mathcal{O}(10^{-12})$ to $\mathcal{O}(10^{-8})$) and the requirement that the $A'$ decays inside the FASER detector volume. Figure~\ref{fig:darkphotonSensitiviy} shows the expected FASER sensitivity to dark photons for different integrated luminosity scenarios, where  the contours are defined such that at least three signal events pass the kinematic and geometrical requirements and the dark photon decays inside the FASER decay volume. 
The contours assume 100\% efficiency and zero background. As will be discussed in Section~\ref{sec:results}, real data analysis is nearly background-free with a signal efficiency of around 50~\% which gives very close to the ideal performance shown here. 
This is because the number of signal events falls off very rapidly at the edge of the sensitivity boundaries of the signal parameter space. The boundaries of the signal sensitivity contours are set primarily by the production rate falling off at too-small couplings and the $A'$s being too short-lived to reach FASER before decaying at too-large couplings. 
More details on the theoretical aspects of the dark photon model are given in Ref.~\cite{FASER:2018eoc}.

\subsection{Neutrino physics programme}
\label{sec:neutrinoPhys}

A huge number of neutrinos are produced in the LHC collisions via hadron decays, and their flux is collimated along the beam collision axis. Table~\ref{tab:neutrino-numbers} summarizes the number of neutrinos expected to traverse, and under-go charged-current (CC) interactions in, the FASER$\nu$ emulsion detector (with a target mass of 1.1~tonnes), assuming a 350~fb$^{-1}$ dataset for the LHC Run~3, as well as showing the average expected neutrino energy. The table includes all three neutrino flavours (summing $\nu$ and $\overline{\nu}$) and also shows the dominant production processes. 

The neutrino numbers reported in the table, and elsewhere in this paper, are obtained using the recipe outlined in Ref.~\cite{FASER:2024ykc}, which is briefly summarized below. Since the neutrinos originate from hadron decay, the neutrino flux is directly related to the far-forward hadron production cross section, which suffers from large uncertainties. Since heavy quark production can be calculated using perturbative QCD, the flux component from charm hadron decay is calculated separately from that arising from light-hadron decay. Light hadron production is estimated using several phenomenological models used for cosmic-ray physics. The central prediction uses \texttt{EPOS-LHC}~\cite{Pierog:2013ria} to simulate the production of forward light-hadrons and the envelope formed by \texttt{EPOS-LHC}, \texttt{SIBYLL~2.3d}~\cite{Riehn:2019jet}, \texttt{QGSJET~2.04}~\cite{Ostapchenko:2010vb} as well as the forward physics tune of \texttt{PYTHIA}~\cite{Fieg:2023kld} is used to define an uncertainty band, with typical uncertainties of $\mathcal{O}$(20\%). 
Using the results of Ref.~\cite{Buonocore:2023kna}, charm hadron production is modeled using \texttt{POWHEG}~\cite{Nason:2004rx, Frixione:2007vw, Alioli:2010xd} matched with \texttt{PYTHIA~8.3}~\cite{Bierlich:2022pfr} for parton shower and hadronization. The uncertainties are described by scale variations, which leads to large uncertainties of $\mathcal{O}$(100\%). 
For neutrinos originating from long-lived hadrons (such as pions and kaons) the trajectory of these hadrons needs to be simulated as they travel through the LHC infrastructure, and the effect of magnetic fields on their direction needs to be considered. This is done using a RIVET routine as described in Refs.~\cite{Kling:2021gos, FASER:2024ykc}. Given the strong magnetic fields in the LHC final focusing magnets, this typically means the neutrinos reaching FASER come from charged hadrons decaying before these magnets (so less than 20~m from the IP).  
The expected neutrino event rates were estimated from the fluxes using the neutrino interaction cross section provided by the \texttt{GENIE}~\cite{Andreopoulos:2009rq} generator. The Bodek-Yang model~\cite{Bodek:2002vp, Bodek:2004pc, Bodek:2010km} employed in \texttt{GENIE} shows good agreement with more recent cross section calculations for high-energy neutrinos~\cite{Candido:2023utz, Jeong:2023hwe} within an uncertainty of $\lesssim 6\%$~\cite{FASER:2024ykc}. 

\begin{figure}[t]
\centering
\includegraphics[width=0.49\textwidth]{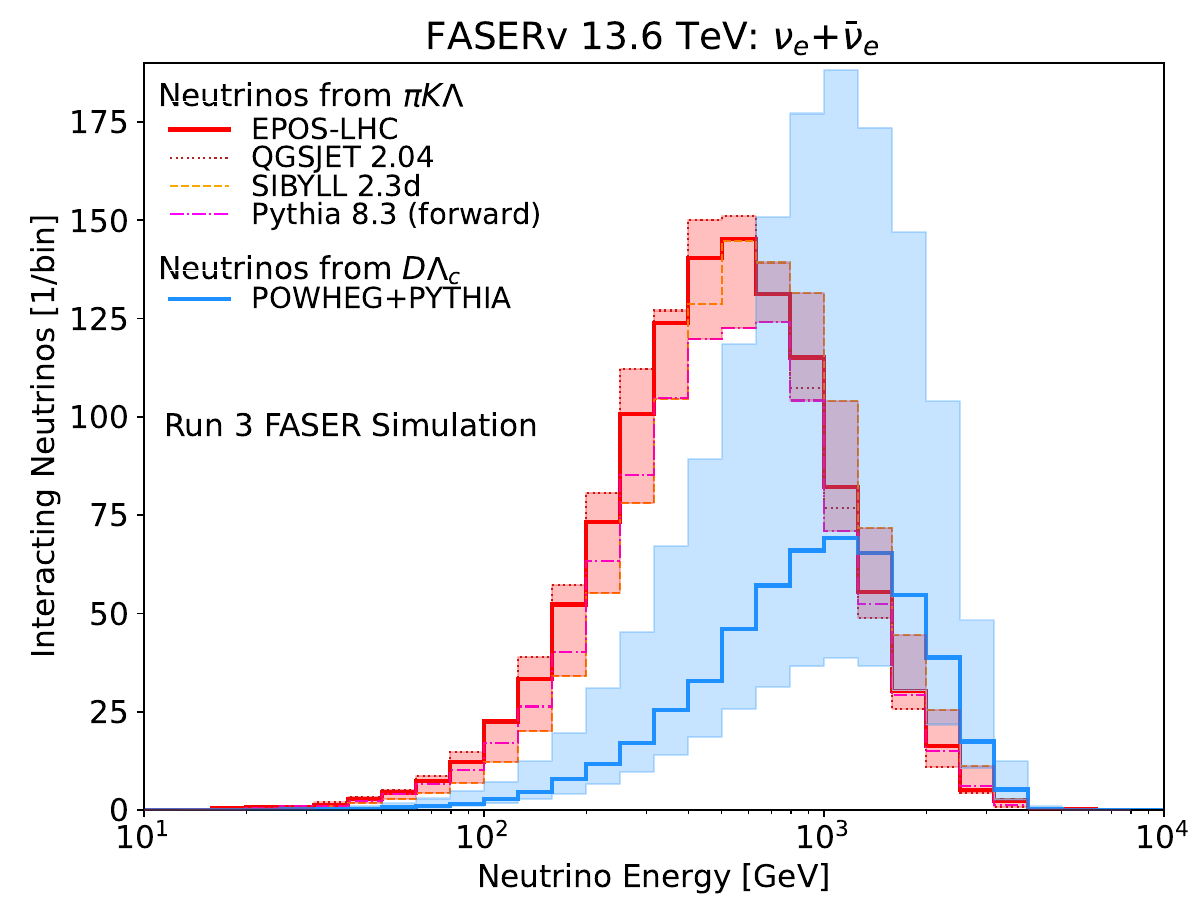}
\includegraphics[width=0.49\textwidth]{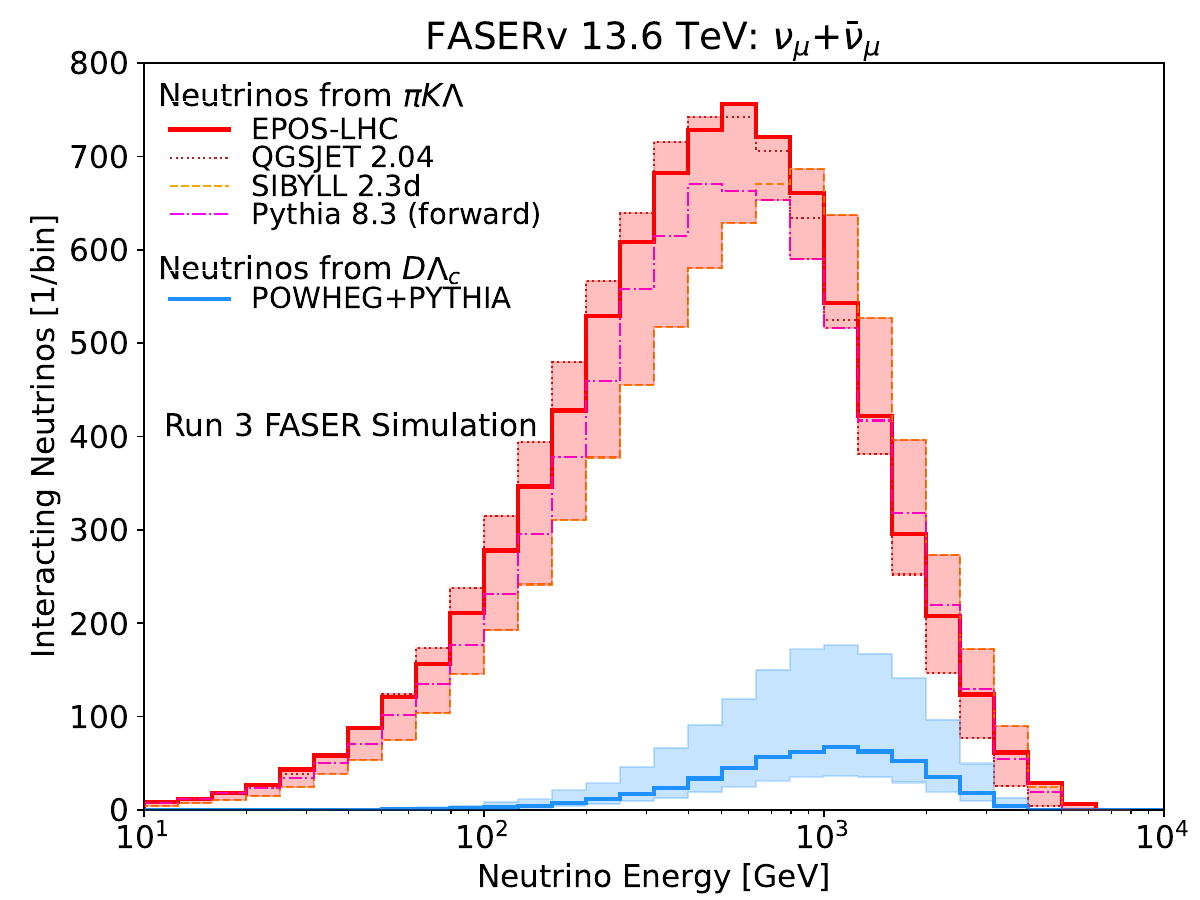}
\caption{Estimated number of charged current neutrino interactions in FASER$\nu$ for LHC Run 3 for electron neutrinos (left) and muon neutrinos (right). The red curve is for neutrinos from light hadron decay, and the blue is from charm hadron decay. (This figure is taken from Ref.~\cite{FASER:2024ykc}.)
}
\label{fig:neutrino-predictions}
\end{figure}

The predicted number of CC neutrino interactions in FASER$\nu$ as a function of energy  can be seen in Fig.~\ref{fig:neutrino-predictions} for both $\nu_e$ and $\nu_\mu$, showing separately the components from the decays of light hadrons and charm hadrons. These plots highlight that the charm component is larger for $\nu_e$, and at higher energy, and also that it has a much larger associated uncertainty.

\begin{table}[thb]
  \centering
  \begin{tabular}{|l|c|c|c|}
  \hline
    - & \bf{$\nu_e$} & \bf{$\nu_\mu$} & \bf{$\nu_\tau$} \\
  \hline
     Dominant production process & 
  Kaon decays & Pion decays & charm decays \\
      & (charm decays at high energy) & (Kaon decays at high energy) & \\
        Number of $\nu$ traversing FASER$\nu$ &
      $\mathcal{O}$($10^{11}$) & $\mathcal{O}$($10^{12}$) & $\mathcal{O}$($10^{9}$) \\
      Number of $\nu$ interacting in FASER$\nu$ &  2331$^{+1227}_{-544}$ &  12014$^{+1145}_{-1636}$ & 46$^{+77}_{-21}$ \\ 
 Average energy  & 785 GeV & 716 GeV & 849 GeV \\
\hline
    \end{tabular}
    \caption{Summary of the number of neutrinos traversing and interacting in FASER$\nu$ assuming a 350~fb$^{-1}$ dataset for the LHC Run~3. All three flavours of neutrinos are considered and the numbers are for neutrinos and antineutrinos combined. The table also shows the dominant production process and the average energy of the interacting neutrinos. Only CC interactions are considered. (This table is modified from Ref.~\cite{Ariga:2025qup}.)
    }
    \label{tab:neutrino-numbers}
\end{table}

\begin{figure}[t]
\centering
\includegraphics[width=0.99\textwidth]{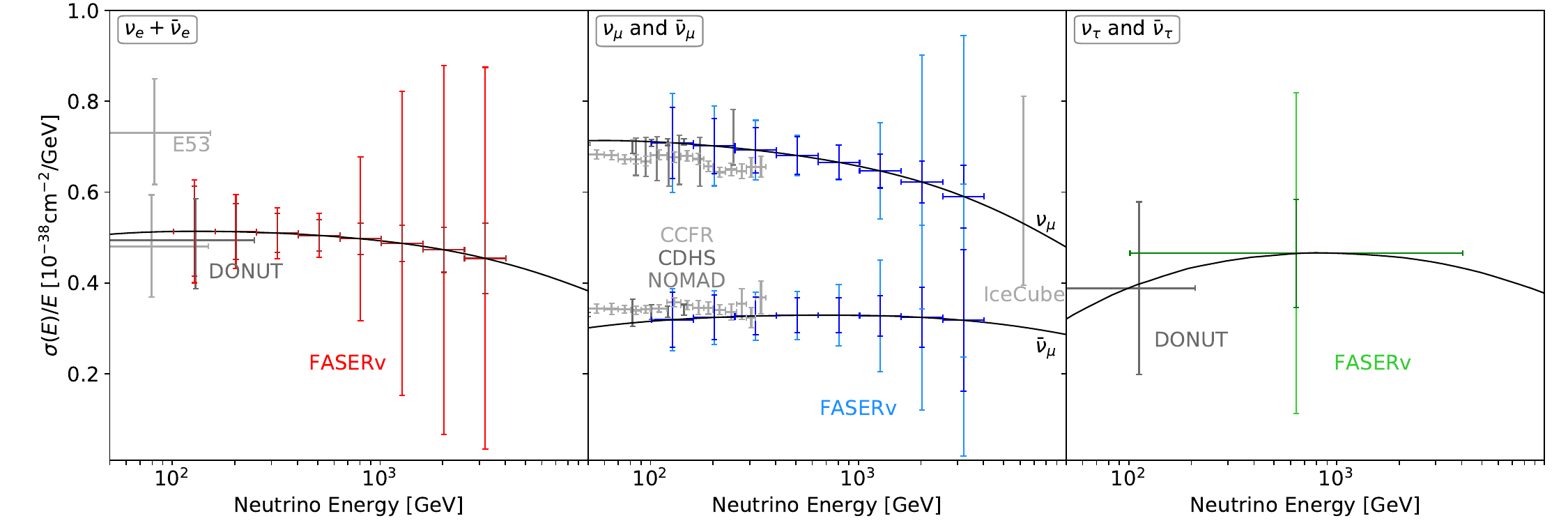}
\caption{FASER$\nu$'s estimated $\nu$-nucleon CC cross section sensitivity for $\nu_e$ (left), $\nu_{\mu}$ (centre), and $\nu_\tau$ (right) for LHC Run 3. Existing constraints are shown in gray~\cite{ParticleDataGroup:2018ovx}. The black curve is the theoretical prediction for the DIS cross section per tungsten-weighted nucleon. The coloured error bars show FASER$\nu$'s cross section sensitivity, where the inner error bars correspond to statistical uncertainties, while the outer error bars show the combination of statistical and flux uncertainties. 
}
\label{fig:xs_measurement}
\end{figure}

Taking advantage of the large number of neutrino interactions, FASER$\nu$ is designed to measure the neutrino CC interaction cross sections for all three neutrino flavours in an uncovered energy regime. Projections of the expected measurement statistical precision from before FASER operations are shown in Figure~\ref{fig:xs_measurement}. 
For $\nu_\mu$, the charge of the muon can be reconstructed in the FASER tracking spectrometer, making it possible to measure both the neutrino and anti-neutrino cross sections separately,~\footnote{Note the FASER spectrometer overlaps with only 42$\%$ of the FASER$\nu$ emulsion detector in the transverse plane, and therefore the number of muon neutrino interactions that can be used to measure the neutrino/anti-neutrino cross sections only corresponds to a target mass of 460~kg.} as shown in the figure. 
Importantly, FASER can measure the cross section in the TeV energy regime where there are no existing measurements.   
Such measurements allow tests of lepton flavour violation in the neutrino sector. 

FASER neutrino measurements can also be used as a probe of forward charged-particle production in an unmeasured region of phase space. These measurements can help to  validate and improve the underlying hadronic interaction models providing valuable input for tuning hadronic shower simulations used to interpret data from ultra-high-energy cosmic ray air-shower experiments.
In addition, measurements of far-forward charm production (via high energy $\nu_e$ measurements in FASER) can constrain the gluon PDF at very low values of x. 
Finally, measurements of Deep Inelastic Scattering (DIS) CC neutrino interactions in the target can be used to constrain proton and nuclear parton distribution functions (PDFs) although large numbers of events are needed for this~\cite{Cruz-Martinez:2023sdv}. More details on the physics opportunities from FASER measurements of neutrinos can be seen in Ref.~\cite{Ariga:2025qup} and the references therein.  

    \clearpage
    \newpage
	\section{Detector}\label{sec:detector}
    
\subsection{Detector requirements}

The most relevant detector requirements needed to fulfill the FASER physics programme on BSM searches and neutrinos measurements are listed below. 
\begin{itemize}
    \item Given its transverse size, fixed by the tunnel and trench constraints, the detector must be centred on the LOS to within about 10 cm to maximise the number of signal events and neutrino energies.
    \item To reject physics background events initiated by high-energy muons, the detector must include a system that vetoes incoming charged particles with an extremely high efficiency. 
    \item The detector must be able to track high-energy charged-particles with good precision. In particular, it needs to separate very closely-spaced charged-particles arising from highly boosted, light BSM particle decays, and to be able to measure the momentum and charge of muons from neutrino scattering up to momenta of $\mathcal{O}$($1$ TeV).
    \item To measure the electromagnetic (EM) energy from boosted BSM particle decays the calorimeter must be capable of measuring $\mathcal{O}$(TeV) EM deposits with a few percent resolution.
    \item In order to be able to maximise the number of recorded neutrino interactions taking into account the space limitations, the neutrino detector must have a high-density leading to a  target mass of $\mathcal{O}$($1$ tonne).
    \item The neutrino detector should have the ability to identify different lepton flavours from neutrino CC interactions. Including having sufficient target material to identify muons, finely-sampled detection layers to identify high-energy EM showers, and excellent position resolution to identify $\tau$-leptons.  
    \item The neutrino detector should be able to estimate the neutrino energy, by measuring muon and hadron momenta, and the energy of EM showers.
    \item To ensure a good efficiency for collecting rare signal events, the trigger must be highly efficient and the data acquisition (DAQ) system needs to operate robustly with little dead-time.  
\end{itemize}

The radiation levels at the FASER location were estimated using FLUKA simulations, and validated using measurements taken during 2018 LHC running~\cite{FASER:2018bac}. Both the high-energy hadron fluence and the thermal neutron fluence were found to be low enough to allow non radiation-hard electronics to be used in the detector. 

\begin{figure}[hbt!]
    \centering
        \includegraphics[trim=0cm 0.8cm 0.5cm 1.5cm, clip=true, width=\textwidth]{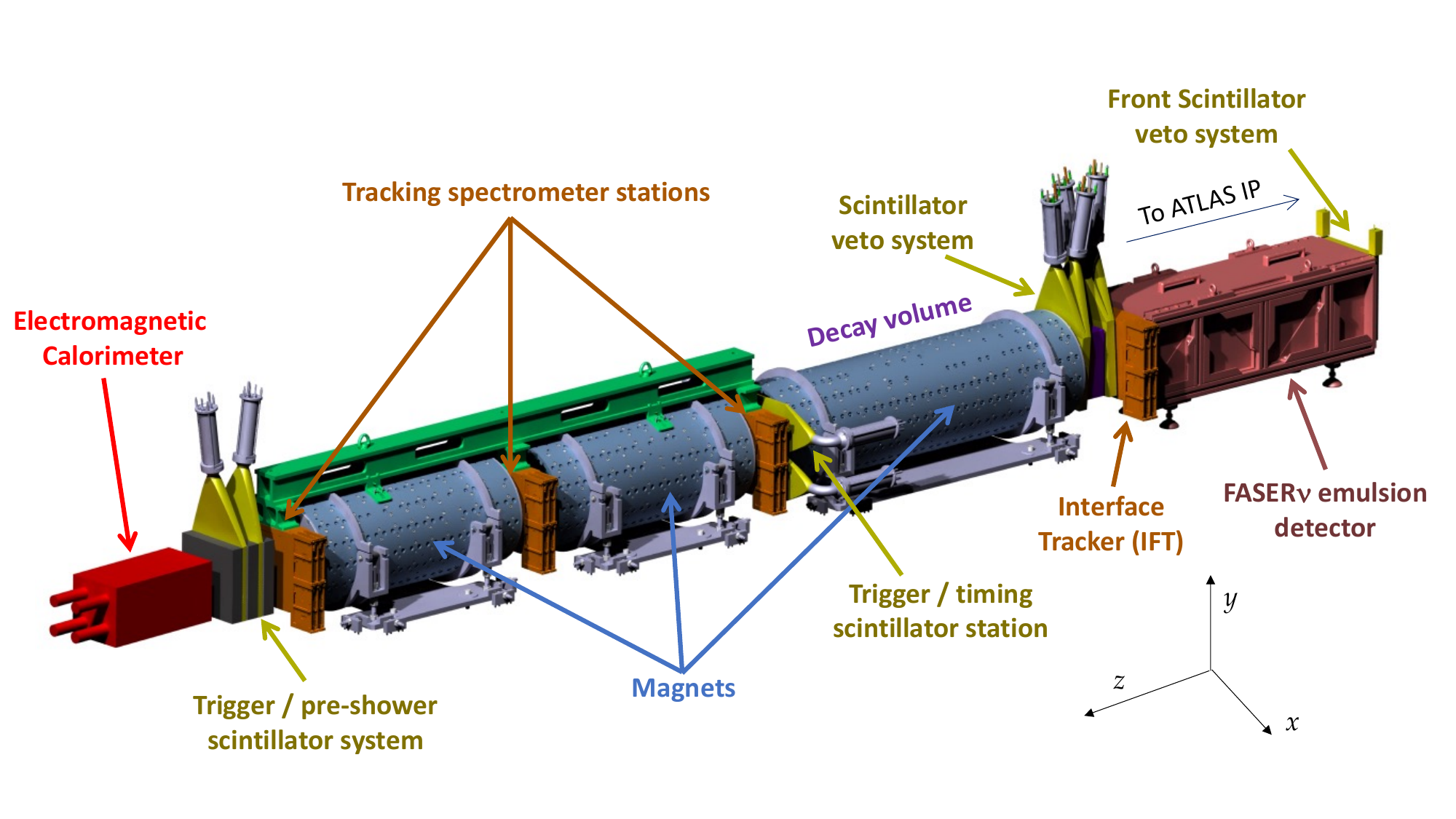}
        \caption{A sketch of the FASER  detector, showing the different sub-detector systems. The FASER coordinate system is also shown. This figure is taken from Ref.~\cite{FASER:2022hcn}.}
    \label{fig:FASER_labels}
\end{figure}

A sketch of the FASER detector is shown in Fig.~\ref{fig:FASER_labels}. It was designed to fulfill the above requirements, but additional constraints were also taken into account, including the short time available to construct and install the detector, the cost, and the need to minimize the required services in the detector location. The detector is about 7~m long and the active area for the electronic detector is a 20~cm diameter circle defined by the aperture of the magnets. 
As shown in the fiugure, FASER uses a cartesian coordinate system with the $z$-axis pointing along the LOS away from IP1, the $y$-axis pointing vertically upwards, and the $x$-axis pointing horizontally towards the LHC machine. The origin of the coordinate system is aligned with the centre of the magnets in the transverse, $x$ - $y$, plane and at the front of the second tracker station in the $z$ coordinate (477.759 m from IP1).  
A more detailed description of the detector can be found in Ref.~\cite{FASER:2022hcn}.

In the rest of this section the detector components are described in more detail.

\subsection{Tracker}
The FASER detector tracking system is composed of two distinct parts, the tracking spectrometer and the interface tracker (IFT),  and a detailed description as well as the results of its commissioning can be found in Ref.~\cite{FASER:Tracker}.
The tracking spectrometer allows the trajectories of charged particles traversing the detector to be reconstructed, and  their position, momentum and charge to be measured. The IFT is placed right after the FASER$\nu$ emulsion detector, and enables  tracks reconstructed in the emulsion to be matched with those in the electronic detectors. This is designed to allow the measurement of the charge and momentum of muons arising from the neutrino interactions reconstructed in the FASER$\nu$ emulsion.

\begin{figure}[th]
\centering
\includegraphics[width=0.45\textwidth]{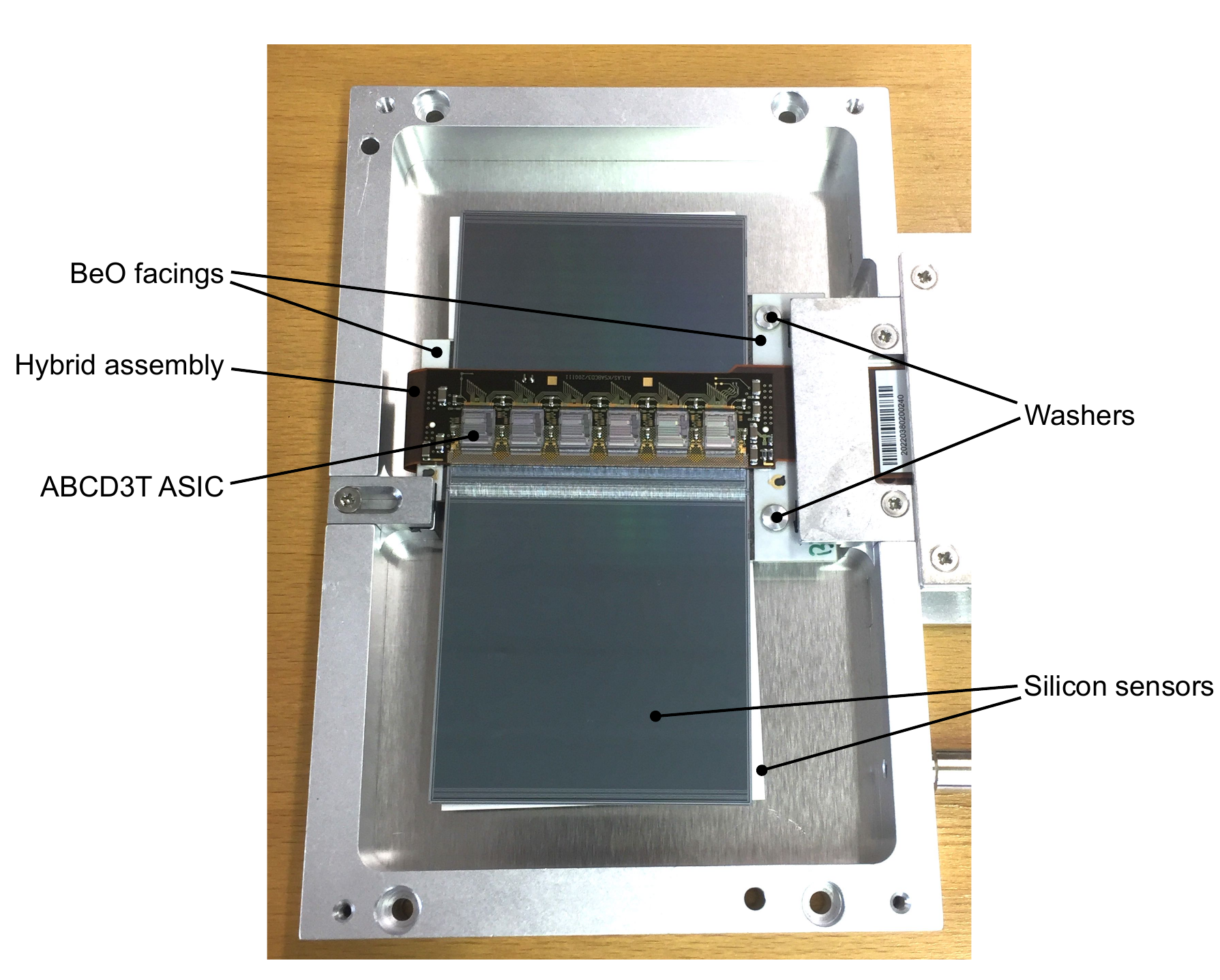}
\includegraphics[width=0.54\textwidth]{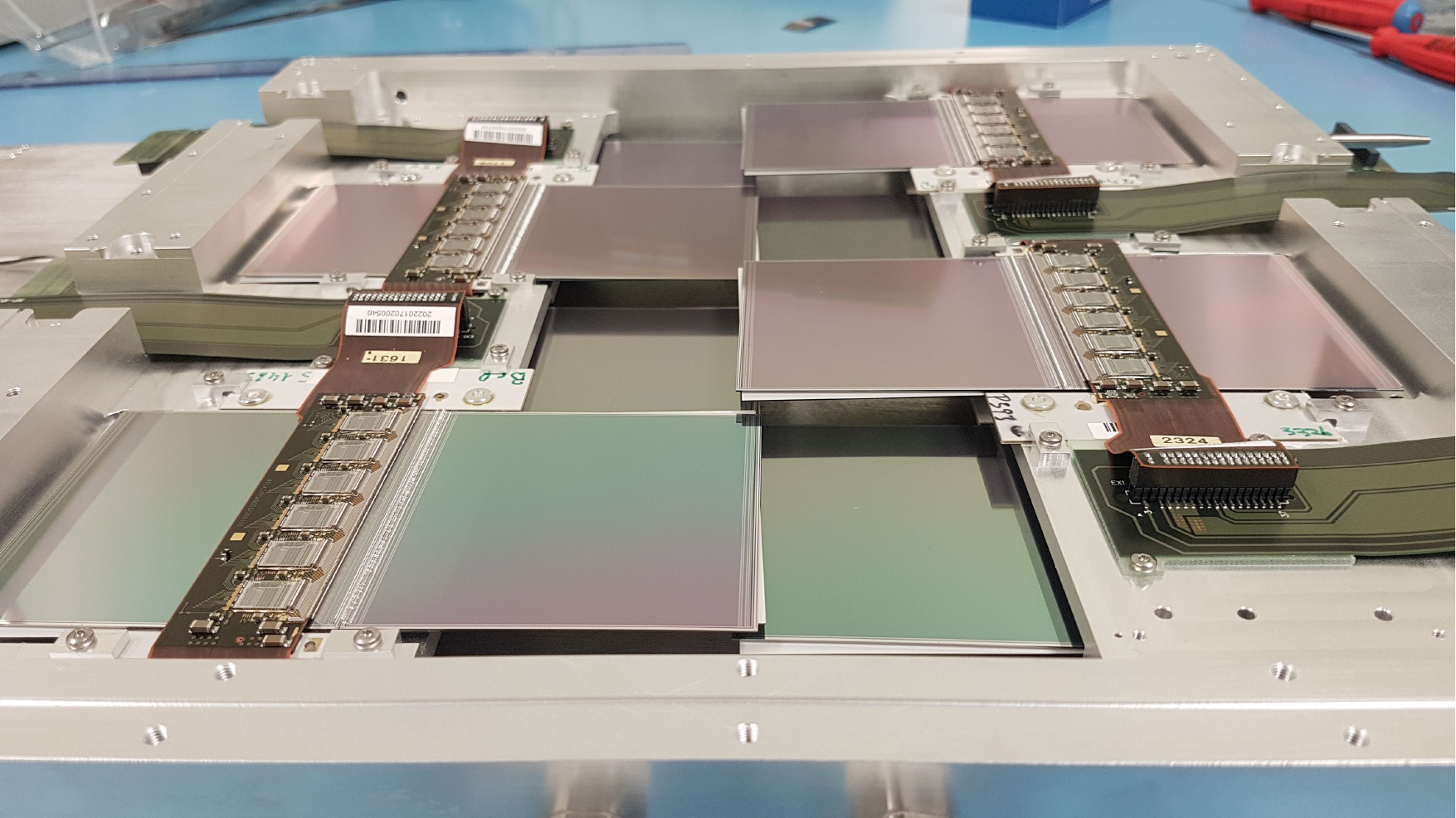}
\caption{(left) Photograph of a SCT barrel strip module. This figure is taken from Ref.~\cite{Abdesselam:2006wt} (right) Photograph of a tracking layer with all eight SCT modules installed. The beam axis is perpendicular to the plane. This figure is taken from Ref.~\cite{FASER:Tracker}.}
\label{fig:SCT-photo}
\end{figure}

The tracking spectrometer consists of three tracking stations, and the IFT is an identical tracking station. Each tracking station consists of three double-layers of single-sided silicon microstrip detectors. A tracking layer is made up of eight silicon strip modules which are spares from the ATLAS experiment’s SCT barrel detector~\cite{Abdesselam:2006wt}, giving a total of 96 modules used in FASER. A module is shown in Fig.~\ref{fig:SCT-photo} (left) and is approximately $6 \times 12$~cm$^2$, and they are arranged in two columns of four modules to give a $24 \times 24$~cm$^2$ active detector area, called a tracking layer, which covers the full aperture of the FASER magnets.  A photo of a tracking layer is shown in Fig.~\ref{fig:SCT-photo} (right). The  modules are operated with a bias voltage of 150~V, providing a hit efficiency well above 99\%. Despite using detector modules from the ATLAS experiment, all other aspects of the tracker system including the mechanics, readout system, cooling system and powering is newly designed and constructed for FASER. 

The SCT modules have a strip pitch of 80~$\mu$m, and a stereo angle between the two sides of 40~mrad, providing a track resolution of order of 20~$\mu$m in the precision coordinate, and 800~$\mu$m in the other coordinate. The modules are aligned such that the precision coordinate corresponds to the magnet bending plane ($y$). 

In order to ensure good alignment between the three tracking stations in the spectrometer, they are all attached to a common precision-machined support beam which is attached to the two spectrometer magnets. Since there is little radiation in the FASER area the silicon can be operated at room temperature, however the on-module readout ASICs produce a total of 450~W of heat which needs to be extracted. This is achieved using a simple water cooling circuit which runs around the outside of each of the tracking layer mechanical frames. The water is cooled to 15~degrees with a simple chiller. Two such chillers are integrated into a cooling unit, with remote control and monitoring functionality, designed and built by the CERN cooling and ventilation group. To avoid problems related to condensation on the tracker modules, each tracker station is flushed with dry-air. In order to protect the tracker, a hardware interlock system powers down the detector in the event of a cooling failure. 

\subsection{Scintillators and Calorimeter}
The FASER detector includes four scintillator stations, used for vetoing and trigger purposes, and an electromagnetic calorimeter,  designed to measure the energy of high-energy electrons and photons and provide redundant triggering for signals with large energy deposits. In addition, the calorimeter in conjunction with scintillator counters placed at the back of the tracking spectrometer can provide simple particle identification.

Each of the four scintillator stations is composed of multiple  scintillator counters, read out by photomultiplier tubes (PMTs), as detailed below:
\begin{itemize}
\item The first two stations are needed to veto charged particles entering the detector. One is placed in front (upstream) of the FASER$\nu$ emulsion detector and consists of two scintillator counters. The other is located in front (upstream) of the FASER decay volume and is made up of four scintillator counters. Each scintillator counter is read out by a single PMT.  
The veto scintillator stations must be able to veto charged particles with very high efficiency, to ensure the experiment remains background free with an expected 10$^9$ muons traversing FASER during Run 3 operations. 
The veto scintillators  are larger than the active transverse size of FASER (given by the aperture of the magnets) in order to be able to veto muons that could enter FASER at an angle with respect to the detector axis.
\item The Trigger/timing scintillator station is placed after the decay volume. It consists of two scintillator counters, each read out by two PMTs. The station is designed to provide a trigger for charged particles exiting the decay volume, and to give a precise time for triggered events, with a resolution of better than 1~ns. 
\item A pre-shower scintillator station is placed at the back (downstream) of the tracking spectrometer. As described in Section~\ref{sec:upgrades}, this station was upgraded with a high granularity silicon/tungsten preshower system in 2025.  It is composed of two scintillator counters, each read out by a single PMT, interleaved with two tungsten absorbers, and graphite blocks. The purpose of the pre-shower is to distinguish between calorimeter signals from neutrino interactions in the calorimeter, and incoming photons. Photon showers  start to develop in the tungsten absorbers (each of about one radiation length), therefore leaving signals in the scintillator counters. The graphite blocks are installed to minimize back-splash from the calorimeter leaving signals in the scintillator counters and the rear tracking station.
\end{itemize}

The calorimeter provides energy measurements with an expected resolution of $\mathcal{O}$($1$ \%) for the energy range of interest. It consists of four spare modules from the LHCb experiment's outer electromagnetic calorimeter (ECAL)~\cite{LHCB:2000ab} and is 25 radiation lengths deep. 

\begin{figure}
    \centering
    \includegraphics[width=0.75\textwidth]{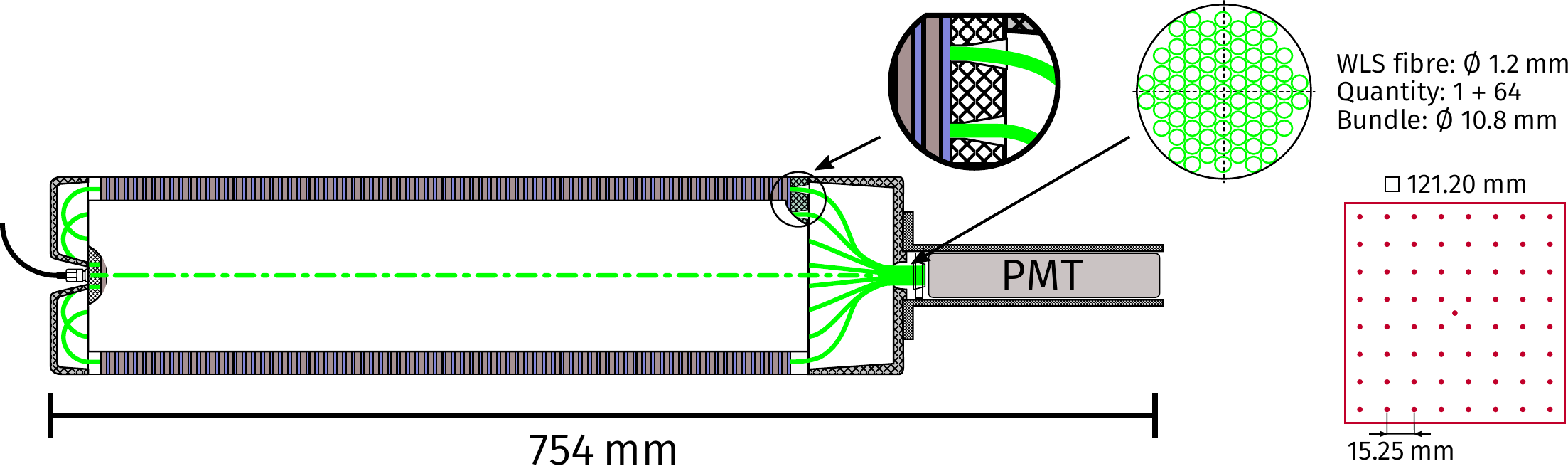}
    \caption{Design of a FASER (LHCb outer) calorimeter module. (This figure is taken from Ref.~\cite{FASER:2022hcn}.) }
    \label{fig:CalorimeterDrawing}
\end{figure}

\begin{figure}
    \centering
    \includegraphics[width=0.7\textwidth]{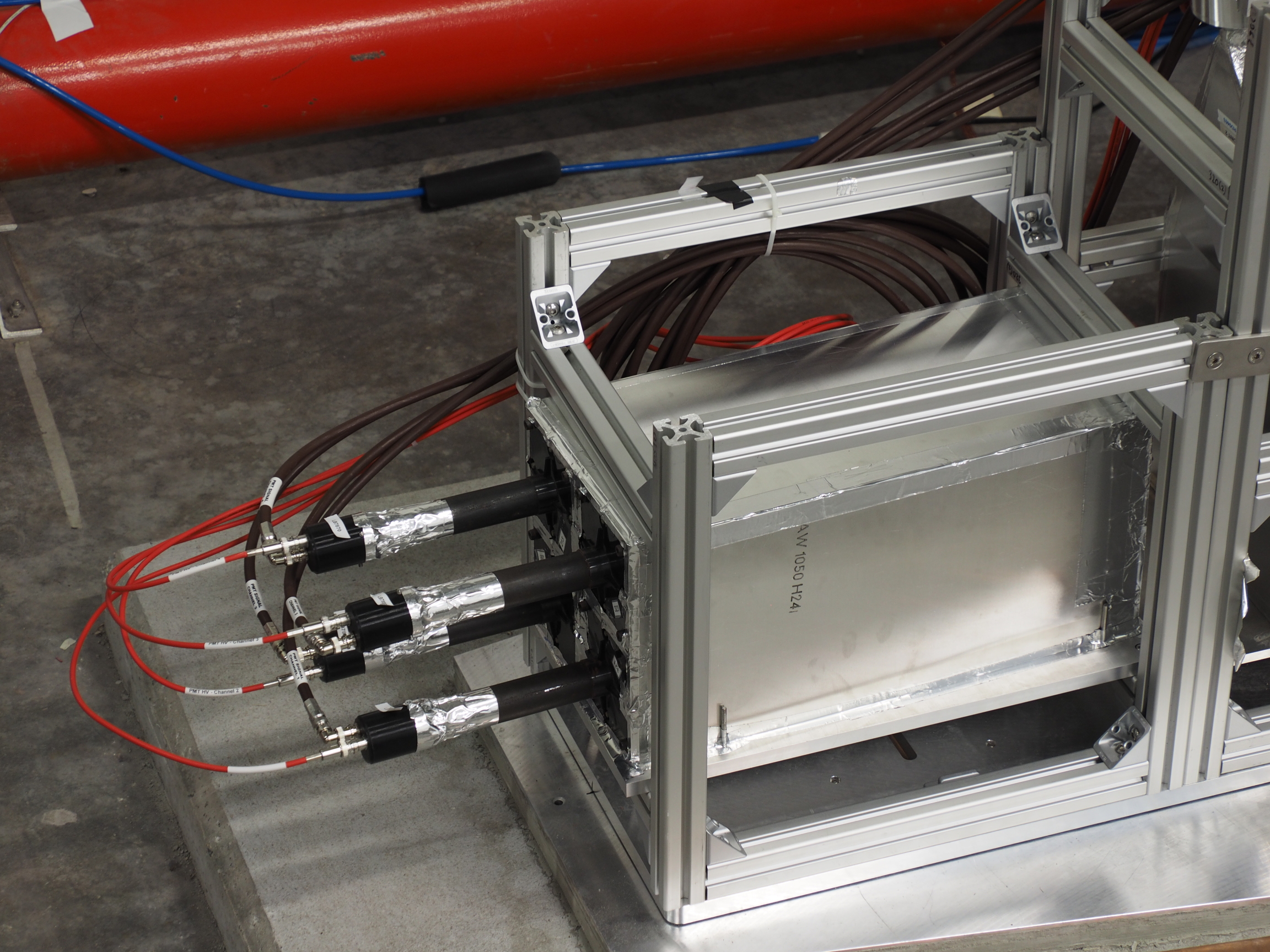}
    \caption{A photo of the calorimeter installed into FASER. (This figure is taken from Ref.~\cite{FASER:2022hcn}.)}
    \label{fig:CalorimeterPhoto}
\end{figure}

Each module is $12 \times 12$~cm$^2$ in the transverse plane and is made up of 66 interleaved lead and plastic scintillator layers with wavelength shifting fibers bringing the collected light to a single PMT at the back of the module. Fig.~\ref{fig:CalorimeterDrawing}, shows a sketch of a calorimeter module. Since there is a single PMT per module, the calorimeter provides no longitudinal segmentation, and has a very coarse transverse segmentation provided by the $2\times 2$ module configuration. The calorimeter can therefore not separate the two closely spaced electron showers expected from a dark photon decay, but instead measures the total electromagnetic energy in the event. As described in Section~\ref{sec:upgrades}, the calorimeter readout was upgraded at the start of 2024 running to give an improved dynamic range. 
Fig.~\ref{fig:CalorimeterPhoto} shows a photo of the FASER calorimeter, with the 4 PMTs protruding out of the back of the four modules.

\subsection{Trigger and Data Acquisition} 
The FASER detector is triggered by signals in any of the scintillator stations or the calorimeter. Before data taking, the expected trigger rate was about 650~Hz for the maximum expected luminosity, where triggers were assumed to be only from through-going muons. In reality the maximum trigger rate in most of Run 3 has been about roughly double this, with the increase mostly due to low-energy noise or beam-background triggers not accounted for before data taking. Here the beam-background is mostly low-energy particles from beam-losses in LHC beam-1 (incoming to IP1 from FASER), for which there is little shielding between the LHC beamline and FASER. The trigger and data acquisition system was designed with sufficient margin to be able to handle the larger than expected rate without problems.

\begin{figure}[ht]
    \centering
    \includegraphics[trim=1cm 0cm 9cm 0cm, clip=true, width=0.9\textwidth]{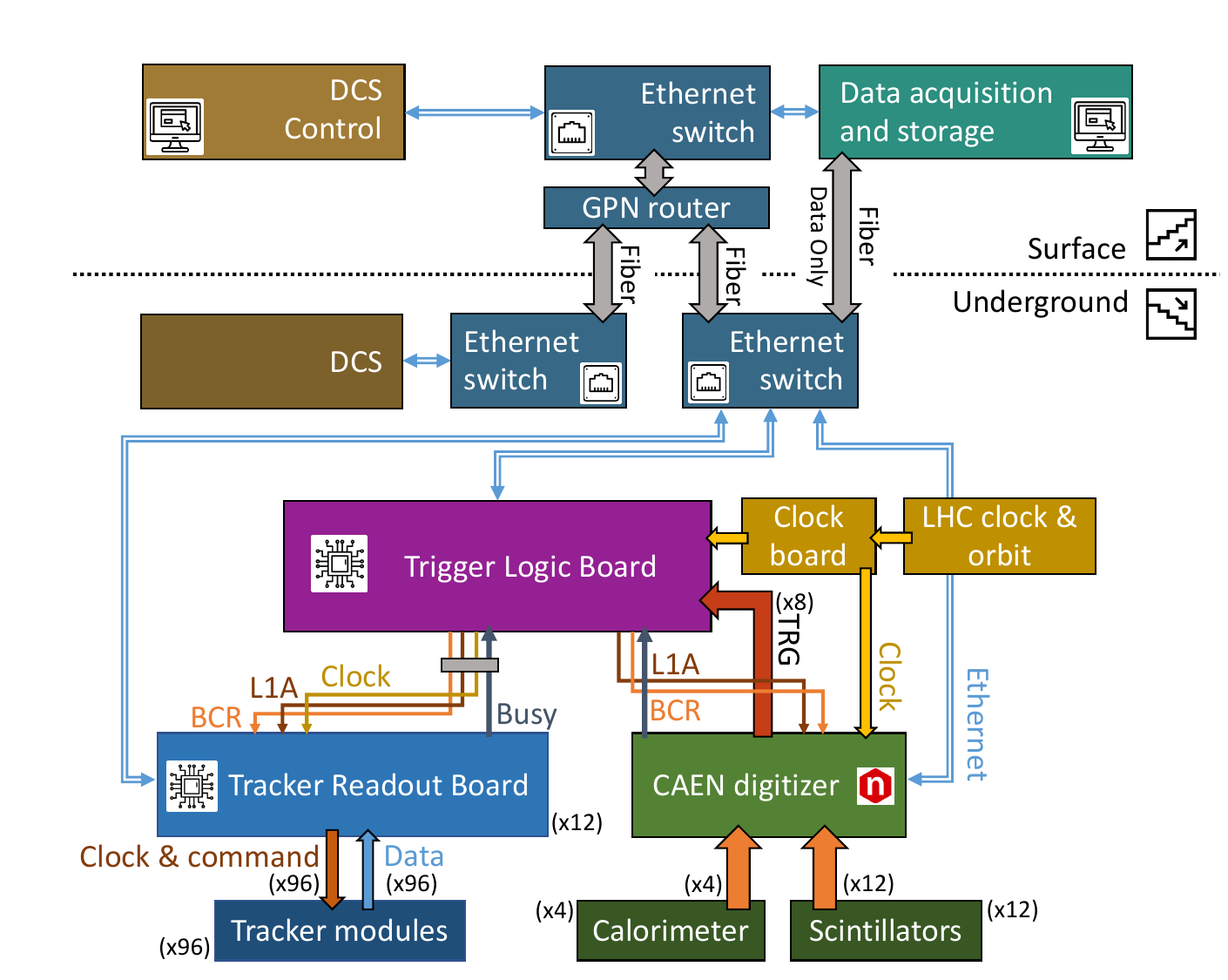}
    \caption{A simple schema of the FASER TDAQ architecture. The numbers in parentheses indicate the number of channels or lines. The blue double-line arrows indicate the connections via Ethernet. The grey thick arrow indicates fibers from the TI12 tunnel to the surface. The digitizer, clock board and TLB are housed in a VME crate. The TRBs are located on the detector. (This figure is taken from Ref.~\cite{FASERTDAQ:2021}.)}
    \label{fig:FASER_TDAQ_schema}
\end{figure}

A sketch of the system is shown in Fig.~\ref{fig:FASER_TDAQ_schema}. 
The raw calorimeter and scintillator counter PMT signals are sent to a commercial digitizer card, which issues trigger signals if the waveforms are above pre-defined thresholds. These digitizer trigger signals are sent to a custom FPGA-based Trigger Logic Board (called the TLB) which can combine inputs to form a global trigger (Level 1 Accept, L1A). The TLB can pre-scale and inhibit triggers and run monitoring algorithms. The L1A signal is distributed to the tracker readout boards (TRBs) and the digitizer, to initiate read out of the detector data. On receiving a L1A signal, the different readout boards send data over a dedicated fiber network to a software event builder DAQ process running on a server situated on the surface (about 600~m away from the detector). The system allows monitoring of the data by software processes running on the DAQ server, while the DAQ electronics are situated in the TI12 tunnel close to the FASER detector. 
The trigger and data acquisition (TDAQ) hardware operates on the LHC clock, which is received, processed and distributed via a dedicated electronics board. 

A detailed description of the FASER TDAQ system, and its commissioning can be found in Ref.~\cite{FASERTDAQ:2021}.

\subsection{Magnets} 

The FASER detector is built around three dipole magnets each with a 0.57~T field. Permanent magnets are used to minimise the required services (power supplies and cooling). Each magnet has a 20~cm diameter aperture, and an outer diameter of 43~cm. The first dipole (1.5~m-long) surrounds the decay volume, and the other two (each 1~m-long) are part of the tracking spectrometer. The three magnets are of identical design except for the length. The primary purpose of the magnets is to separate closely spaced charged-particles produced in the decay of boosted, light BSM particle decays, and to measure the charge of muons arising from neutrino interactions.

\begin{figure}[hbt!]
    \centering
    \includegraphics[width=0.64\textwidth]{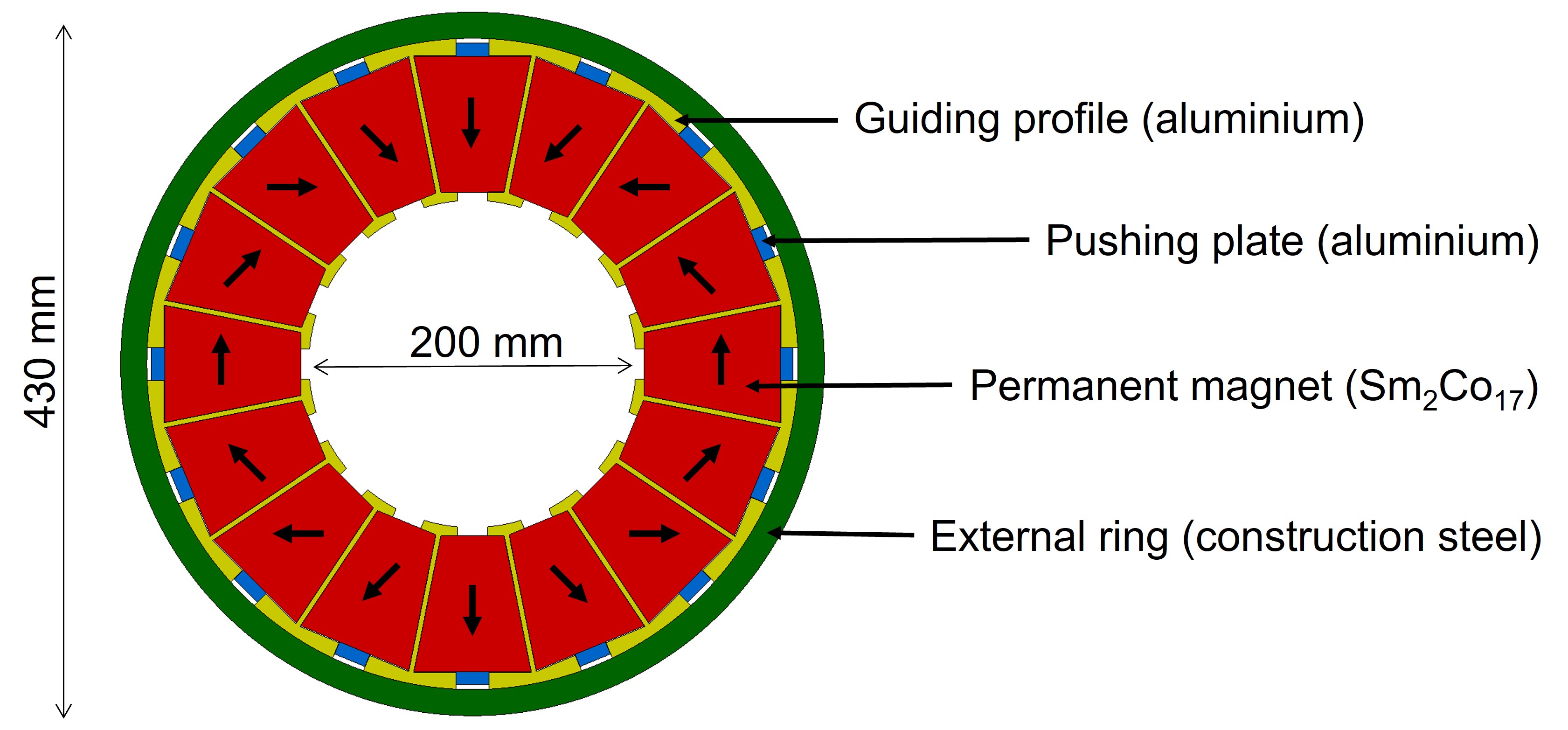}
    \includegraphics[width=0.35\textwidth]{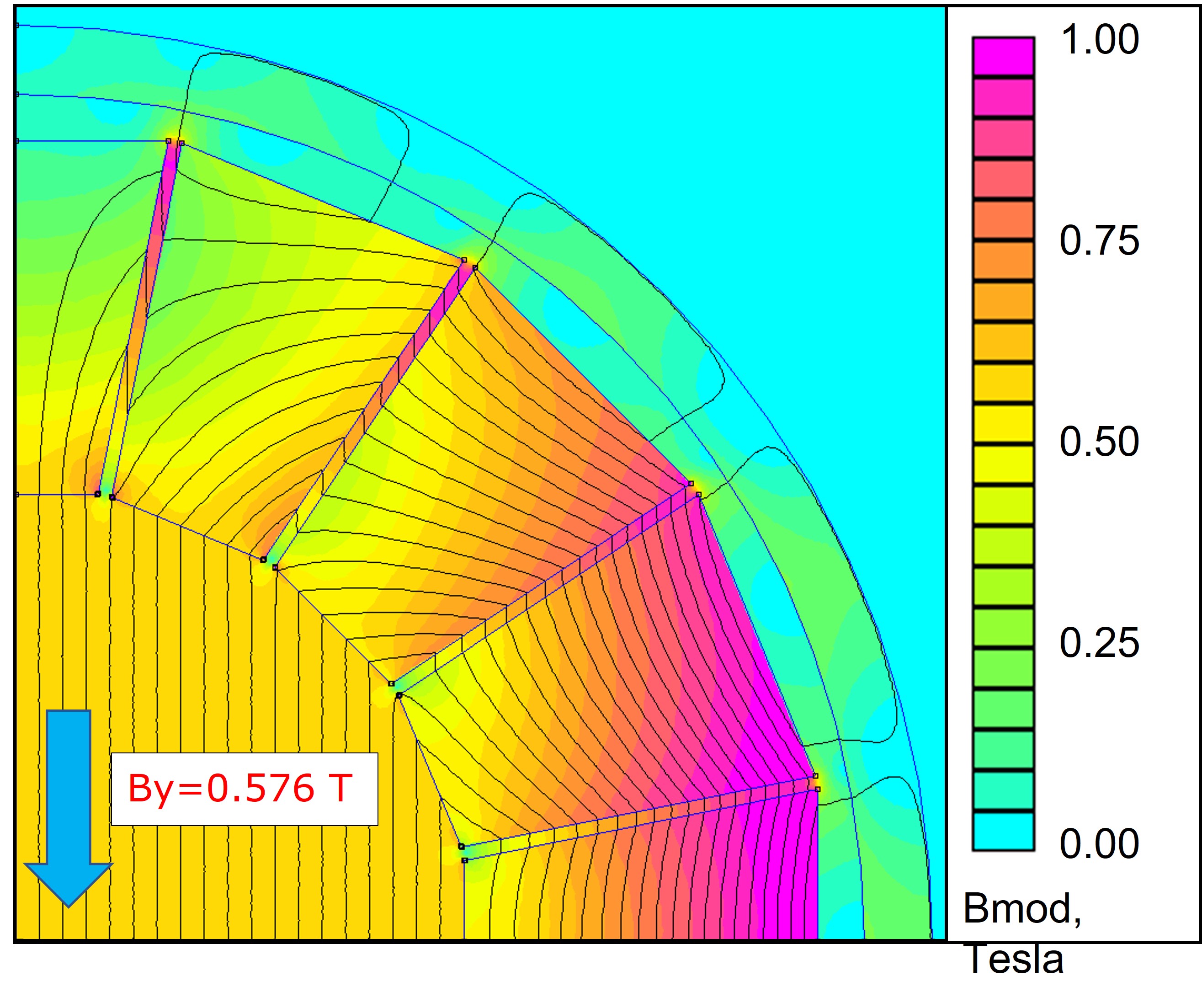}   
    \caption{(left) Cross section of one of the FASER dipole magnets. The arrows indicate the direction of the magnetic field in each of the permanent magnetic blocks. (right) 2D magnetic field distribution. (These figures are taken from Ref.~\cite{FASER:2022hcn}.)}
    \label{fig:mag}
\end{figure}

The magnets are based on a Halbach array design, incorporating several sections, each made up of 16 permanent magnet blocks arranged in a cylindrical array as shown in Fig.~\ref{fig:mag} (left). Each block has the axis of magnetisation defined such that the field in the central aperture is a pure dipole field, as can be seen in Fig.~\ref{fig:mag} (right). The magnets were designed,  constructed and had the field mapped by the CERN magnet group.

\subsection{Detector integration} 
\begin{figure}[hbt!]
    \centering
    \includegraphics
    [trim=3cm 3.5cm 4cm 3cm,clip,width=1.0\textwidth]{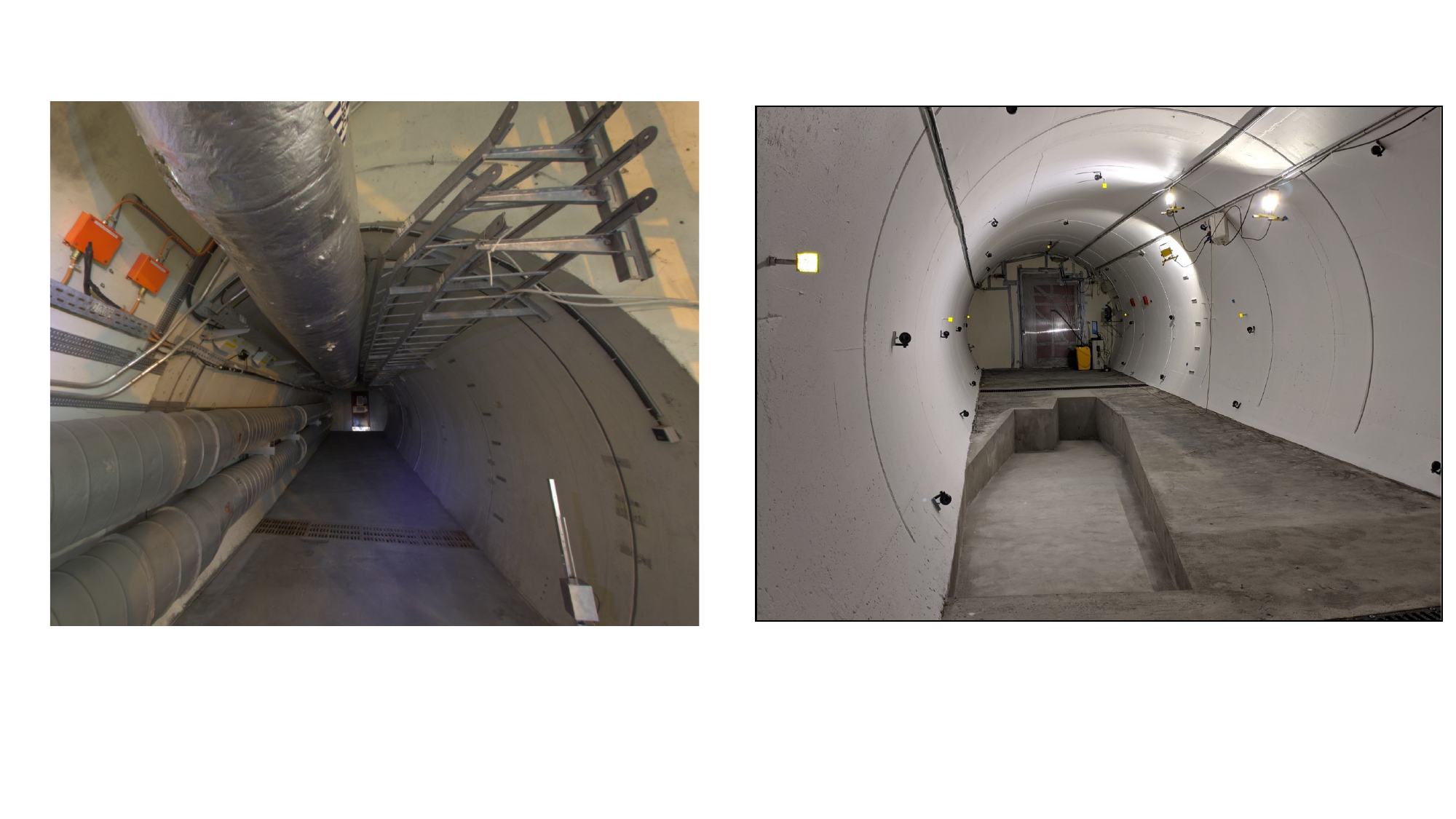}
    \caption{The TI12 tunnel (left) before the FASER preparation work, and (right) after the tunnel has been cleared out and the FASER trench dug in the tunnel floor. (This figure is taken from Ref.~\cite{FASER:2022hcn}.)}
    \label{fig:TI12beforeafter}
\end{figure}

In order to align the detector with the LOS a small trench was excavated in the TI12 tunnel. The main trench is 50~cm deep at the upstream end and 20~cm deep at the downstream end and is 5.5~m long and 1.4~m wide. At the front a narrower part of the trench extends by 1.2~m upstream of the main detector which is used to place the FASER$\nu$ emulsion detector, described later. The trench was designed by the CERN civil engineering group and the excavation work was carried out by a contractor in February - March 2020. Fig.~\ref{fig:TI12beforeafter} shows a photo of the TI12 tunnel in 2018 before any FASER work was carried out (left) and after the civil engineering work was completed (right).

\begin{figure}[hbt!]
    \centering
    \includegraphics[trim=4cm 1cm 4cm 3cm,clip,width=1.\textwidth]{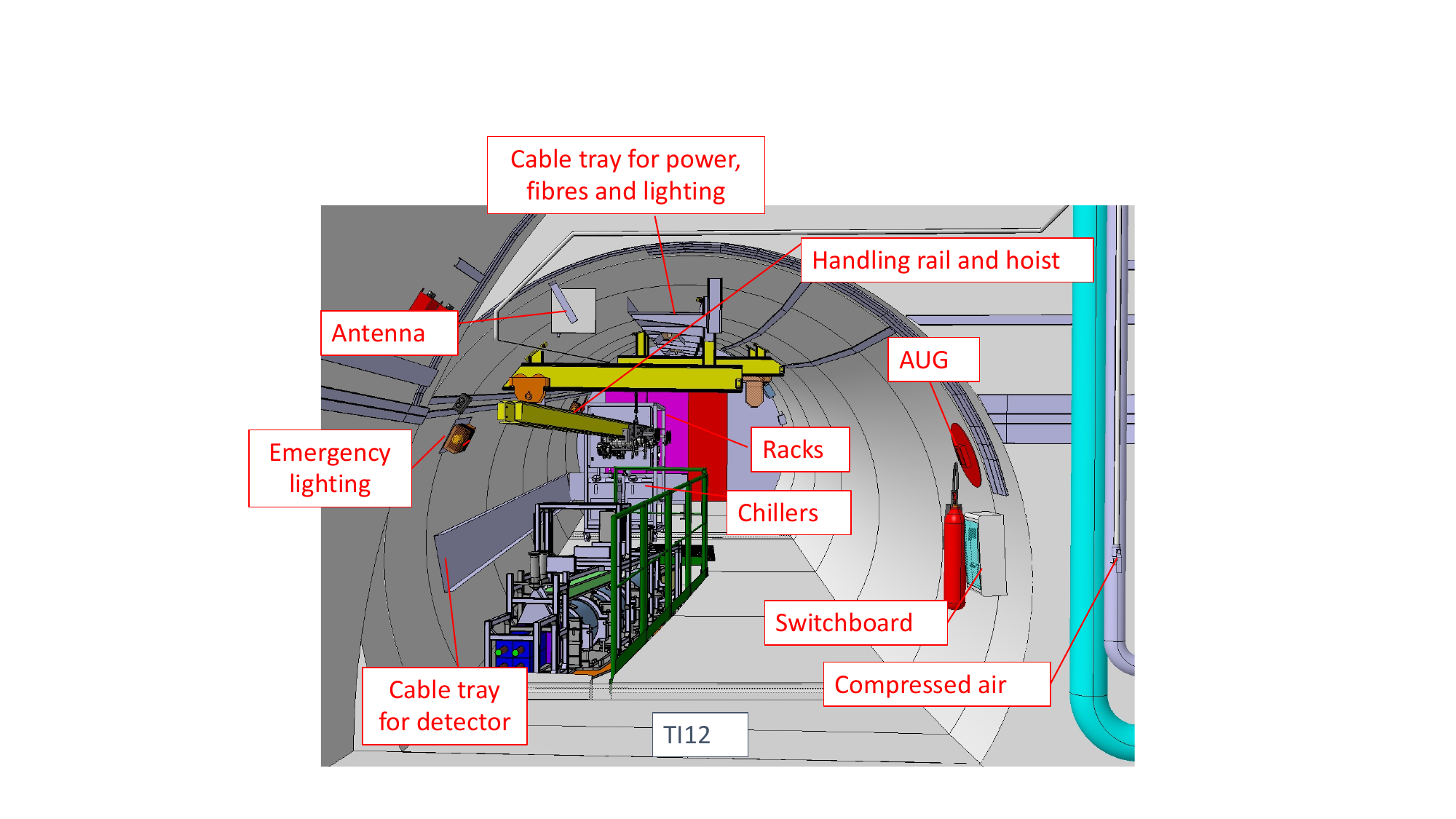}
    \caption{An integration picture of the TI12 tunnel showing the different infrastructure installed for FASER. (This figure is taken from Ref.~\cite{FASER:2022hcn}.)}
    \label{fig:TI12-infrastructure}
\end{figure}

Following the civil engineering work the detector services and technical infrastructure were installed, as summarized by the integration picture shown in Fig.~\ref{fig:TI12-infrastructure}.

The FASER detector is mechanically supported by two aluminium baseplates, one which is grouted to the bottom of the trench, and the other which sits on top of this and has the detector and magnets attached to it. The design allows the upper baseplate to be pushed over the lower baseplate to move the detector sideways, to correct for changes in the direction of the beam crossing angle at IP1, which could move the LOS by around 7~cm (such changes would only happen during technical stops at the end of each year). 

\begin{figure}[hbt!]
    \centering
    \includegraphics[width=0.9\textwidth]{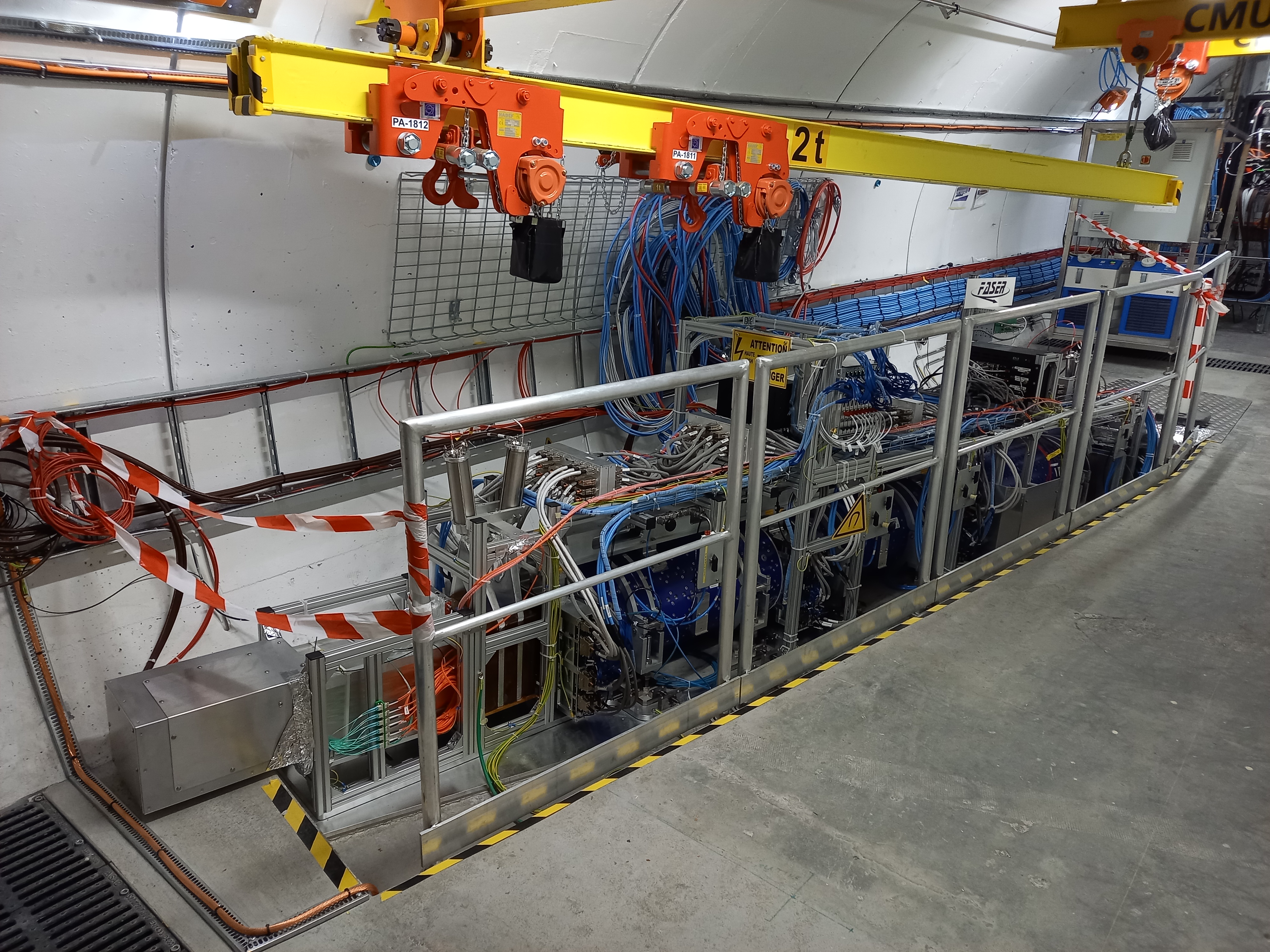}
    \caption{The FASER detector after installation in TI12. (This figure is taken from Ref.~\cite{FASER:2022hcn}.)}
    \label{fig:FASER-afterInstallation}
\end{figure}

The detector was installed into TI12 in March 2021, and a picture of the final detector can be seen in Fig.~\ref{fig:FASER-afterInstallation}. 
Following the installation, the magnets and tracker were surveyed by the CERN survey team. The results are used for both the global alignment of the detector with respect to the LOS, and to help constrain the internal alignment of the detector components. 

\subsection{FASER\texorpdfstring{$\nu$}{nu} emulsion detector} 
\begin{figure}[hbt!]
\centering
\includegraphics[width=0.7\textwidth]{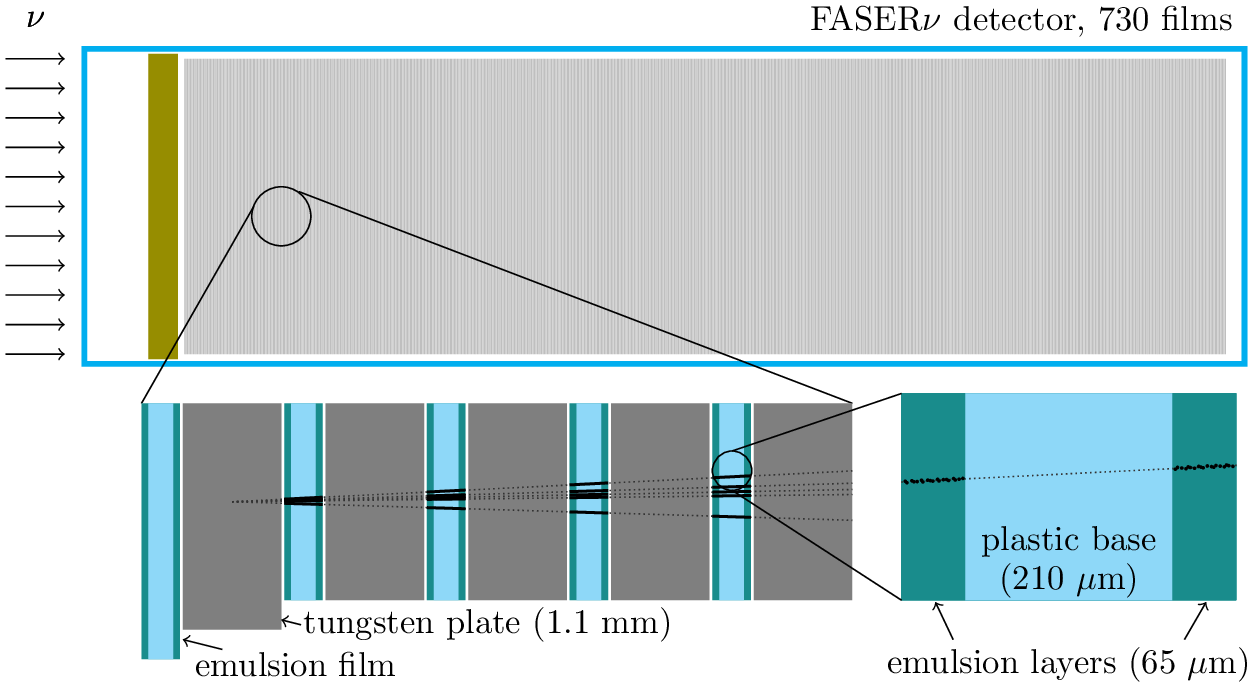}
\includegraphics[width=0.6\linewidth]{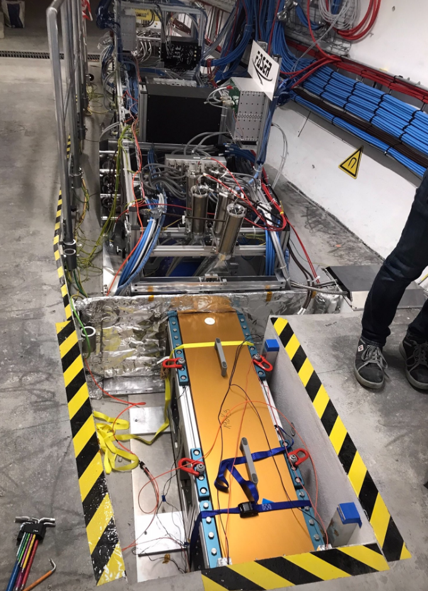}
\caption{ (top) A sketch showing the structure of the FASER$\nu$ emulsion/tungsten detector. (This figure is taken from Ref.~\cite{FASER:2025qaf}.) 
(bottom) A photograph of the FASER$\nu$ box installed in the FASER trench. The cover of the trench where FASER$\nu$ is placed was not installed when this picture was taken. (This figure is taken from Ref.~\cite{FASER:2022hcn}.)
}
\label{fig:FASERnu}
\end{figure}

The FASER$\nu$ emulsion detector is made up of 730 1.1-mm-thick tungsten plates interleaved with emulsion films, as shown in Fig.~\ref{fig:FASERnu} (top). The tungsten acts as the target for the neutrino to interact with, and the emulsion films record the trajectories of charged particles produced in the interaction with sub-micron position resolution and with an angular resolution at the level of 50 $\mu$rad. The detector has a transverse size of $25 \times 30$~cm$^2$ with a total target mass of 1.1 tonnes. Given that the emulsion is a passive detector, the detector needs to be extracted, and the emulsions developed and scanned before particle trajectories and corresponding neutrino interaction vertices can be reconstructed and used in data analysis. More information of the FASER$\nu$ operational workflow is given in Section~\ref{sec:operations}.  

The detector can be used to identify leptons produced in CC neutrino interactions. Muons are identified as long tracks penetrating through the up to eight interaction lengths of the detector. Electrons are identified by the EM shower they produce in the detector, and their energy can be estimated by the shower size. Finally, tau leptons are identified by observing the short $\tau$ track, with either a kink (for 1-prong $\tau$ decays) or by a secondary vertex (for hadronic $\tau$ decays). It is also possible to estimate the momentum of charged particle tracks by the effect on the measured trajectory due to multiple Coulomb scattering in the detector.

Since the detector records all charged particle trajectories passing through the emulsion films, it must be replaced before the track multiplicity becomes so high that the track and vertex reconstruction performance is significantly degraded. Given the exceptional track resolution, a multiplicity of less than about  $\mathcal{O}(10^6)$ tracks per cm$^2$ is acceptable, meaning that the detector needs to be replaced after exposure to 30~fb$^{-1}$ of collision data.  The operation can be done during scheduled technical stops of the LHC, which are typically two or three times per year and each lasting a few days. 

In order to obtain the best tracking performance the alignment stability between the emulsion films must be stable at the sub-micron level over the period that the FASER$\nu$ detector is installed (up to two months).  The alignment is fixed by vacuum packing submodules each containing ten emulsion films and tungsten plates, and by then applying a strong pressure between the installed submodules using a screw pusher system. In order to mitigate possible misalignments caused by temperature variations during the exposure time the FASER$\nu$ detector is kept at a constant temperature (typically within 0.1 degrees) using an active cooling system. The system uses the 15 degree water provided by the cooling unit used for the tracker cooling, running through a heat exchanger equipped with a set of fans to keep a constant temperature, with a feedback system controlling the fan speeds. The FASER$\nu$ trench has thermal insulation installed around it to improve the temperature stability.

A picture of the FASER$\nu$ detector installed in the trench in front of the main FASER detector can be seen in the bottom panel of Fig.~\ref{fig:FASERnu}.

    \clearpage
    \newpage
    \section{Detector Operations and Upgrades}\label{sec:operations}
    \subsection{FASER electronic detector operations}
FASER has a light-weight operational model, which has proven to work well since the start of data taking. The experiment does not have a control room, but is controlled and monitored remotely by shifters. 
During data taking periods the experiment has two weekly shifters, the Run Manager and the Monitoring shifter. 
The Run Manager is responsible for the operations, whose tasks include  controlling the data taking via the run control GUI and the detector state using the DCS Finite State Machine (implemented in WinCC), as well as organizing any interventions on the detector and chairing the weekly Operations Meeting. During regular physics data taking the system automatically stops and starts runs based on the LHC fill number and no intervention is needed. During periods with no beam the Run Manager can initiate calibrations of the different detector systems, which is typically done once every two weeks.
The Monitoring Shifter's job is to monitor both the DCS detector data (temperatures, voltages etc..) and the detector data. This is typically checked a few times per day, using Grafana monitoring dash boards. The system includes many automated alerts based on both DCS and detector data which allow problems to be caught quickly without a large shift crew.

The model works very well, with only a small number of operational incidents during the operations to date (end of 2025 running). Rare issues which have occurred, and have been quickly spotted, include a crash of the digitizer, a failure of the cooling unit, and running out of disk space for the raw detector data.

\begin{figure}[h]
    \centering
    \includegraphics[width=0.52\textwidth]{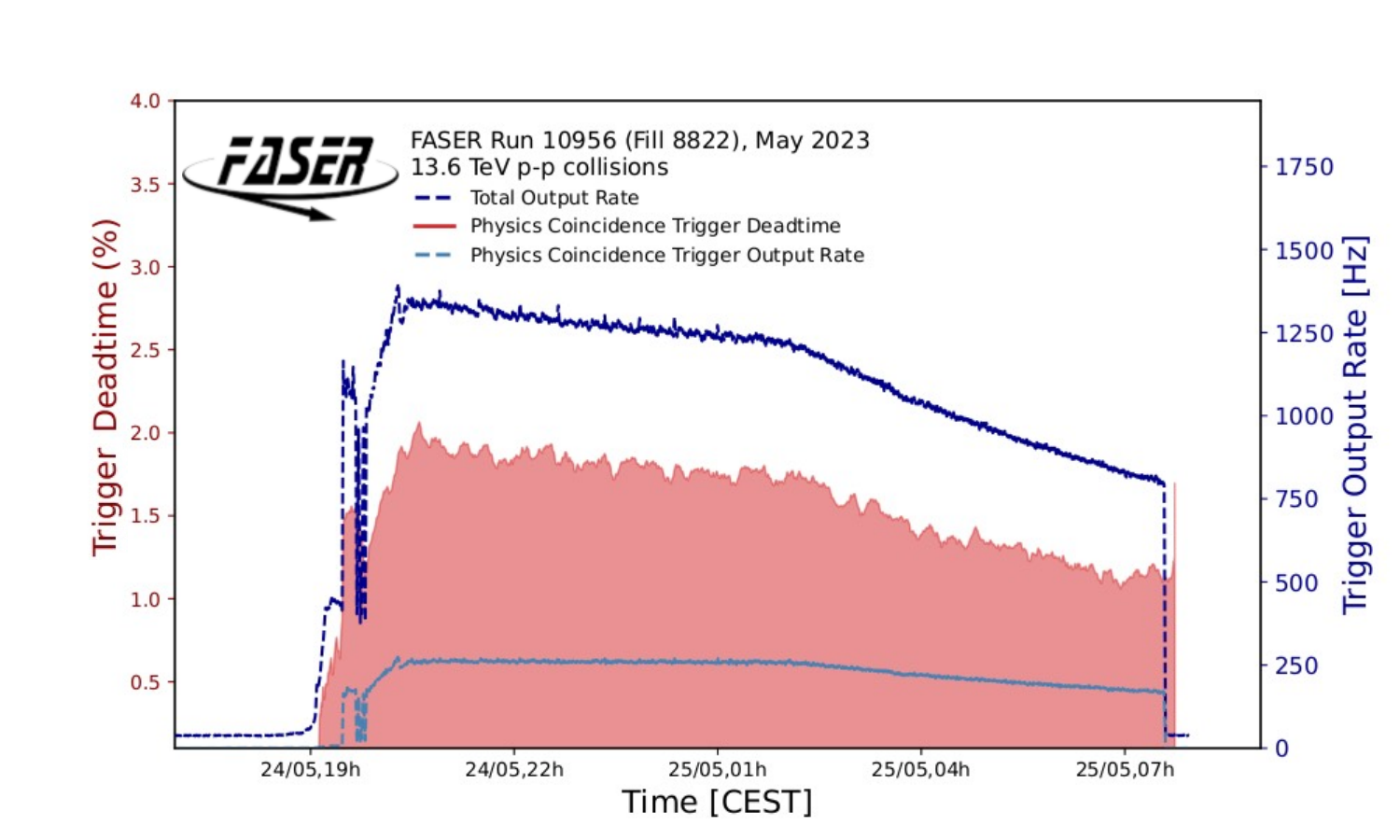}
    \includegraphics[width=0.47\textwidth]{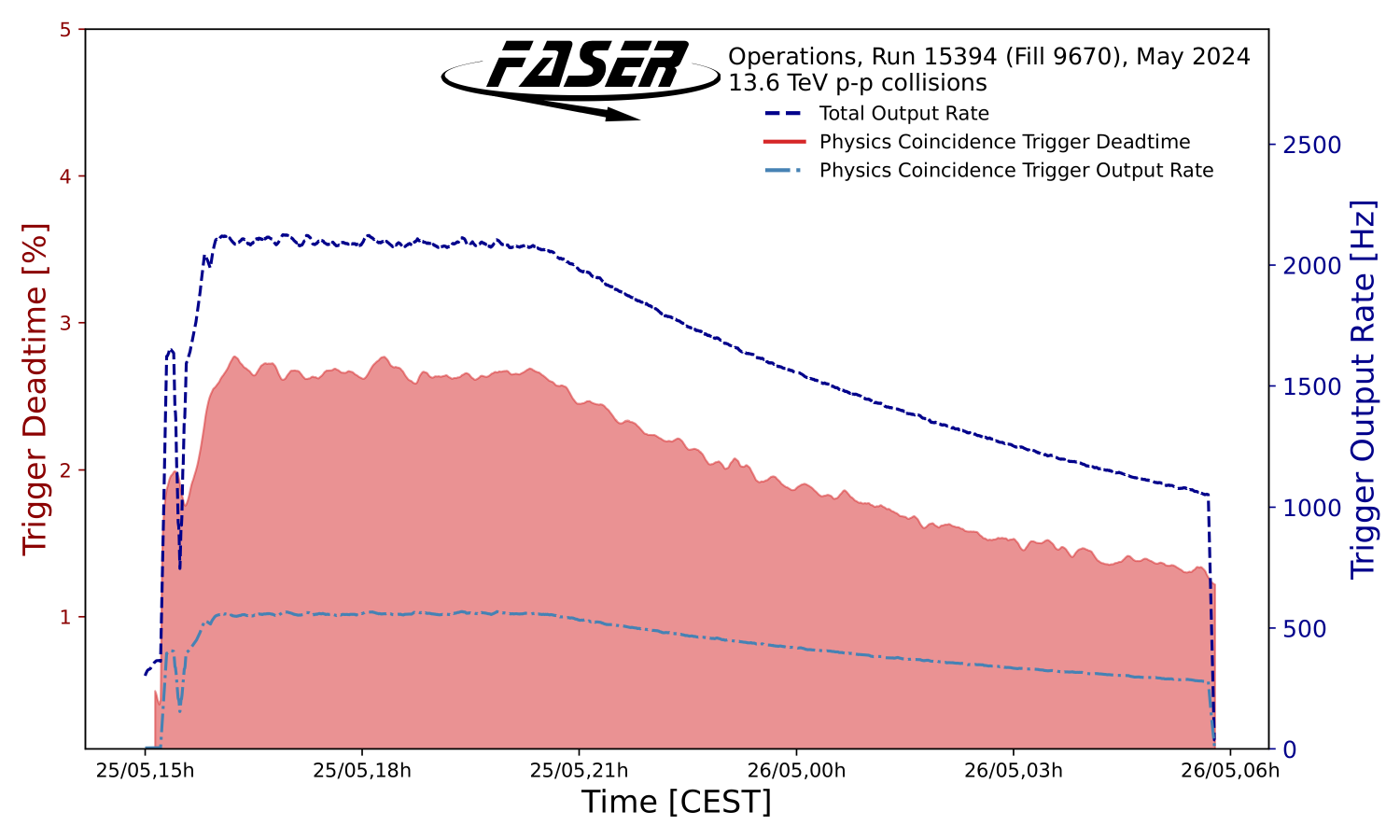}
    \caption{ The trigger rate and deadtime during a typical LHC fill in 2023 (left) and 2024 (right). }
    \label{fig:FASER-trgOps}
\end{figure}

The typical trigger rate at maximum luminosity (of around 2 $\times$ 10$^{34}$ cm$^{-2}$ s$^{-1}$) was about 1.3~kHz in 2022,2023 and 2025 running. In 2024 the LHC beam optics were changed which led to a significant increase in the background rate, and a maximum trigger rate of more than 2~kHz was observed, the optics were reverted back for 2025 running. With these conditions the trigger system induces a maximum deadtime of around 2\% (3\% in 2024), as shown in Fig.~\ref{fig:FASER-trgOps}. 

\begin{figure}[h]
    \centering
    \includegraphics[width=0.98\textwidth]{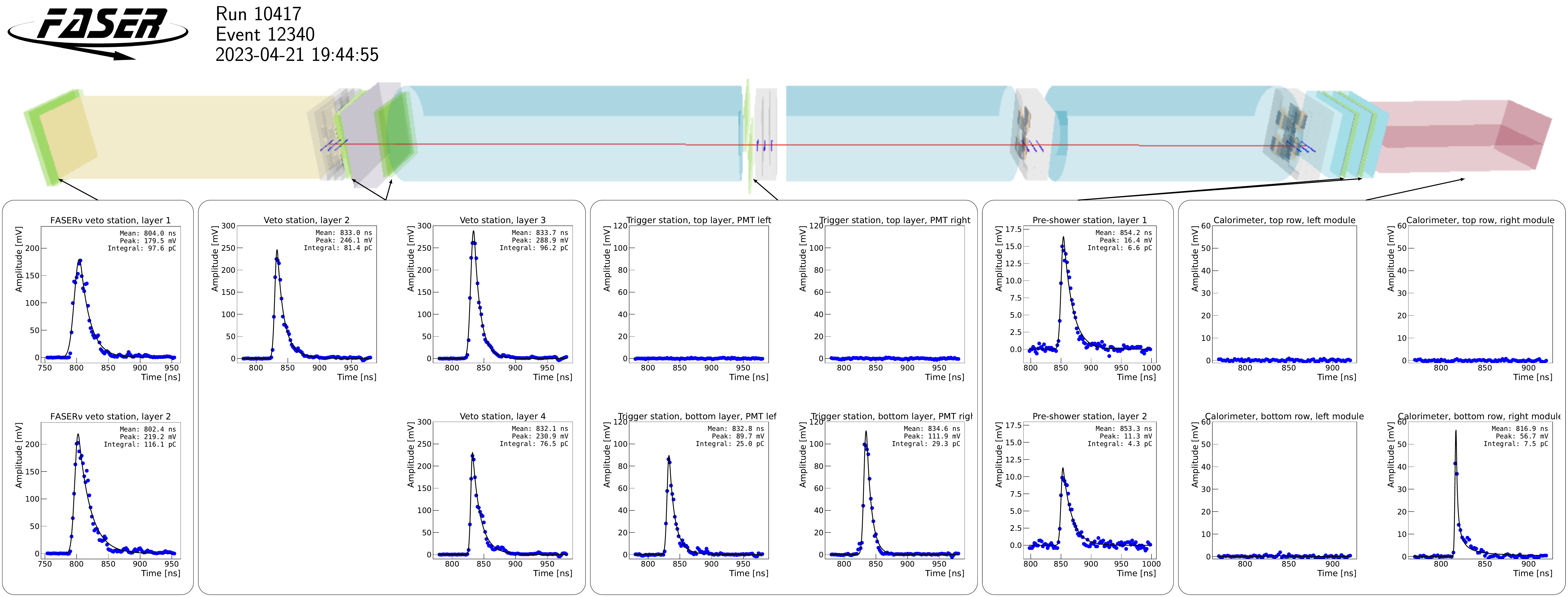}
    \caption{ An event display of a typical event in FASER, showing a muon track traversing the detector from left to right. The bottom part of the figure shows the PMT waveforms from the different scintillator and calorimeter modules. }
    \label{fig:EVD}
\end{figure}

An event display of a typical triggered event is show in Fig.~\ref{fig:EVD}, showing a muon traversing the detector. The display shows the strips that field in the tracking stations and the reconstructed track. In addition the PMT waveforms for the different scintillators and calorimeter modules are shown. These show a clear signal of a minimum-ionising-particle (MIP) passing through the full detector. In the calorimeter one of the four calorimeter module shows a small waveform peak corresponding to the expected energy deposited by a MIP.

During the LHC cycle the TCL6 collimators, situated 220~m from the IP on the outgoing beam, are inserted to a position very close to the beam. These collimators are designed to reduce radiation problems from forward collision debris on the downstream LHC components. In both 2022 and 2023 there was one LHC fill in which these collimators were not inserted due to technical problems, and this had a significant impact on the FASER trigger rate. For these two fills the trigger rate was about two times higher than in other fills, and analysis of the background tracks showed a large increase in negatively charged tracks with an angle in the range of -5~mrad $<$ Tan $\theta_X$ $<$ -1~mrad~\cite{muonPaper}.

\begin{figure}[h]
    \centering
    \includegraphics[width=0.75\textwidth]{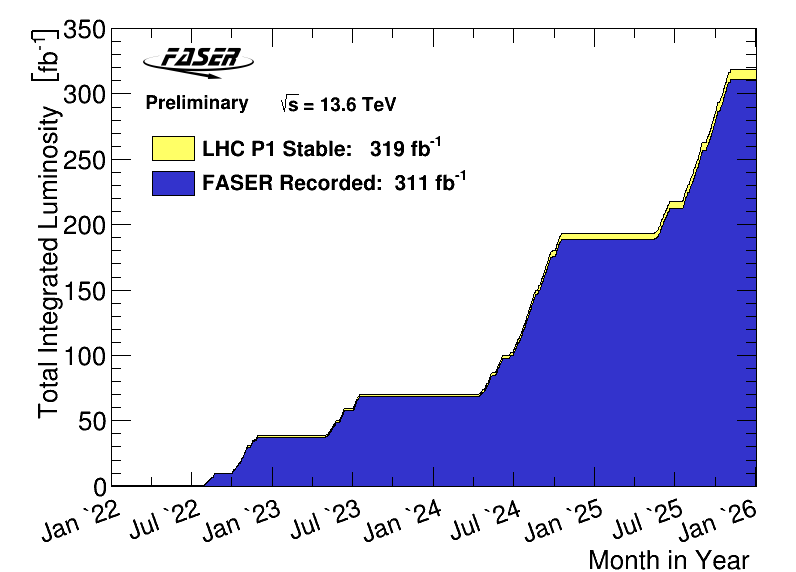}
    \caption{ The delivered (yellow) and recorded (blue) luminosity as a function of time from 2022 - to the end of 2025 running.}
    \label{fig:FASER-operations}
\end{figure}

FASER uses the luminosity measurement from the ATLAS experiment~\cite{ATL-DAPR-PUB-2024-001} with an uncertainty of 1.9\%.  
The delivered and recorded luminosity as a function of time is shown in Fig.~\ref{fig:FASER-operations}, highlighting that 97\% of the delivered luminosity has been recorded by FASER. A total luminosity of 311~fb$^{-1}$ had been recorded by FASER by the end of the 2025 run.

For the first 9.5~fb$^{-1}$ delivered to FASER in 2022, the calorimeter was deliberately set up to have good sensitivity to small signals from muons (MIPs), meaning the calorimeter readout would saturate for high energy signals above a few hundred GeV. For this reason this data was not used in some physics analyses (those which relied on having sensitivity to high energy calorimeter deposits). An upgrade to the calorimeter readout was installed at the end of the 2023 data taking to improve the dynamic range of the readout as discussed in the next section.

\begin{table}[thb]
  \centering
  \begin{tabular}{|l|c|c|c|}
  \hline
    {\bf Year} &  {\bf Dates} & {\bf Luminosity} \\
  \hline
    2022 & 5/7/22 - 28/11/22 & 38~fb$^{-1}$\\
    2023 & 21/4/23 - 17/7/23 & 31~fb$^{-1}$\\
    2024 & 7/3/24 - 17/10/24 & 125~fb$^{-1}$\\
    2025 &  22/4/25 - 31/11/25 & 125~fb$^{-1}$ \\
        \hline
        {\bf Total} & & {\bf 319~fb$^{-1}$} \\
\hline
    \end{tabular}
    \caption{Summary of the delivered luminosity to FASER between 2022 - 2025. 
    }
    \label{tab:FASER-lumi}
\end{table}

\subsection{FASER\texorpdfstring{$\nu$}{nu} operations}
As discussed in Section~\ref{sec:detector} the FASER$\nu$ detector needs to be periodically replaced to keep the occupancy at a manageable level. For most of the years in Run 3 the maximum luminosity the detector could be exposed to was around 30~fb$^{-1}$, however in 2024 running the LHC changed to a different beam optics (with reversed polarity of the final focusing quadrupole magnets in IP1) which led to a significantly larger background rate that required the FASER$\nu$ detector to replaced every 10~\fb$^{-1}$. 
Due to limited resources only three FASER$\nu$ detectors could be installed in 2024, and for the rest of the time a FASER$\nu$ box with tungsten but no emulsion was installed. Table~\ref{tab:FASERnu-lumi} shows the different FASER$\nu$ detectors installed to date (up to the end of 2025 running), showing that the total luminosity recorded with emulsion installed was about 200~fb$^{-1}$ compared to the 319~\fb$^{-1}$ delivered to IP1 and recorded by the electronic FASER detector.

\begin{figure}[h]
    \centering
    \includegraphics[width=.55\textwidth]{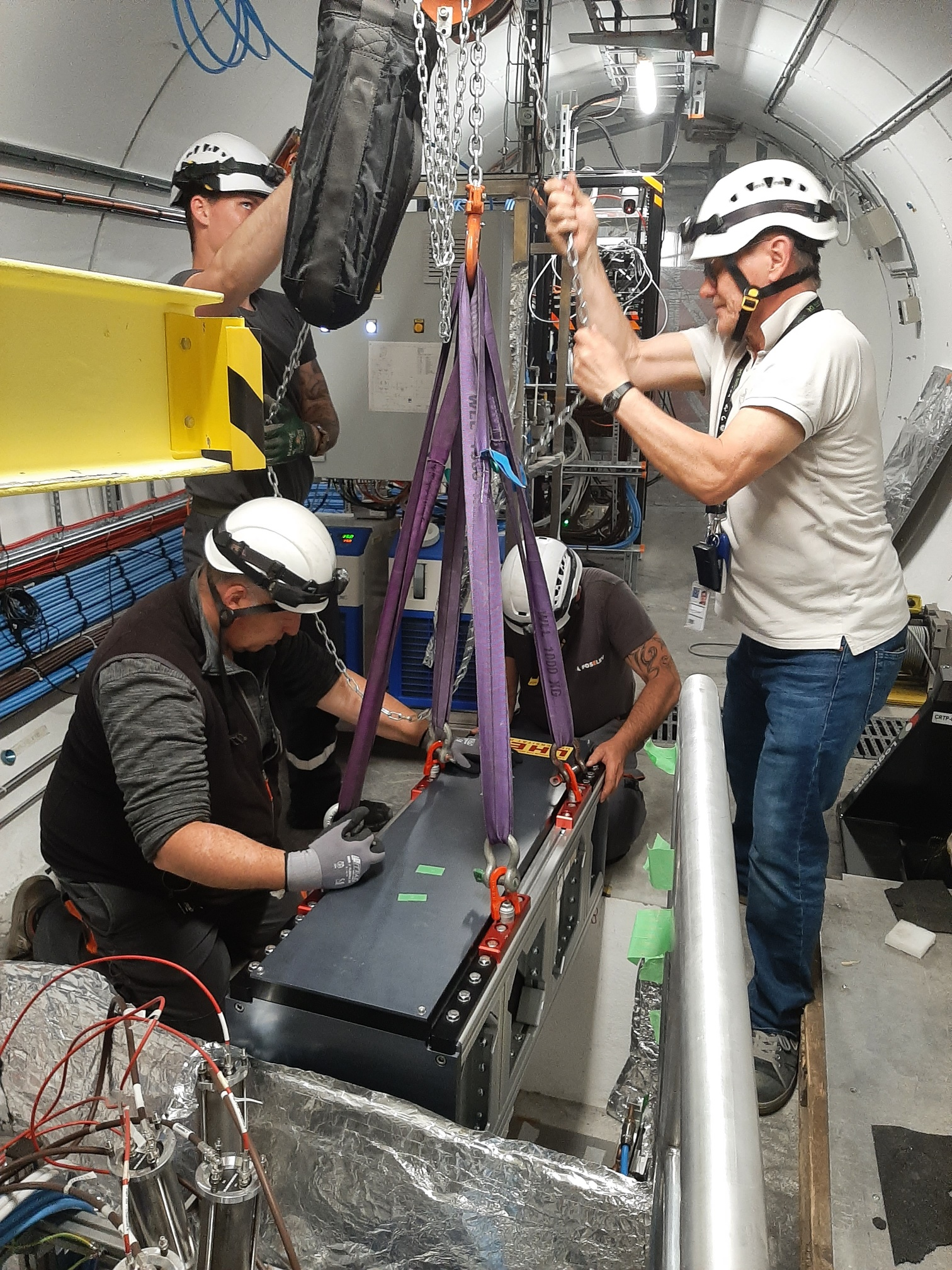}
    \caption{ A photograph of the FASER$\nu$ detector being installed into the trench using the dedicated manual crane, during the installation of the F242 detector in May 2024.  }
    \label{fig:FASERnu-exchange}
\end{figure}

The exchange of a FASER$\nu$ detector is usually planned during scheduled Technical Stops of the LHC, but in some cases the exchange had to be done during dedicated LHC accesses. The procedure takes about 3 hours and requires the CERN transport team to lift the incoming and outgoing FASER$\nu$ detectors in and out of the FASER$\nu$ trench and over the LHC. Figure~\ref{fig:FASERnu-exchange} shows the F241 detector being lowered into the trench using the dedicated manual crane in May 2024.

\begin{figure}[h]
    \centering
    \includegraphics[width=1.\textwidth]{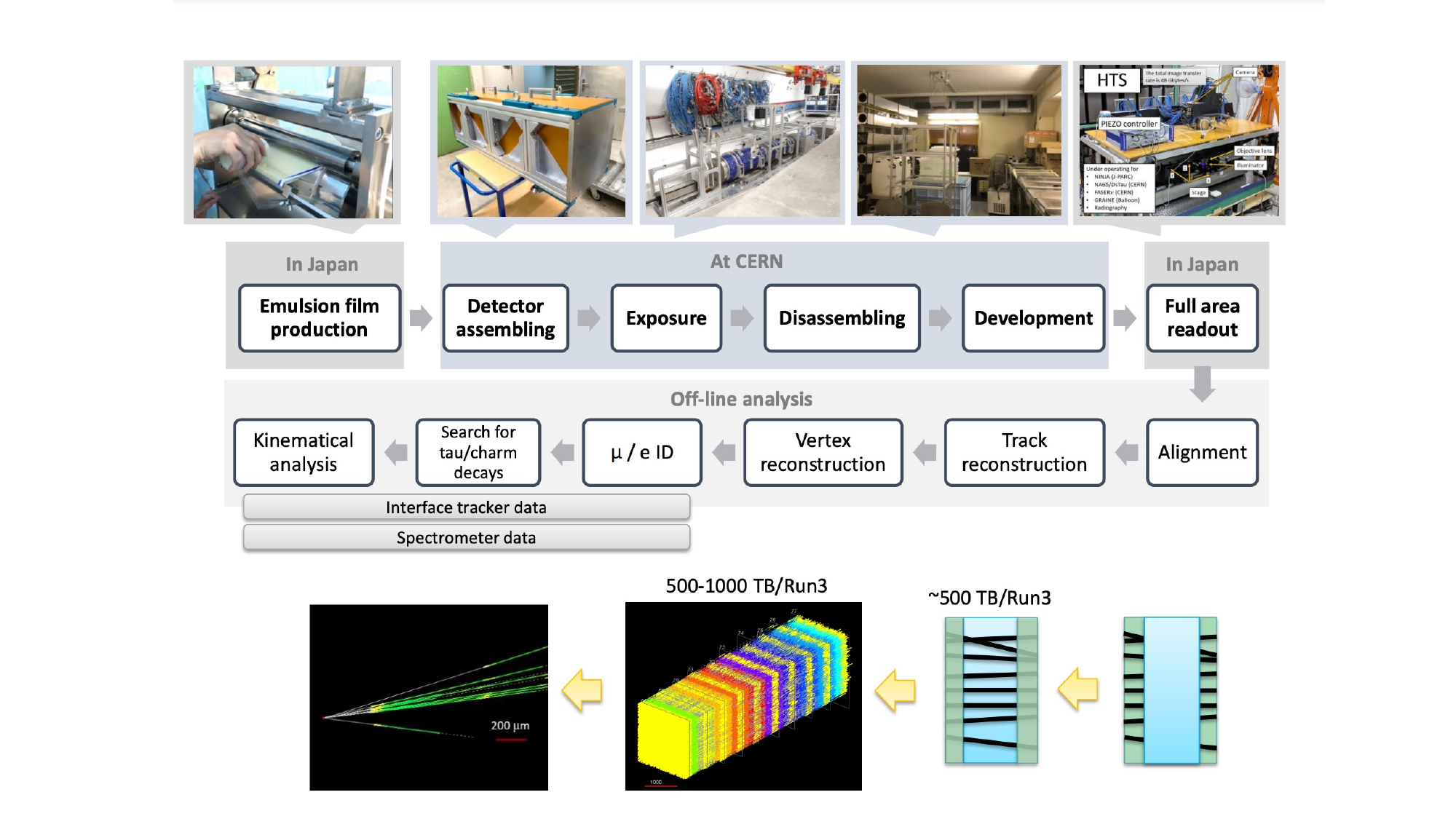}
    \caption{ A schematic of the FASER$\nu$ emulsion workflow, including the operational activities carried out in Japan and at CERN, as well as the conceptual steps of the offline data analysis.  }
    \label{fig:FASERnu-workflow}
\end{figure}

\begin{figure}[hbt!]
\begin{center}
\includegraphics[height=8cm]{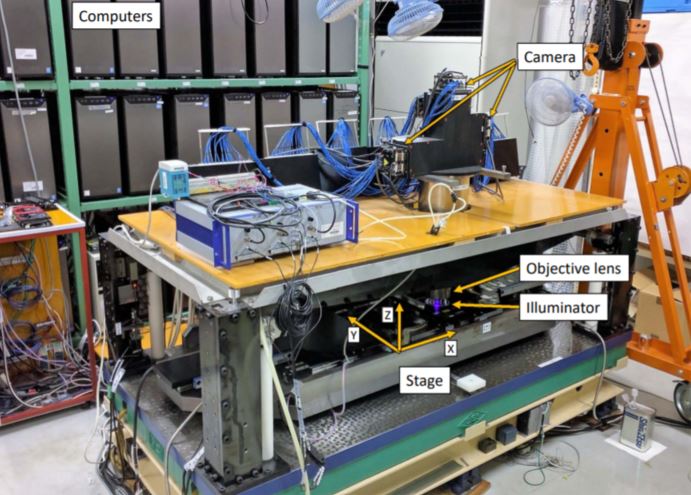}
\caption{The fast emulsion readout system HTS~\cite{Yoshimoto:2017ufm}, with a readout speed of 0.45~m$^2$/hour/layer. (This figure is taken from Ref.~\cite{FASER:2022hcn}.)
}
\label{fig:hts}
\end{center}
\end{figure}

The FASER$\nu$ operational workflow has several steps both before and after the exposure, as shown in Fig.~\ref{fig:FASERnu-workflow}. The pre-installation  workflow includes production of the emulsion gel and films in Japan and the assembly of the FASER$\nu$ detector in the dark room at CERN. After the detector has been removed from FASER following exposure to the beam, it is disassembled and the emulsion films developed in the dark room at CERN (this takes about 2 weeks), before the developed films are sent to Japan for scanning. The scanning process uses the HTS~\cite{Yoshimoto:2017ufm} (shown in Fig.~\ref{fig:hts}) and HTS-2 systems to digitize the track segment data from each film. The scanning process is slow and it takes 30 - 65 minutes to scan each film, depending on the scanning system and the film occupancy. The digitized data is further processed with several alignment processes and track reconstruction  as described in Ref.~\cite{FASER:2025qaf}.

\begin{table}[thb]
  \centering
  \begin{tabular}{|l|c|c|c|c|}
  \hline
    {\bf Year} & {\bf Name} & {\bf Dates} & {\bf Luminosity} & {\bf Fraction of films scanned} \\
  \hline
    2022 & F222 & 6/7 – 14/9 & 9.5~fb$^{-1}$ & 100\% \\
     & F223 & 14/9 – 29/11  & 28.6~fb$^{-1}$ & 100\% \\
  \hline
      2023 & F231 & 20/3 – 20/6 & 21.2~fb$^{-1}$ & 100\% \\
     & F232 & 20/6 – 3/8  & 10.7~fb$^{-1}$ & 100\% \\
       \hline   
      2024 & F241 & 20/3 – 6/5 & 11.6~fb$^{-1}$ & 0\%  \\
     & F242 & 12/6 – 8/7 & 9.9~fb$^{-1}$ & 0\% \\
    & F243 & 4/10 – 22/10  & 11.9~fb$^{-1}$ & 0\% \\
       \hline   
      2025 & F251 & 27/3 – 24/6 & 24.2~fb$^{-1}$ & 70\% \\
     & F252 & 24/6 - 15/8  & 25.6~fb$^{-1}$ & 0\% \\
    & F253 & 5/9 - 1/10 & 22.7~fb$^{-1}$ & 0\% \\
        & F254 & 1/10 - 30/10 & 28.5~fb$^{-1}$ & 0\% \\    
        \hline
        {\bf Total} & & & {\bf 204.4~fb$^{-1}$} & \\
\hline
    \end{tabular}
    \caption{Summary of the different FASER$\nu$ detectors installed into FASER up to the end of 2025 running. The table also shows for each detector the dates it was installed for, the luminosity it was exposed to, and the fraction of the data that has been scanned to date. 
    }
    \label{tab:FASERnu-lumi}
\end{table}

\clearpage
\subsection{Detector upgrades}
\label{sec:upgrades}

{\bf Calorimeter readout upgrade }

\begin{figure}[h]
    \centering
    \includegraphics[width=.44\textwidth]{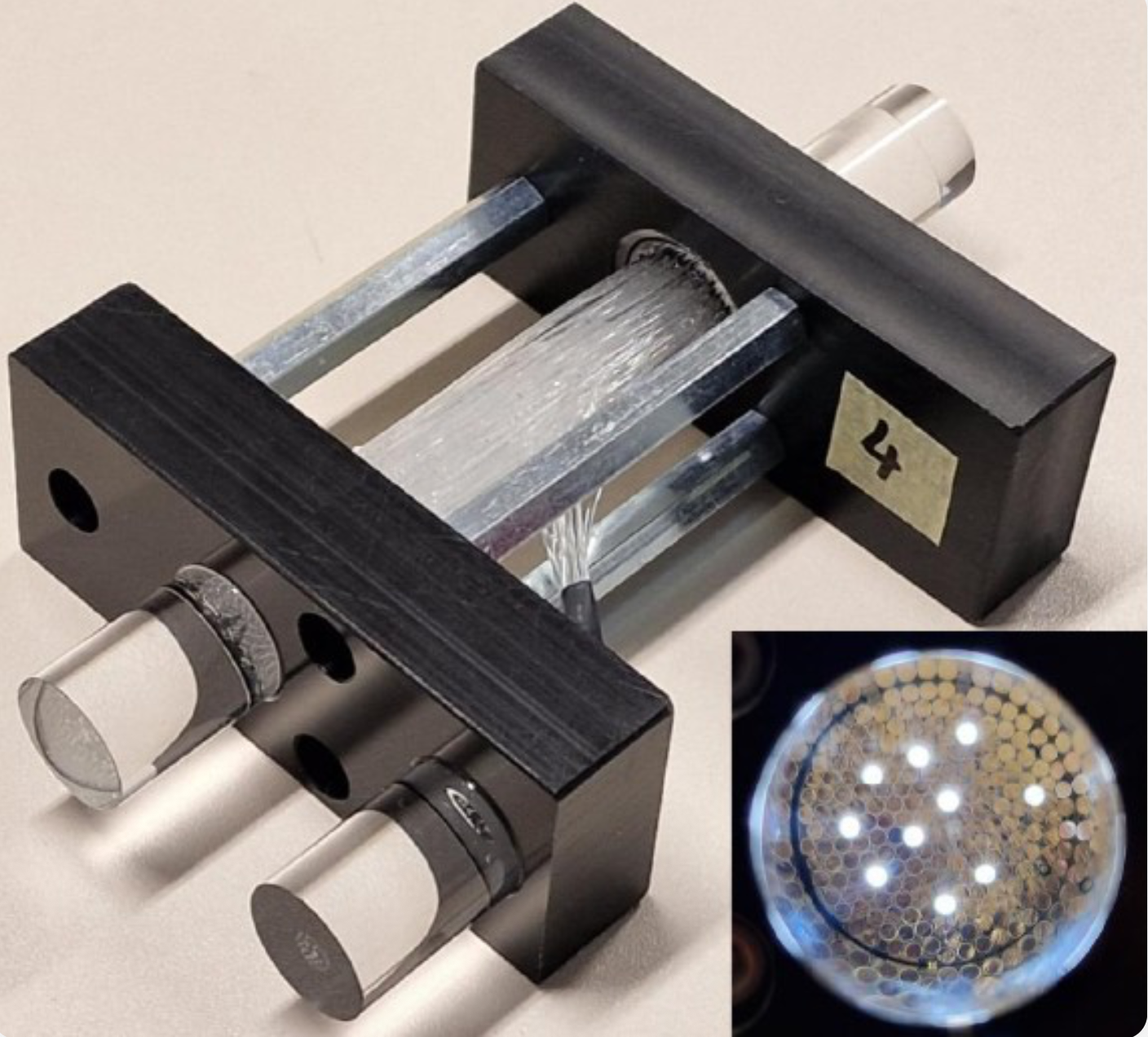}     
    \includegraphics[width=.54\textwidth]{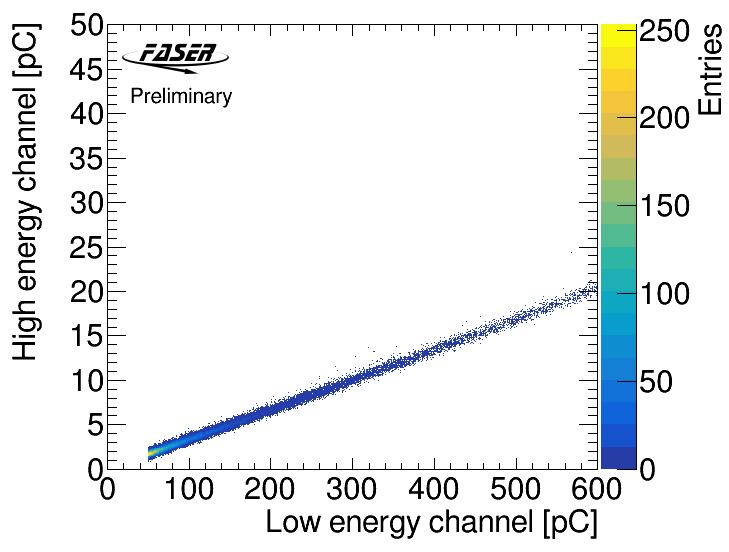}
    \caption{ (left) The optical splitter used to distribute the light between the HE and LE readout PMTs. (right) The charge readout by the HE channel versus the LE channel for one module of the calorimeter, showing a strong correlation for particles in the  energy range where the two readout channels overlap. (These figures are taken from Ref.~\cite{Jakobsen:2025fpf}.)}
    \label{fig:calo-upgrade}
\end{figure}

In 2022 and 2023 running each calorimeter module was readout by a single PMT. This system does not provide sufficient dynamic range for the calorimeter to be able to measure TeV electromagnetic energy deposits, as can be present in the BSM signatures relevant for FASER, while at the same time being sensitive to the 330~MeV energy deposited by MIPs traversing the calorimeter modules. The calorimeter was therefore configured to be sensitive to high energy signals and could not use MIP signals for calibration and performance studies.  In the technical stop after the 2023 run, the calorimeter readout was upgraded to improve this situation. The light produced in the calorimeter was split between two PMTs with 97\% going to the low-energy (LE) readout PMT and the remaining 3\% to the high-energy (HE) readout PMT. The splitting was done by using a bundle of optical fibers with the relevant fraction of fibers going to each of the PMTs, as can be seen in Fig~\ref{fig:calo-upgrade} left. The light ratio and PMT gains were chosen so that the LE channels have sensitivity to energy deposits in the range 0.1 $<$ E $<$ 100 GeV whereas the HE channels cover the range 10 $<$ E $<$ 3000 GeV, such that the range 10 $<$ E $<$ 100 GeV can be used to cross calibrate the HE readout using the LE data. The system is described in more detail in Ref.~\cite{Jakobsen:2025fpf}.
The calorimeter readout upgrade has performed as expected and data can be used to show the correlation between the energy measurements in the LE and HE channels (Fig~\ref{fig:calo-upgrade} right), the expected saturation of the LE channels at above 100 GeV is confirmed with studies of the data.

{\bf Preshower upgrade }

\begin{figure}[h]
    \centering
    \includegraphics[width=.9\textwidth]{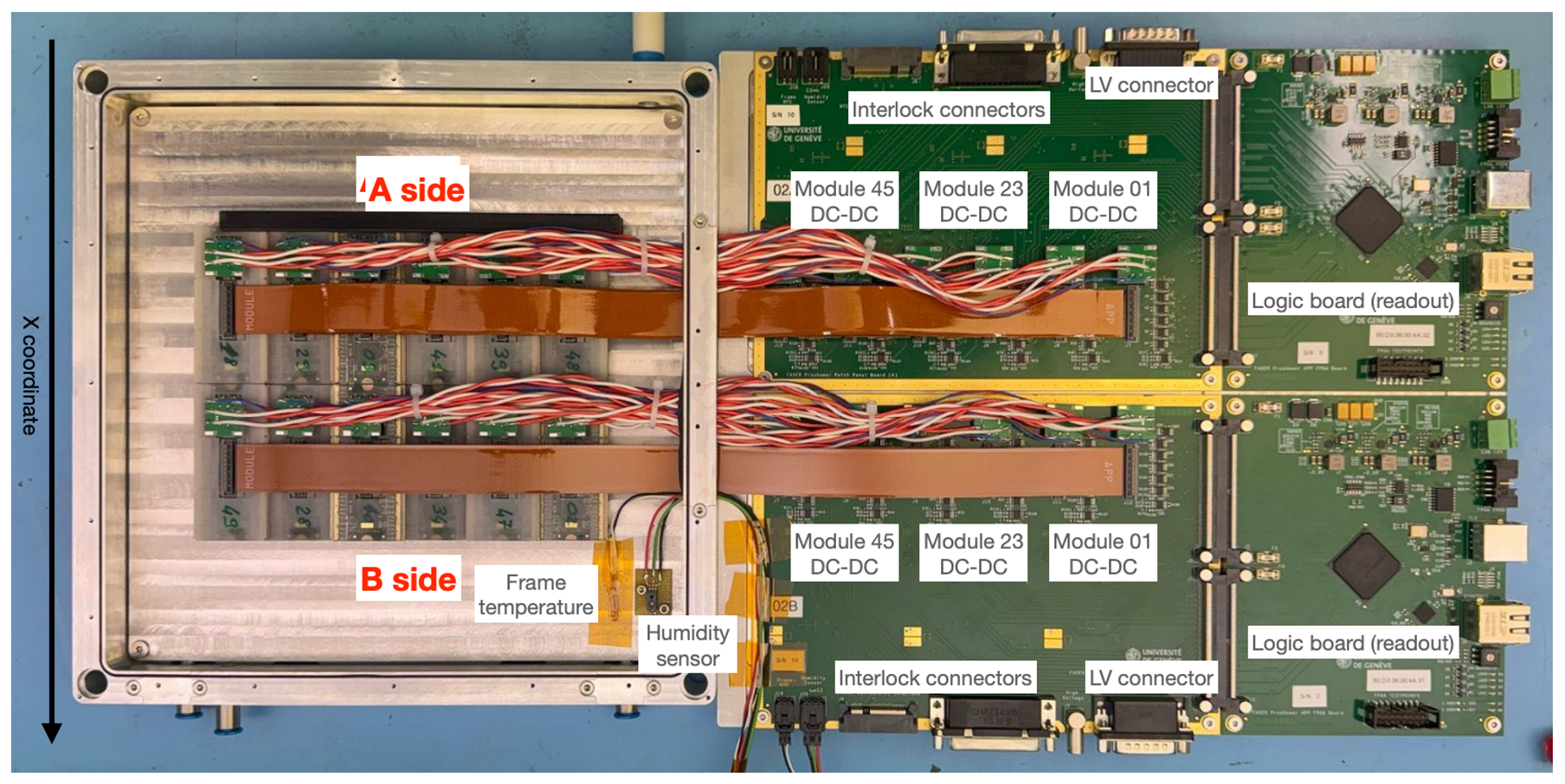}
    \includegraphics[width=.9\textwidth]{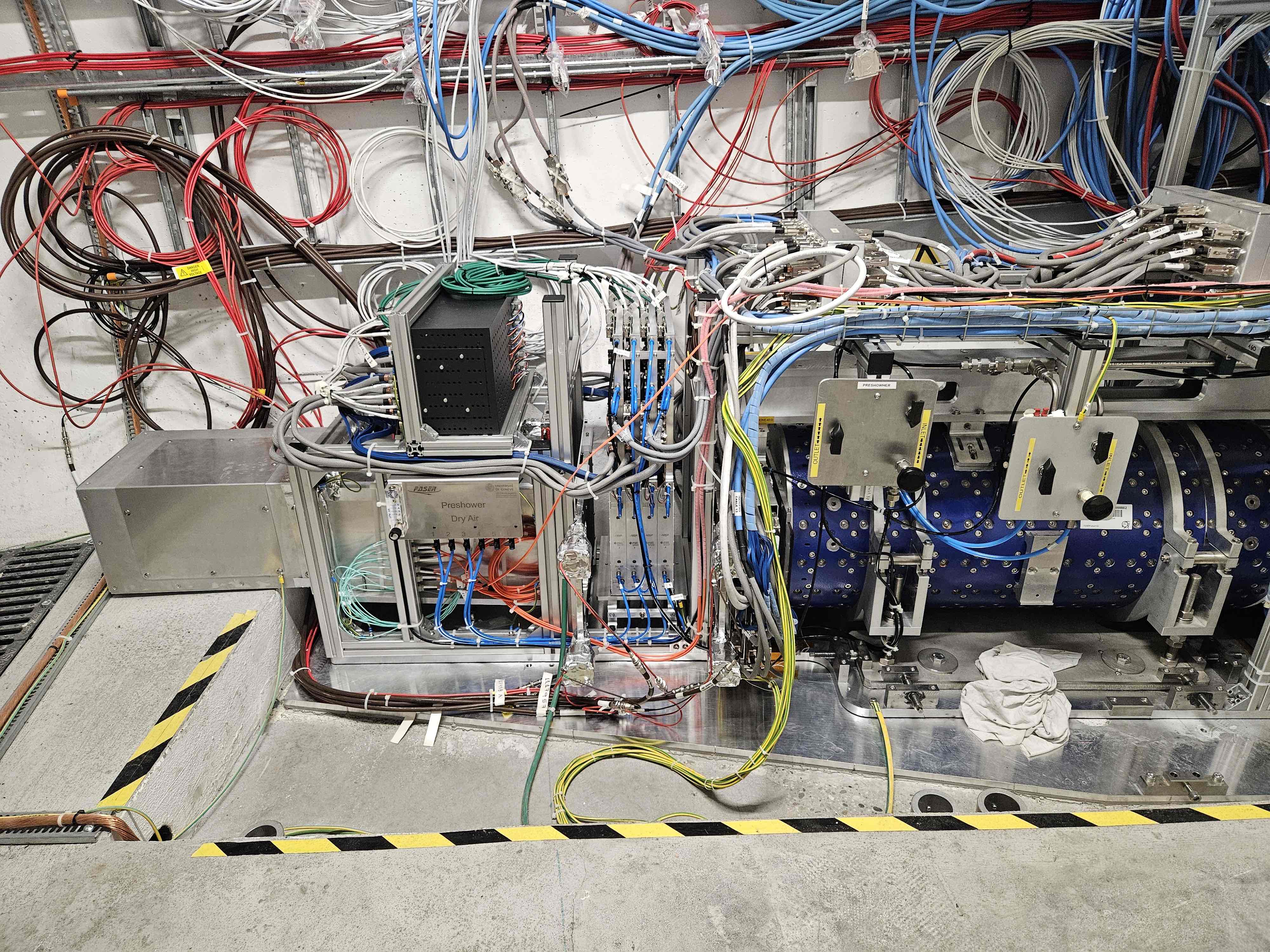}    
    \caption{ (top) One of the preshower layers. The instrumented region is on the left side where the 12 modules are visible. To the right is the on-detector readout board. (bottom) A photograph of the back of the FASER detector with the upgraded preshower detector installed.  }
    \label{fig:preshower}
\end{figure}

In early 2025 an upgrade of the FASER preshower was installed into the detector. The upgrade replaced the simple scintillator-based preshower described in Section~\ref{sec:detector} with a high-granularity silicon-pixel / tungsten preshower. The new system was designed to be able to separate two high-energy photons separated by as little as 200-$\mu$m, as is the signature for axion-like-particles (ALPs) decaying to photons inside the FASER detector. A technical proposal~\cite{Boyd:2803084} for this upgrade was submitted to the LHCC in March 2022 and the project was then approved by CERN~\cite{approval3-RBminutes}. The detector is based on monolithic pixel silicon detector layers with pixel size of 100~$\mu$m interleaved by tungsten plates. The original detector design had 6 instrumented layers, but the ASIC yield was lower than foreseen so the final detector was constructed with 4 instrumented layers. However, simulation studies demonstrate that good physics performance is still expected with the reduced number of layers.  
The detector ASICs design was optimized to have a large dynamic range for measuring the charge per pixel, as it is important to be able to very precisely measure the position of the core of the high-energy electromagnetic showers in the signal, while also keeping sensitivity to the much lower energy deposited in the pixels at the edge of the showers. Simulation studies with simplified reconstruction algorithm show the efficiency to reconstruct closely spaced showers in signal ALP events is in the range 65\% to 90\%, with fake rates between 2\% and 4\%, depending on the photons energy and separation, and it is expected this performance can be improved with more sophisticated algorithms. 
The preshower ASICs were designed at the University of Geneva and fabricated by IHP Microelectronics. The ASICs were thoroughly tested before being assembled into modules (with each module containing 6 ASICs), and then into planes (with each plane containing 12 modules). The system was tested in two testbeams at the CERN SPS in Autumn 2024, before installation into FASER in February 2025. Fig.~\ref{fig:preshower} shows an assembled preshower plane (top) and the final detector installed into FASER (bottom). The detector has been operating well during 2025 running, with stable electrical and thermal performance throughout the year. Calibrations tag 1.8\% of pixels as noisy. The working point of one of the planes has been optimized to improve low-energy shower reconstruction and increase the overall sensitivity to ALPs. An ALP search using the 2025 and 2026 preshower data is expected to be released in the coming years.

    \clearpage
    \newpage    
    \section{Detector Performance}
    \label{sec:performance}
    Some of the key detector performance parameters are discussed in this section, and these are then used in the physics analyses described in Section~\ref{sec:results}. 
The performance studies with the electronic detector data are carried out after event reconstruction, which is performed using FASER's Calypso~\cite{calypso} offline software system, based on the open-source Athena framework~\cite{ATL-PHYS-PUB-2009-011,athena} from the ATLAS experiment. Charged particle track reconstruction is performed using the combinatorial Kalman filter from the ACTS library~\cite{ACTS}. Extracting the PMT charge from the scintillator and calorimeter modules is done by summing the digitised waveform values after pedestal subtraction.
Monte Carlo simulation samples are produced to compare with the data. 
These samples are simulated using \texttt{GEANT4}~\cite{G4} with a detailed description of the detector geometry, including passive material. The samples include a realistic level of detector noise, and are reconstructed in the same way as the data.

\subsection{Scintillator system performance}

\begin{figure}[htbp]
\includegraphics[width=0.49\textwidth]{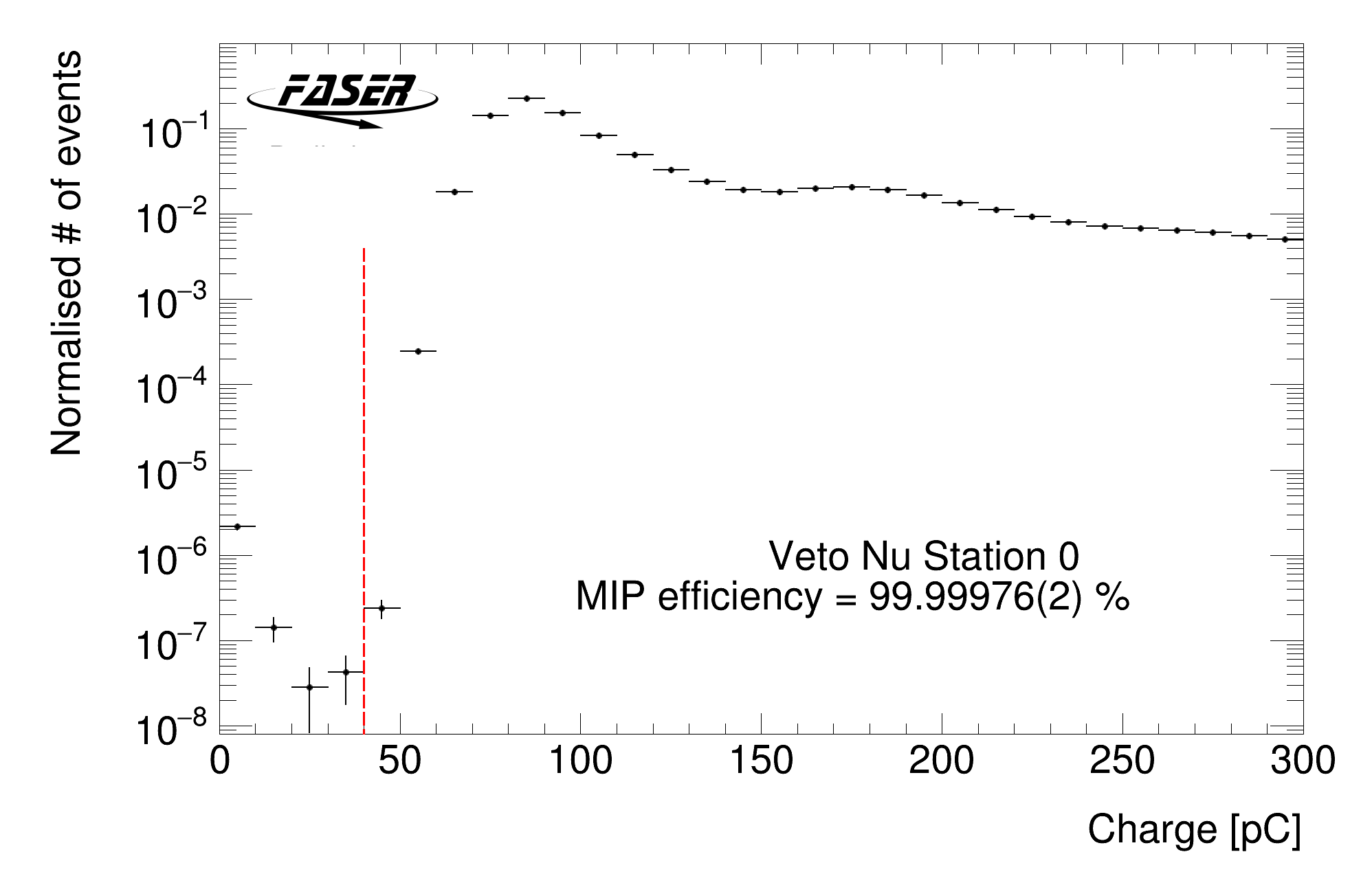}
\includegraphics[width=0.49\textwidth]{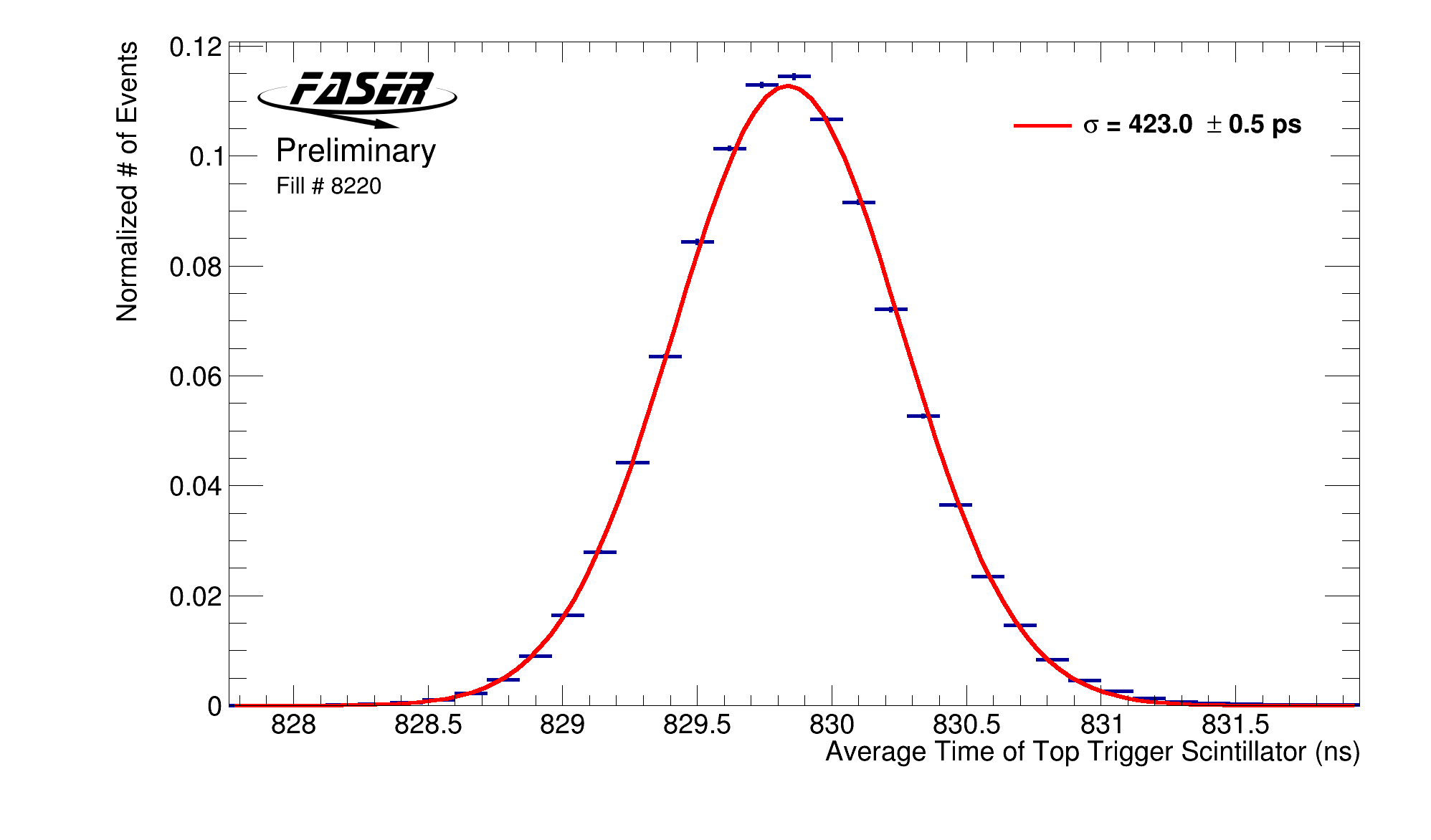}
\caption{(left) Example of the scintillator efficiency measurement, showing the charge distribution for events with good fiducial tracks for the most downstream Veto scintillator. The efficiency is calculated as the ratio of the number of events with charge above the 40~pC threshold (indicated on the figures) to the total number of events. (This figure is taken from Ref.~\cite{FASER:2023tle}.) (right) The timing distribution of the top Trigger/timing scintillator, with a Gaussian fit applied. }   
\label{fig:scint-eff}
\end{figure}

The veto scintillator system plays a crucial role in both the BSM and neutrino physics programme of FASER. The key performance metric is to have an extremely high efficiency for rejecting incoming muons for each of the scintillators.
The efficiency is measured using collision data, by selecting events which require a signal in all of the veto scintillators except for the one under test, and with a well reconstructed charged-particle track which when extrapolates passes through the central region of the scintillators. The fraction of these events for which the scintillator under test has a measured charge above the defined threshold for detecting a MIP is defined as the efficiency. Figure~\ref{fig:scint-eff} (left) presents an example of this measurement, showing the measured charge distribution for the most downstream veto scintillator where the red line indicates the charge threshold for a MIP. In this case the measured inefficiency is less than $10^{-5}$, and similar performance is seen for all of the veto scintillators.

The scintillators can also give a measurement of the time of the incident charged particles with sub-nanosecond precision. Figure~\ref{fig:scint-eff} (right) shows the time distribution of signals above a MIP in the top Trigger/timing scintillator, for an LHC fill in 2022 running. Since the Trigger/timing scintillator has a PMT on both ends of the scintillator, the average time of both PMTs is used to correct for time walk, leading to a timing resolution of better than 500~ps. The timing resolution of the bottom Trigger scintillator is similar.

\subsection{Calorimeter performance}

\begin{figure}[htbp]
\includegraphics[width=0.57\textwidth]{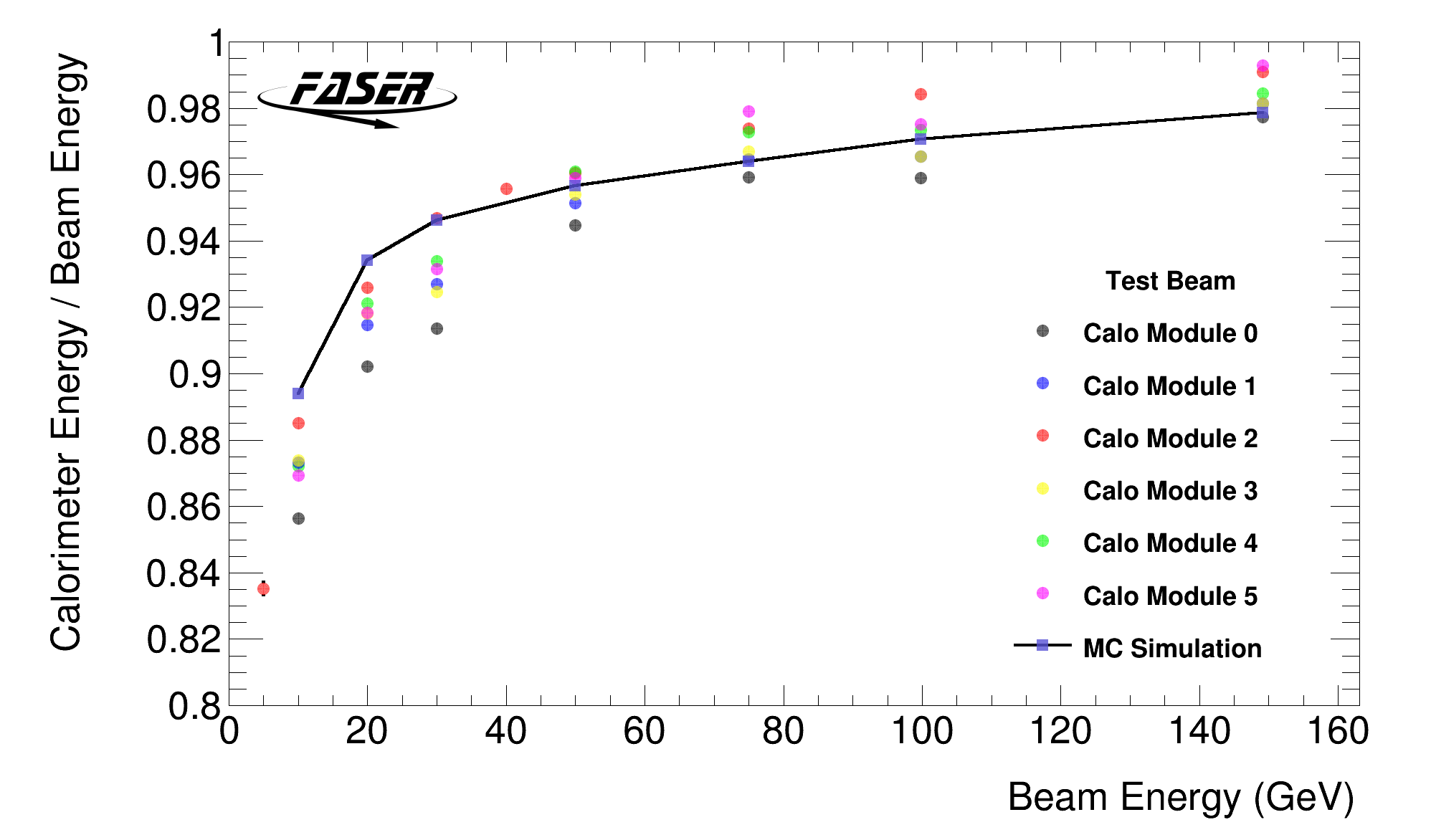}
\includegraphics[width=0.42\textwidth]{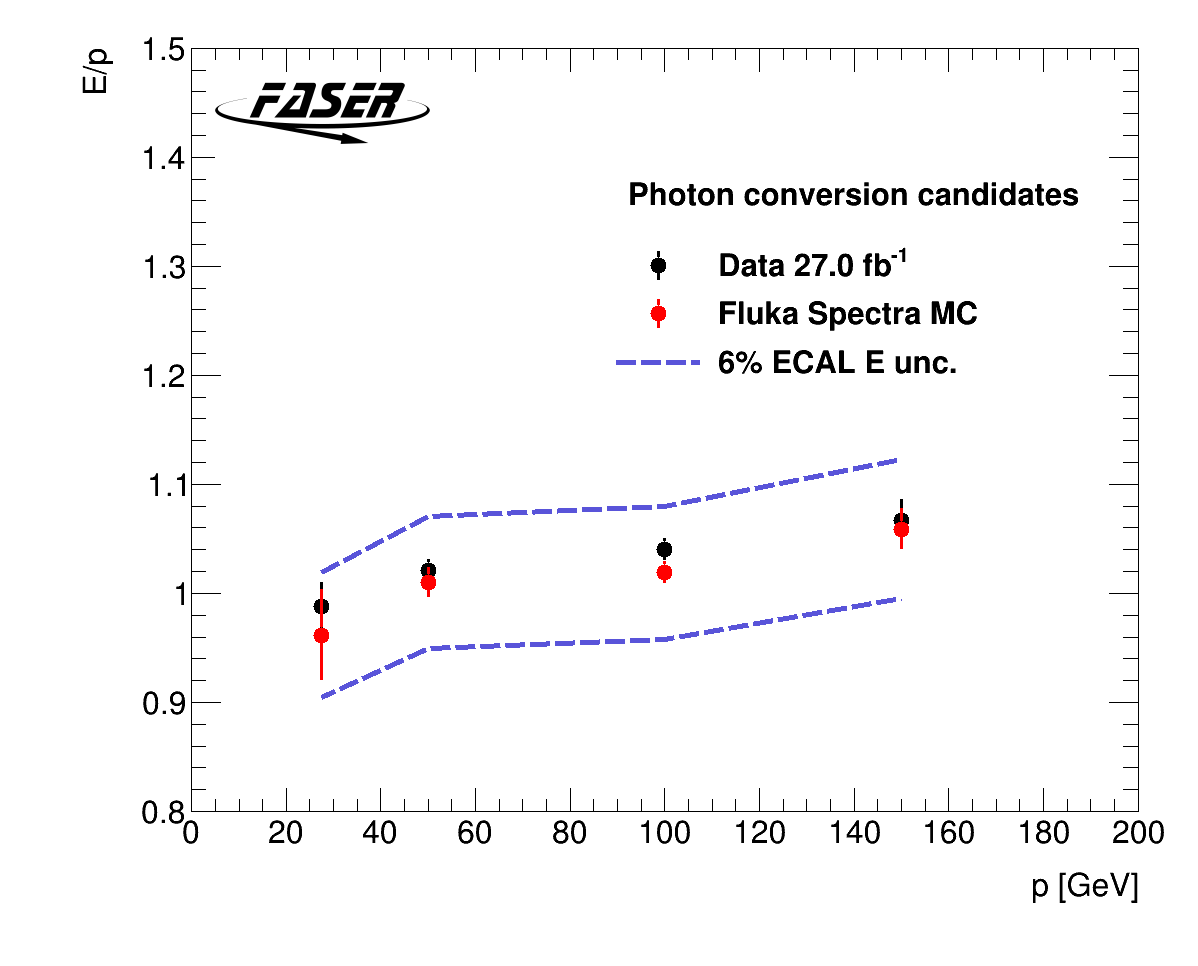}
\caption{ (left) The calorimeter response as a function of energy, for six calorimeter modules used in a testbeam at the SPS in summer 2021. The response in the MC simulation is also shown.
(right) The Gaussian-fitted peak position of the $E/p$ in data and MC simulation as a function of the momentum of photon conversion candidates. (These figure are taken from the auxiliary material in Ref.~\cite{FASER:2023tle}.)} 
\label{fig:EoP}
\end{figure}

In Summer 2021, the calorimeter energy response was measured in a testbeam at the CERN SPS~\footnote{For this testbeam the four calorimeter modules installed in FASER were temporarily removed, and used along with two spare calorimeter modules.}
using electron beams with energies of between 50~\gev and 250~\gev~\cite{testbeampaper}. Fig~\ref{fig:EoP} (left) shows the measured response for 6 calorimeter modules and for the Monte Carlo (MC) simulation, as a function of the beam energy.  
The energy scale of the calorimeter is calibrated using the simulation and an in situ measurement from muon data, where the simulation is used to estimate the most probable value of the energy deposited by a muon of 330~\mev . This procedure leads to an uncertainty on the energy scale of 6\%. 

The energy scale is validated with a sample of radiative muon events, where the radiated photon converts to an $e^+e^-$ pair, leading to a three track-signature in the detector. In these events an $E/p$ variable is constructed as the total calorimeter energy ($E$) divided by the scalar sum of the momentum of the lowest momentum two tracks $p$ (since the highest momentum track is assumed to be the parent muon, and its contribution to the calorimeter energy is assumed to be negligible). Figure~\ref{fig:EoP} (right) shows the $E/p$ peak position, as a function of the momentum of the photon conversion candidate for data and MC simulation. The peak position is well compatible with the expected value of 1 within the 6\% uncertainty on the energy scale.  

\subsection{Spectrometer performance}

\begin{figure}[htbp]
\begin{center}
\includegraphics[width=0.4\textwidth]{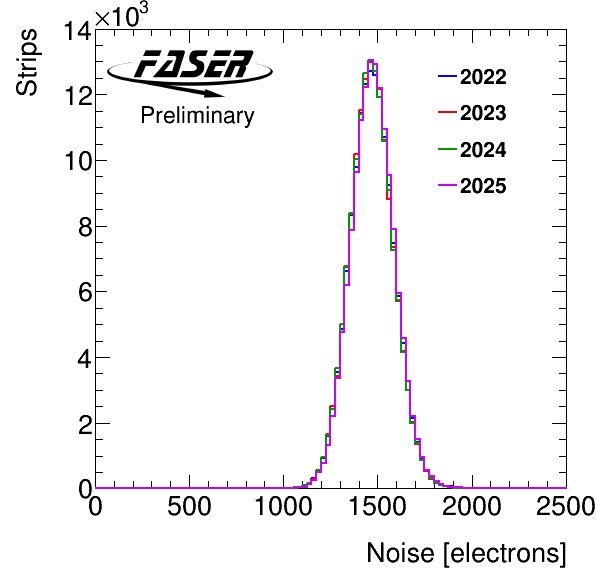}
\includegraphics[width=0.4\textwidth]{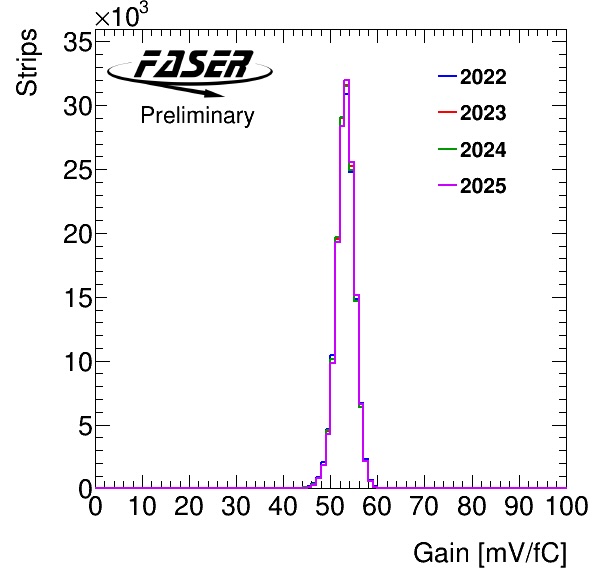}
\caption{The distribution of the noise (left) and gain (right) from tracker calibrations performed in 2022, 2023, 2024 and 2025 data taking. Showing the stability of the performance across these years.}   
\label{fig:tracker-stability}
\end{center}
\end{figure}

The FASER tracking spectrometer performance is monitored with regular calibrations which measure the noise and gain of the tracking modules, as well as the number of badly performing channels. The system has been extremely stable since FASER operations started, with more than 99.7\% of the channels operational. Example distributions of the detector noise (left) and gain (right) are shown in Fig~\ref{fig:tracker-stability}, where the results from calibrations taken in 2022, 2023, 2024 and 2025 are presented, showing very stable detector performance. The number of non-operational or badly performing channels has also been extremely stable during the full time of physics operations. 

\begin{figure}[htbp]
\begin{center}
\includegraphics[width=0.95\textwidth]{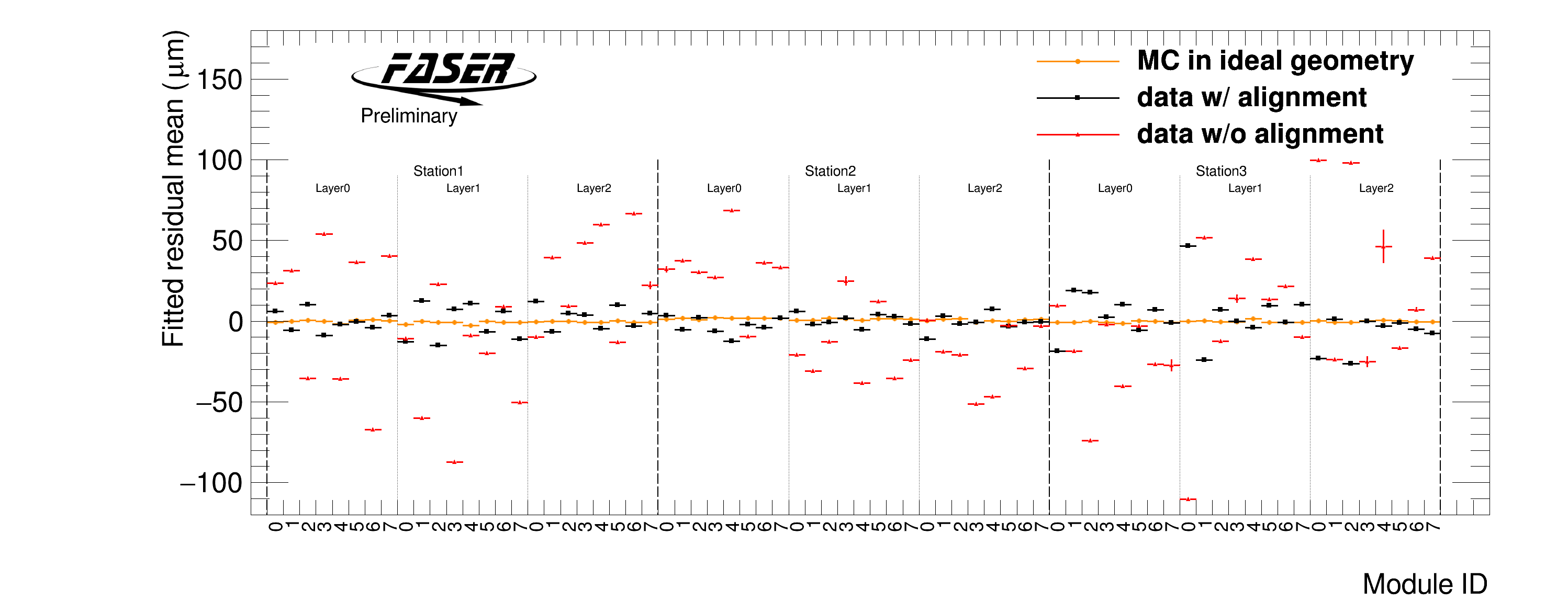}
\includegraphics[width=0.95\textwidth]{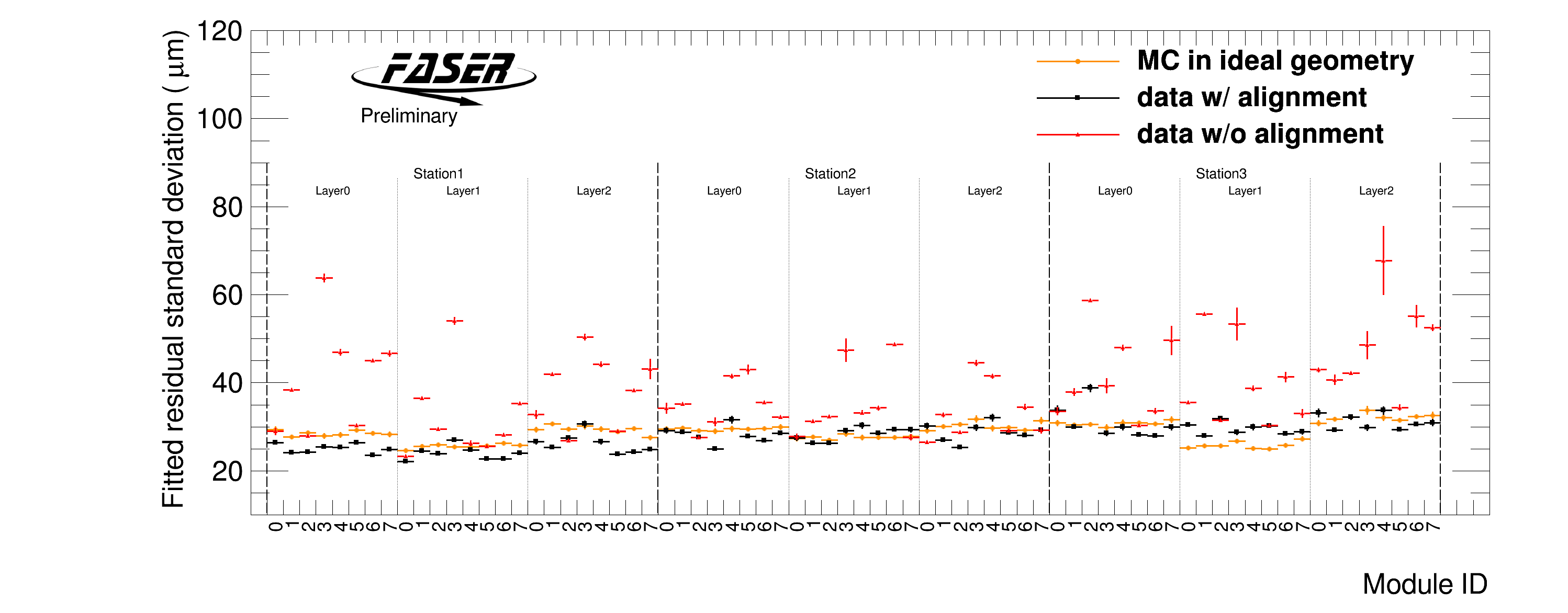}
\caption{The quality of the alignment of the FASER tracker is assessed by comparing the residual distribution mean (top) and width (bottom) before (red) and after (black) data-driven alignment corrections are applied. Also shown is the performance with perfectly aligned MC simulation (orange). }   
\label{fig:tracker-alignment}
\end{center}
\end{figure}

The reconstructed track momentum resolution and charge measurement performance is mostly defined by the alignment of the tracking modules. 
A track-based alignment of the tracking detector is performed using an iterative local $\chi^2$ alignment method, and shows an improved agreement in the hit residual and track $\chi^2$ distributions when comparing to the perfectly aligned MC. The alignment only considers the most sensitive distortions, translations in the transverse plane and rotations around the longitudinal axis, at both the individual module and tracking layer level. Figure~\ref{fig:tracker-alignment} shows the mean (top) and standard-deviation (bottom) of the residuals in the different modules of the tracker, for before alignment corrections (red), and after these corrections (black) have been applied. The expectation from perfectly aligned MC simulation is shown in orange.

 The effect of the alignment on the momentum resolution and charge ID was studied by creating 40 MC simulation geometries with mis-alignments applied to the tracker modules at the level of the measured residuals. With these samples the momentum resolution was measured to be around 20\% for a 1~\tev\ muon (compared to the expectation of 15\% with a perfectly aligned detector).
\newpage

\subsection{FASER\texorpdfstring{$\nu$}{nu} performance}

\begin{figure}[ht]
\centering
\includegraphics[width=0.39\textwidth]{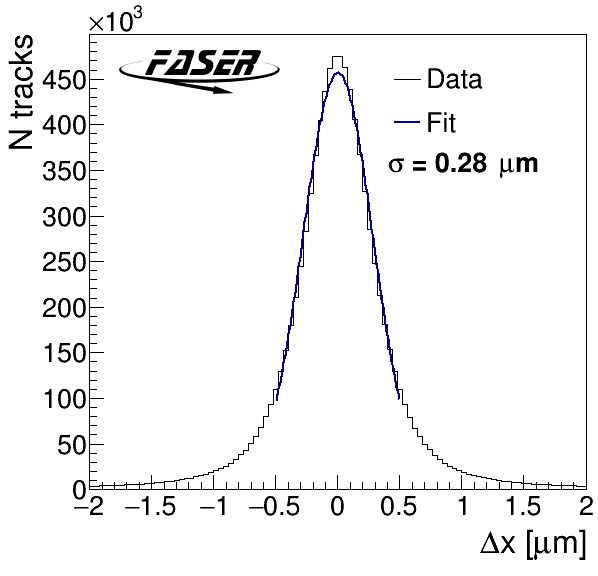}
\includegraphics[width=0.5\textwidth]{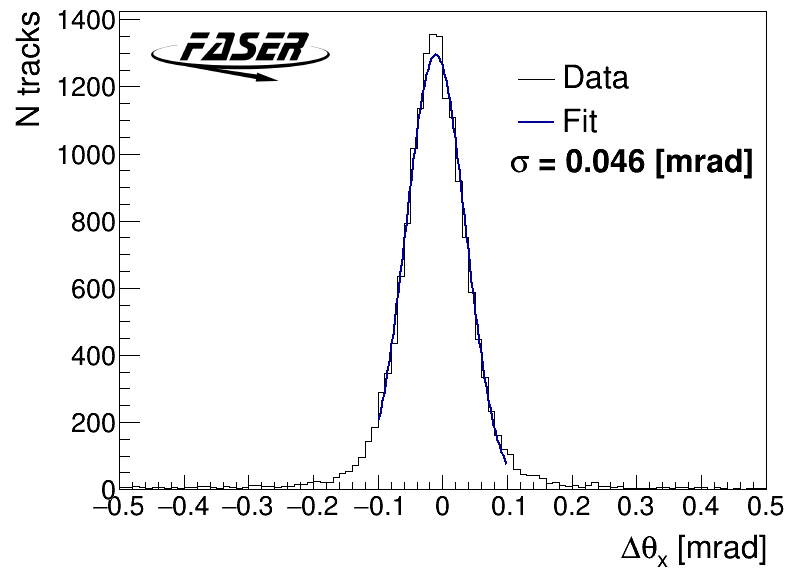}
\caption{ 
The measured track position (top left) and angular (top right) resolution in the FASER$\nu$ emulsion detector. These figures are taken from Ref.~\cite{FASER:2025qaf}. 
}
\label{fig:fasernu-plots}
\end{figure}

The FASER$\nu$ emulsion detector can reconstruct charged-particle tracks with unprecedented position and angular resolution, but since the detector is passive it records all through going charged-particle tracks while the detector is in place. In order to keep the track multiplicity manageable for analysis the detector is therefore replaced every 20-30~fb$^{-1}$, corresponding to a track multiplicity of less than $10^6$~tracks/cm$^2$. The detector data reconstruction and performance are detailed in Ref.~\cite{FASER:2025qaf}. Fig~\ref{fig:fasernu-plots} shows the measured position resolution (top left) of 0.3~$\mu$m and angular resolution (top right) of 0.05~mrad.

\begin{figure}[ht]
\centering
\includegraphics[width=0.43\textwidth]{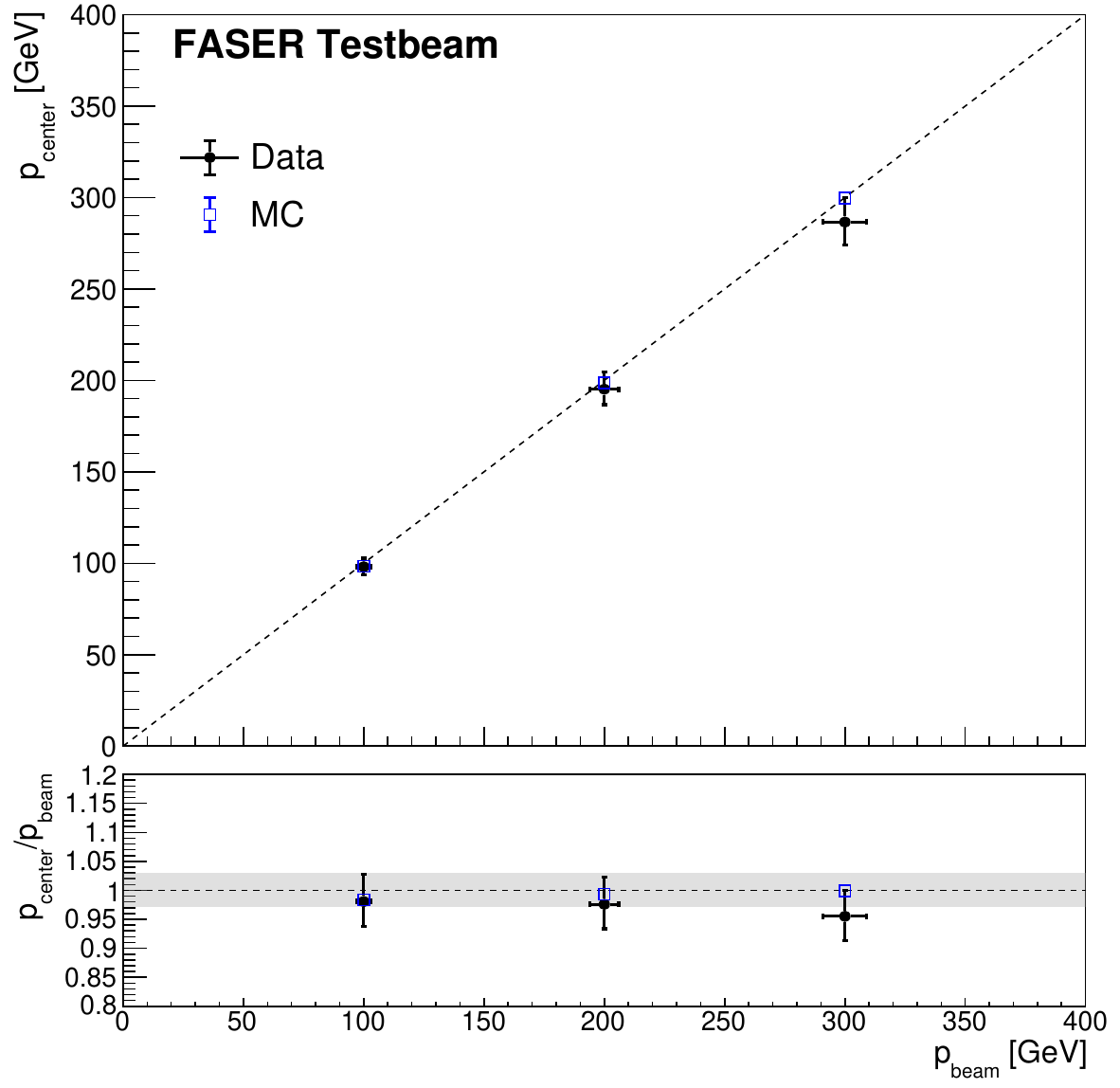}
\includegraphics[width=0.43\textwidth]{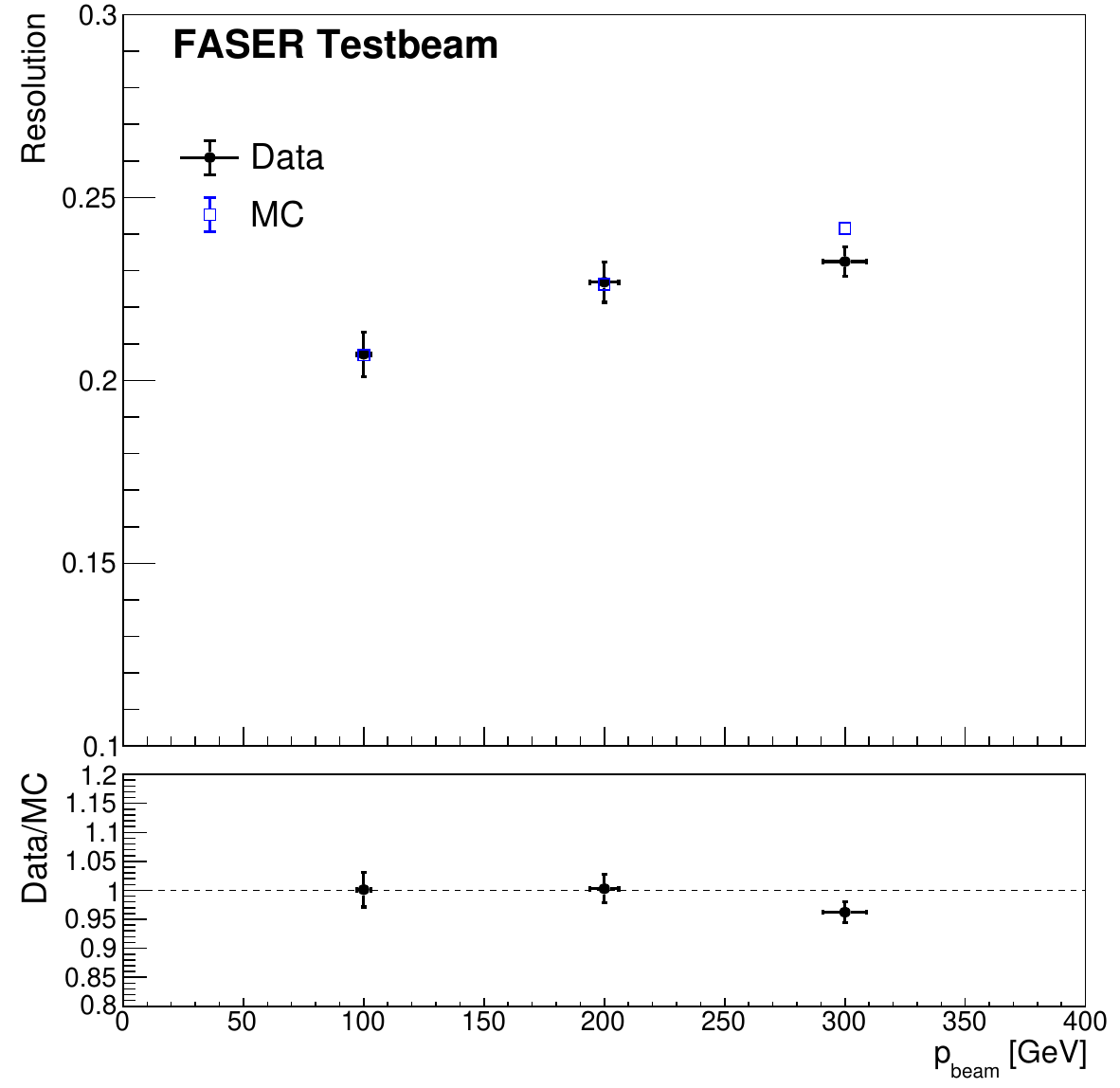}
\includegraphics[width=0.45\textwidth]{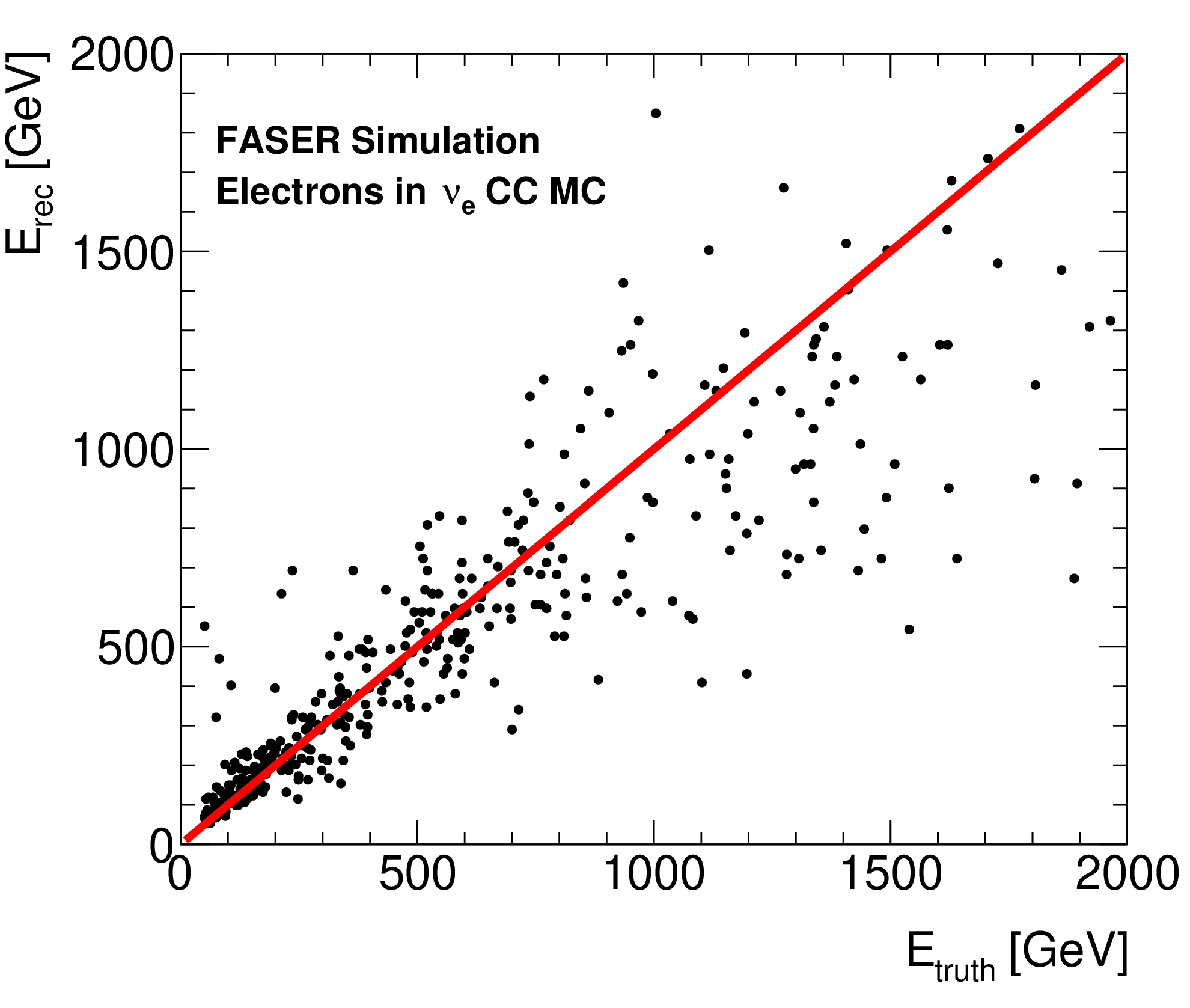}
\caption{ 
(top) The FASER$\nu$ muon momentum measurement performance from testbeam, and compared with MC simulation: (left) showing the central value of the reconstructed muon momentum versus the muon beam momentum; (right) showing the reconstructed muon momentum resolution versus the beam momentum. (Both plots are taken from Ref.~\cite{FASER:2026hzm}). (bottom) The electron energy measurement performance in FASER$\nu$ from simulation (taken from Ref.~\cite{FASER:2024hoe}).
}
\label{fig:fasernu-lep-rec}
\end{figure}

The precise track reconstruction in the emulsion can be used to measure the momentum of muons via the effect of multiple coulomb scattering on the track trajectory, with a resolution measured in testbeam of around 20 - 25\%~\cite{FASER:2026hzm} as can be seen in Fig.~\ref{fig:fasernu-lep-rec} (top plots). Electromagnetic showers can also be reconstructed in the emulsion with the energy measured by counting the track segment multiplicity at the shower maximum~\cite{FASER:2026jhh}. Fig.~\ref{fig:fasernu-lep-rec} (bottom plot) shows the electron energy measurement performance in Monte Carlo simulation.

    \clearpage
    \newpage    
	\section{Physics Results}\label{sec:results}
    \subsection{Searches for new particles}

As discussed in Section~\ref{sec:physicscase}, FASER was originally designed to search for light and weakly-coupled new particles that could be produced in the LHC collisions and travel 480~m along the LOS, before decaying in the FASER detector. 
Since particles produced along the LOS have a large boost, this would leads to a spectacular high-energy appearance signature in the detector.

To date, FASER has released results related to searching for new particles i) using charged particle final states to target a dark-photon ($A'$) signal~\cite{FASER:2023tle,FASER:DarkPhoton2026}, and ii) using neutral final states, to target an axion-like-particle (ALP) signal decaying to two photons~\cite{FASER:2024bbl}.
There are additional ongoing searches, including updates to the analyses presented below, with larger dataset and improved analysis methods, as well as a search for quirks based on precise detector time measurements, as proposed in Ref.~\cite{Feng:2024zgp}. 

\subsubsection{Charged particle final states}
{\bf First FASER search for dark photons}

The search for dark photons using the data collected in 2022, corresponding to an integrated luminosity of 27.0~fb$^{-1}$, was released in 2023~\cite{FASER:2023tle}. 
The analysis selects events with no significant signal in any of the five veto scintillators~\footnote{The detector shown in Fig.~\ref{fig:FASER_labels} shows two veto scintillators upstream of FASER$\nu$ and four upstream of the decay volume. However, in 2022 and 2023 one of the veto scintillators upstream of the decay volume was not read out due to a limitation in the number of digitizer channels. With the installation of the calorimeter readout upgrade (Section~\ref{sec:upgrades}) an additional digitizer was installed and all of the veto scintillators could therefore be used.}, two reconstructed charged-particle tracks and a large energy deposit in the calorimeter of more than 500~\gev . To minimize background from incoming muons that miss the front vetoes, the charged particle tracks are required to have more than 20~\gev\ of momentum, and when extrapolated to the front veto system, are required to pass through the central region (defined as a fiducial volume of radius 95~mm centered on the detector axis). 
The  analysis selections have a typical efficiency of around 50\% on signal (where the efficiency is defined for signal events which decay in the FASER decay volume).

Several possible backgrounds that could pass the event selection, were considered as discussed below. 
\begin{itemize}
    \item  As shown in Section~\ref{sec:performance}, the inefficiency of each of the veto scintillators was measured with data, by selecting events with a track and where all veto scintillators fired except for one scintillator that was under test. The resulting inefficiencies were all below $10^{-5}$ per scintillator, leading to a total veto inefficiency of smaller than $10^{-20}$ (since the different planes are uncorrelated). Given that we expect $\mathcal{O}$($10^{8}$) muons entering FASER in the 2022 dataset, the probability of a muon entering through the vetoes without being vetoed is completely negligible.
    \item The possible background from incoming muons that miss the front vetoes is mitigated by the requirement that the extrapolated tracks are in the fiducial volume at the front of the detector. Large MC simulation samples, and a data driven method are used to confirm that this background is negligible~\cite{FASER:2023tle}.
    \item A possible background from neutral hadrons, produced by muon interactions in the rock in front of FASER, entering the decay volume and producing a signature that passes the event selection, was estimated using a combination of data control regions and MC simulation~\cite{FASER:2023tle}. This background is heavily suppressed by the need for the parent muon to scatter sufficiently to miss the veto scintillators, and for the produced neutral hadron to pass through the full 8 interaction lengths of FASER$\nu$, and to lead to a large energy in the calorimeter. The final estimate for this contribution was $(8.4 \pm 11.9) \times 10^{-4}$ events.
    \item The dominant background was found to be from neutrino interactions occurring in the detector material after the veto system. Such a background is mitigated by the small amount of material in the detector volume before the calorimeter. It was estimated using very large MC simulation samples of neutrino interactions in FASER. Using a sample of 300~$\iab$ ($\sim10000\times$ larger than the data used in this analysis) the background was estimated as (1.5 $\pm$ 0.5 (stat.) $\pm$ 1.9 (syst.))$\times10^{-3}$ events.
    \item A possible non-collision background from cosmic-rays or beam-halo particles entering FASER was studied by using data taken with no collisions, and was found to be negligible. 
\end{itemize}
The total background estimate is therefore $(2.3 \pm 2.3) \times 10^{-3}$ events.

Various sources of systematic uncertainties on the expected signal yield were considered, as outlined in Table~\ref{tab:DP-systs}. This table shows the largest source of uncertainty (at the level of 50\%) comes from the modeling of the signal production, which was parameterized as a function of the $A'$ energy ($E_{A'}$) as discussed in~\cite{FASER:2023tle}. The experimental uncertainties are  much smaller with the largest of 7\%. 

\begin{table}[tbh]
    \centering
    \begin{tabular}{|c|c|c|}    
    \hline
    Source & Value & Effect on signal yield \\
    \hline
    
    Signal Generator & $\frac{0.15 + (E_{A'} / 4 \text{TeV})^3}{1 + (E_{A'} / 4 \text{TeV})^3}$ & 15-65\% (15-45\%) \\
    Luminosity & 2.2\% & 2.2\% \\
    MC Statistics & $\sqrt{\sum{W^2}}$ & 1-3\% (1-2\%) \\

    Track Momentum Scale & 5\% & $<$ 0.5\% \\
    Track Momentum Resolution & 5\% & $<$ 0.5\% \\
    Single Track Efficiency & 3\% & 3\% \\
    Two-track Efficiency & 7\% & 7\% \\ 

    Calo E scale & 6\% & 0-8\%  ($<$ 1\%)\\
    
    \hline    

\end{tabular}
\caption{Summary of the systematic uncertainties on the signal yield for the $A'$ search. For each of the sources of uncertainty, the source and size of the uncertainty is shown. The effect on the signal yield across the full signal parameter space probed is also shown. The numbers in parenthesis indicate the effect on the signals within the parameter space for which this analysis is sensitive. (This table is taken from Ref.~\cite{FASER:2023tle}.)}
\label{tab:DP-systs}
\end{table}

\begin{figure}[ht]
\centering
\includegraphics[width=0.32\textwidth]{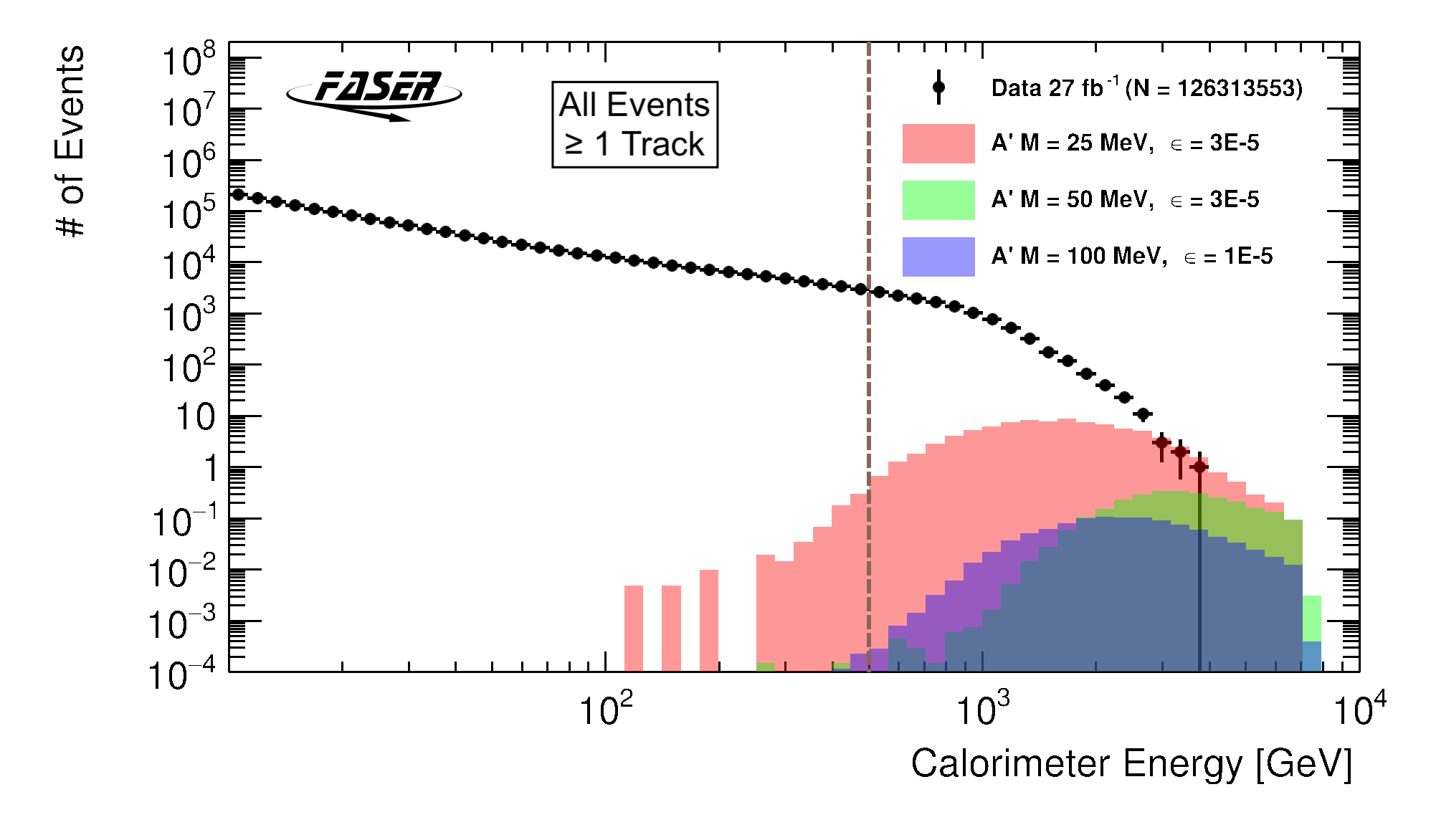}
\includegraphics[width=0.32\textwidth]{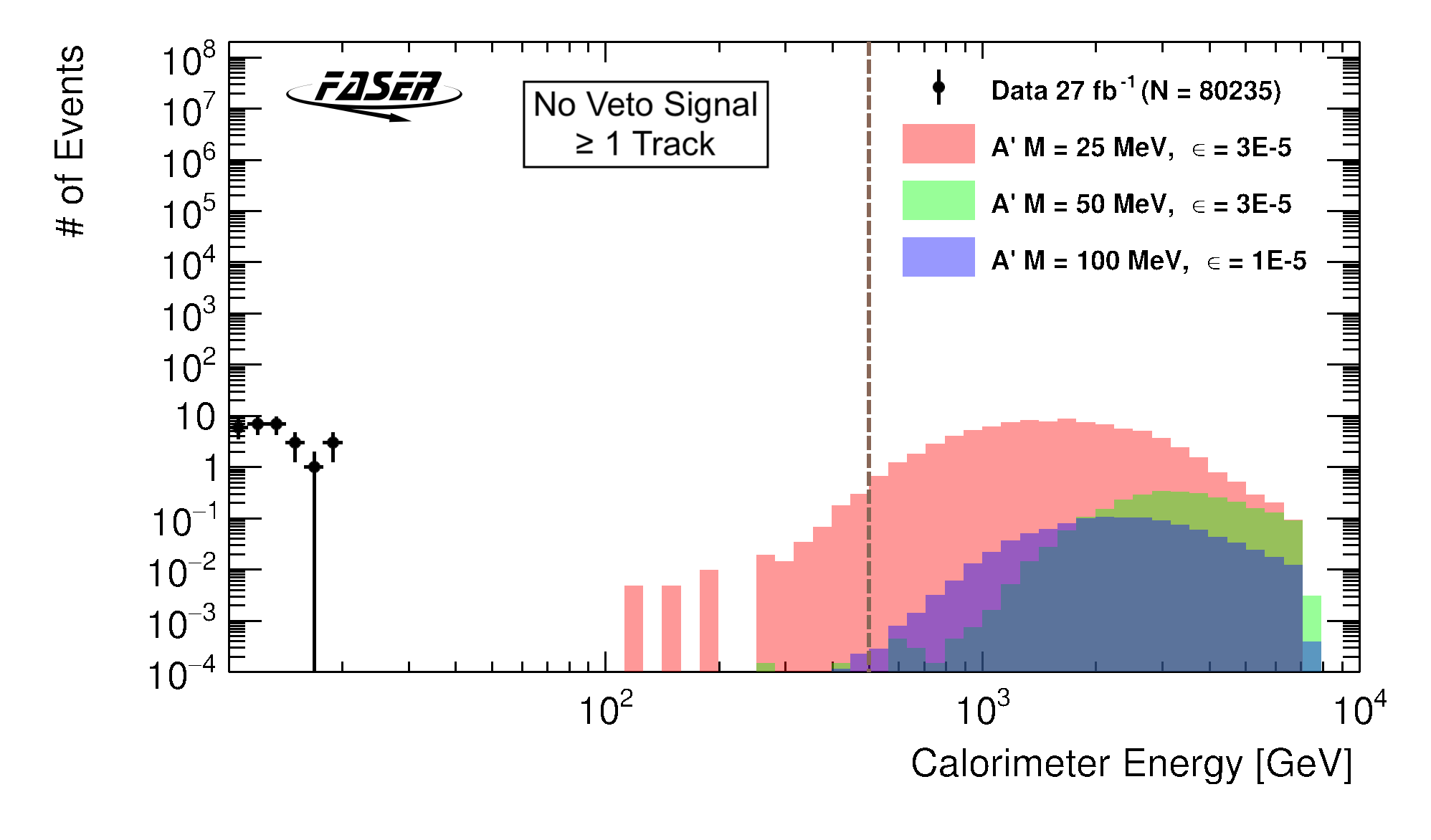}
\includegraphics[width=0.32\textwidth]{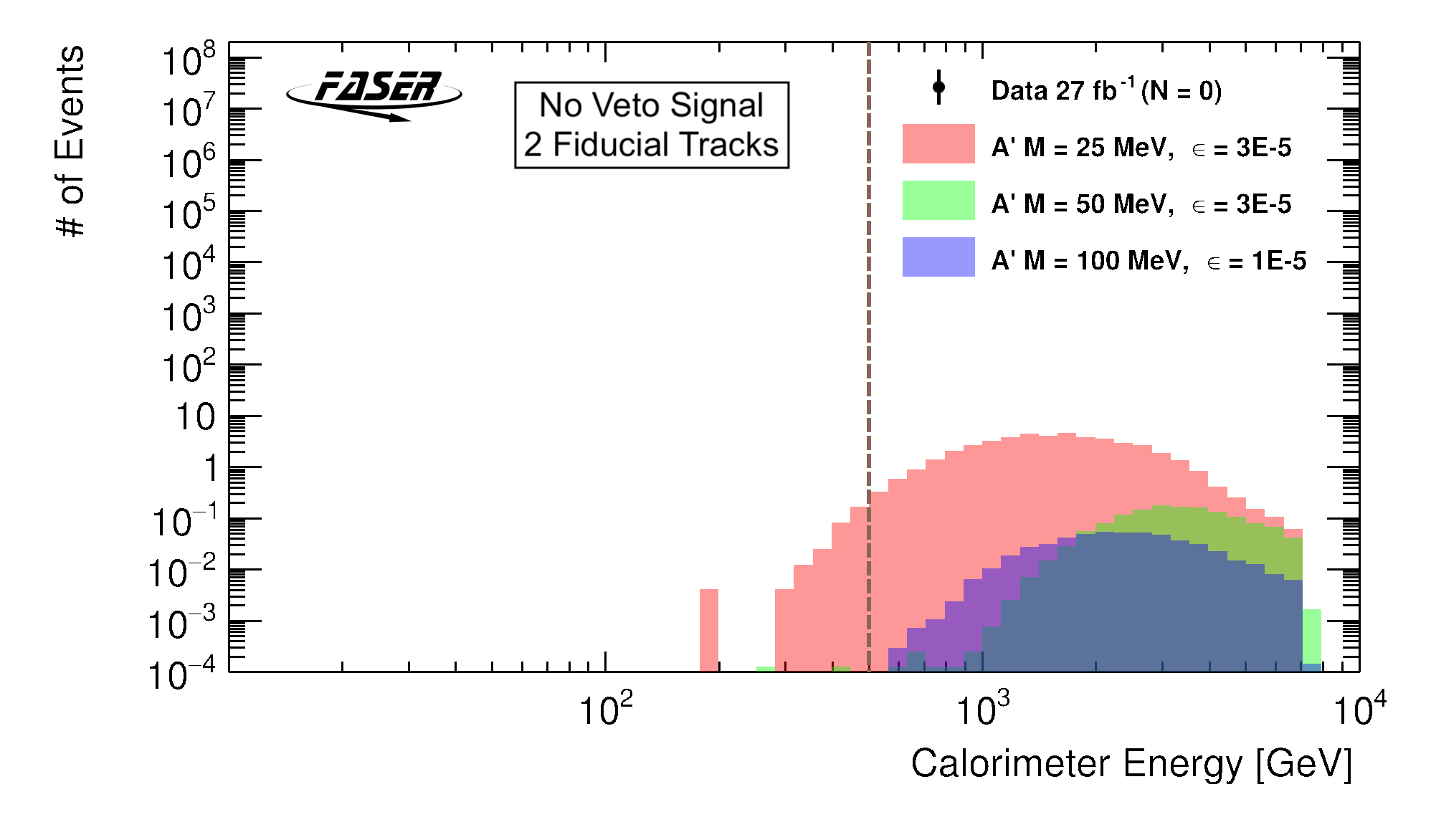}
\caption{The number of observed events (black points) as a function of the calorimeter energy, and the distribution for three $A'$ signal models (coloured histograms), for three different event selections. (left) No event selection; (middle) A veto requirement, and one reconstructed track; (right) The full analysis selection. (These figures are taken from Ref.~\cite{FASER:2023tle}.) }
\label{fig:Aprime-SR}
\end{figure}

\begin{figure}[ht]
\centering
\includegraphics[width=0.49\textwidth]{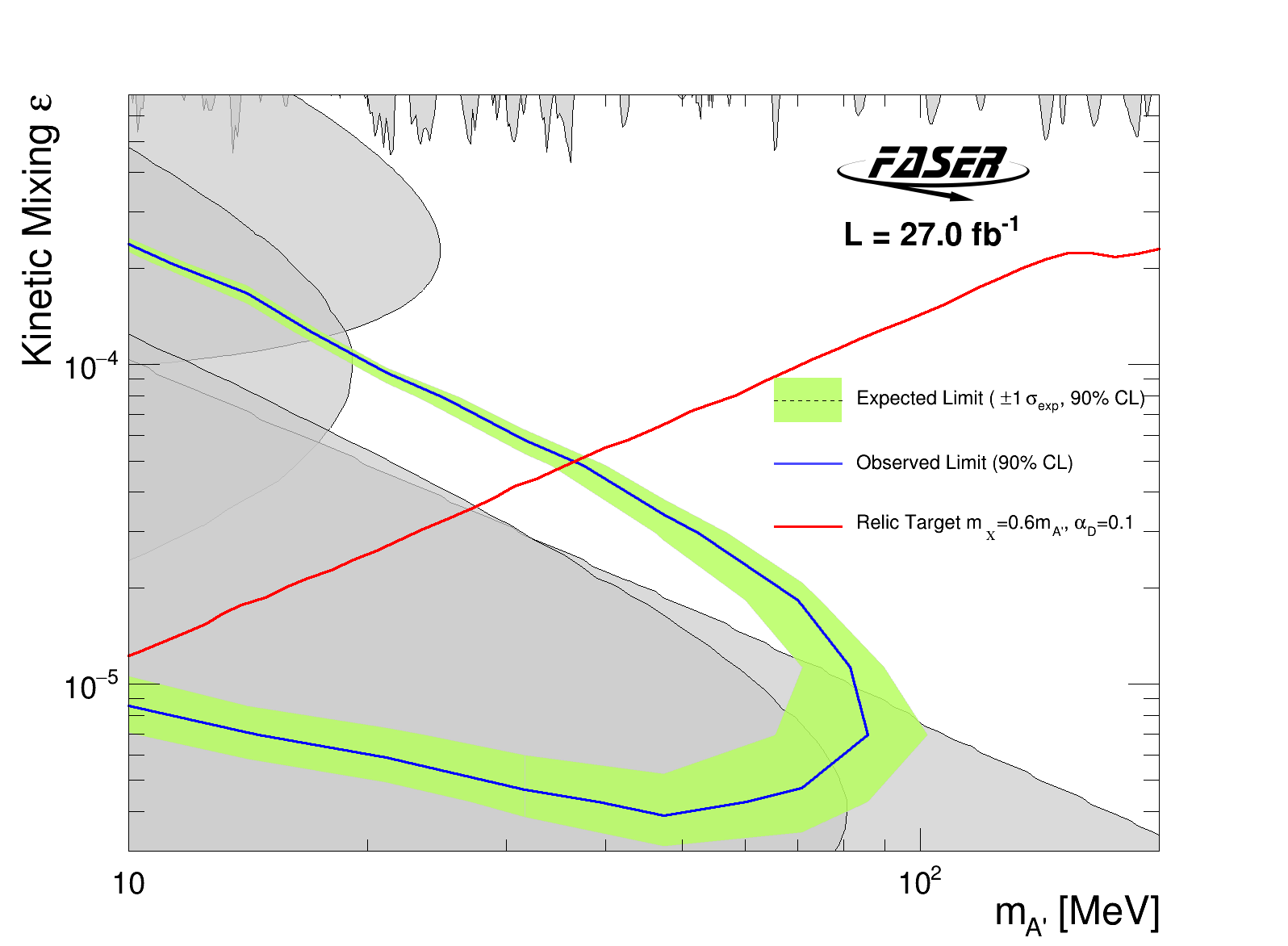}
\includegraphics[width=0.49\textwidth]{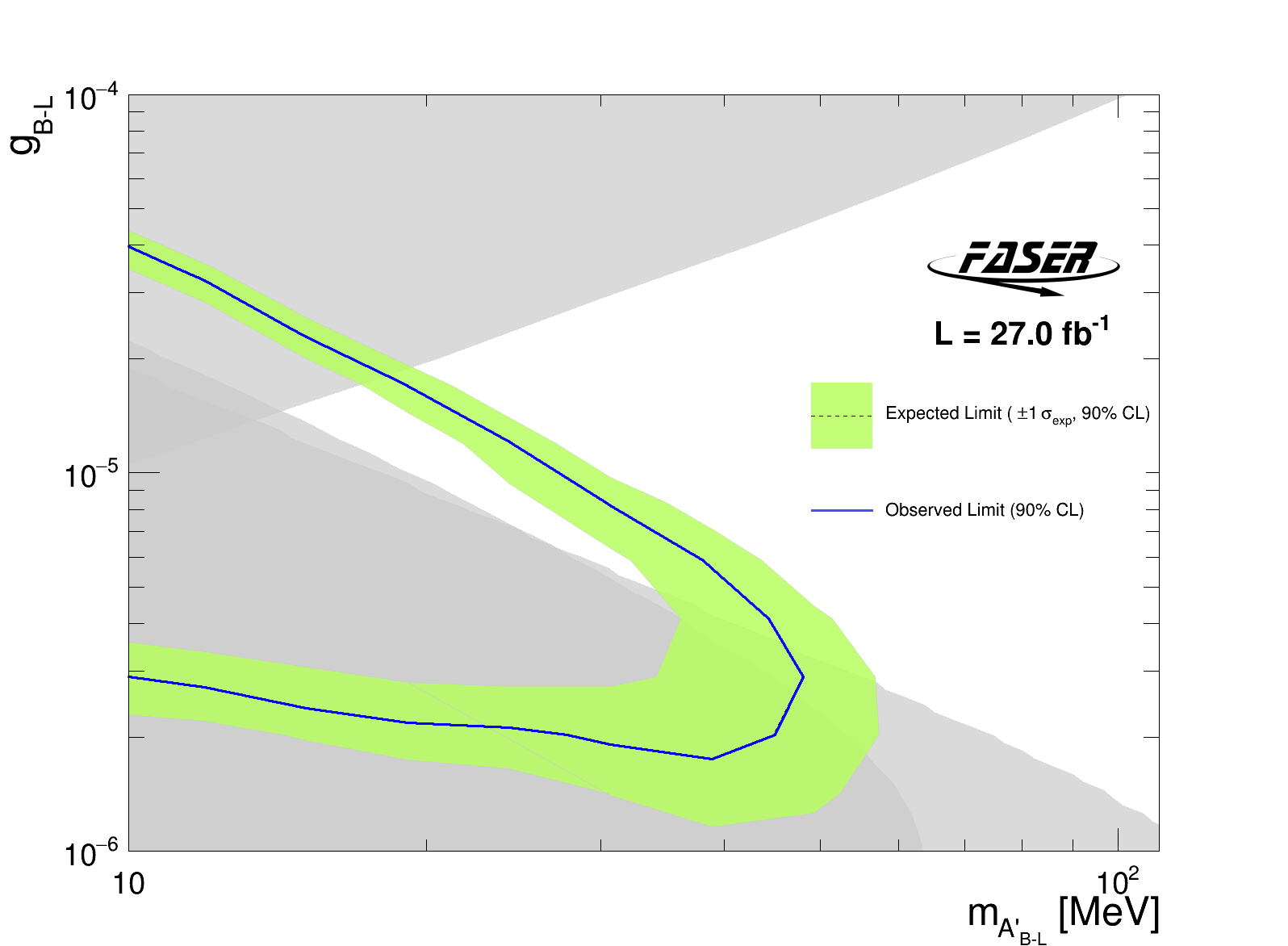}
\caption{Results from the first FASER $A'$ search. (left) The observed (expected) exclusion for dark photons as a function of mass and coupling ($\epsilon$); (right) The observed (expected) exclusion for the B-L gauge boson model as a function of mass and coupling ($\epsilon$). (These figures are taken from Ref.~\cite{FASER:2023tle}.)
}
\label{fig:Aprime-exlcusion}
\end{figure}

To avoid unconscious bias, a blind analysis technique was used. The data in the signal region was not looked at until after the analysis selections, background estimates and systematic uncertainties had been finalized. After unblinding, zero events were observed in the signal region. Figure~\ref{fig:Aprime-SR} shows the number of observed events as a function of the calorimeter energy, where the left figure shows the data with no selection and the right figure with the full signal region selection, whereas the central panel shows after an intermediate selection. Also shown in the figures are the distributions from three signal dark-photon models.
The unblinded data was used to set exclusion limits on the signal parameter space at 90\% confidence level using a profile likelihood test statistic~\cite{Cowan:2010js} and the CLs method~\cite{CLsMethod:2002Read}. The resulting exclusion contour is shown in Fig.~\ref{fig:Aprime-exlcusion} (left). 
The analysis excludes signal models in the range $\epsilon \sim 4 \times 10^{-6} - 2 \times 10^{-4}$ and $m_{A'} \sim 10~\mev - 80~\mev$, covering a region of parameter space motivated by the observed dark matter relic density.  
The analysis has also been interpreted in the B-L gauge boson model~\cite{FASER:2023tle}, with the resulting exclusion limit shown in Figure~\ref{fig:ALPs-exlcusion} (right).

{\bf Latest dark photon search}

An updated version of the dark photon search was released for the winter conferences in March 2026~\cite{FASER:DarkPhoton2026}. This analysis used a 6.5 times larger dataset than the previous search, corresponding to 177~
fb$^{-1}$ of data (the dataset taken in 2022-2024). Several innovations and improvements to the analysis strategy were implemented as briefly described below. 
\begin{itemize}
    \item The previous analysis suffered from an inefficiency caused by the two-track requirement. This was particularly  relevant for $A'$ models with larger couplings, for which the $A'$ needs to be very boosted to decay in FASER, and therefore the two $A'$ decay products are very closely spaced, and can be reconstructed as just 1 track. To improve the efficiency, the updated analysis includes a signal region requiring at least 1 track. 
    \item In addition, to increase the acceptance, an orthogonal signal region is used which allows the $A'$ to decay after the decay volume, but within the tracking spectrometer. In this case the selection doe not require any reconstructed tracks, but rather has requirements on the number of track segments reconstructed in the back two tracking stations, with selections also applied to the track segment position and angles. 
\end{itemize}
The combination of these two signal regions means FASER is twice as sensitive to $A'$ decays compared to the previous selections for the same luminosity. 

\begin{figure}[ht]
\centering
\includegraphics[width=0.6\textwidth]{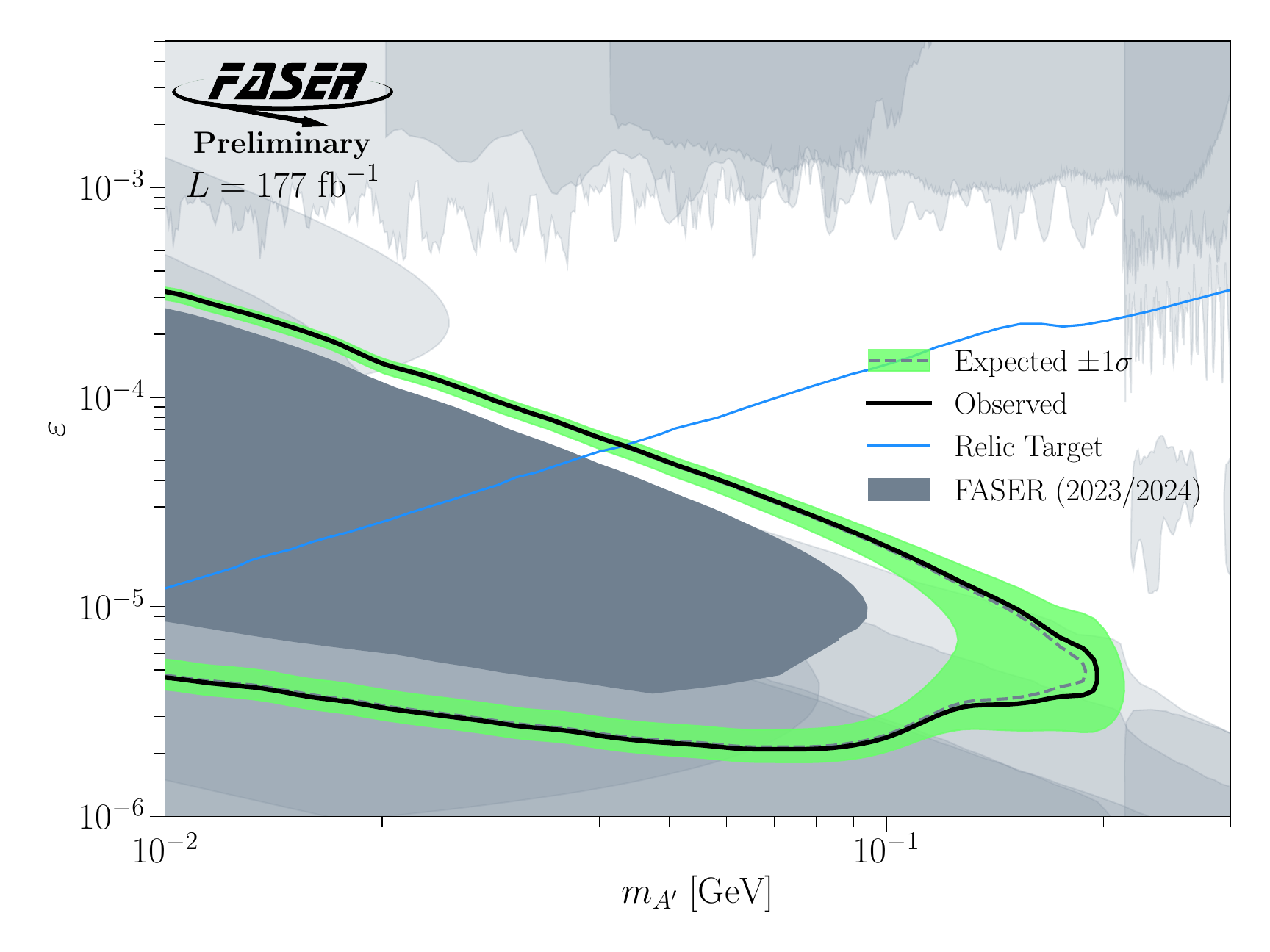}
\caption{Result from the 2026 $A'$ search, showing the observed (expected) exclusion for dark photons as a function of mass and coupling ($\epsilon$) (This figure is taken from Ref.~\cite{FASER:DarkPhoton2026}.)
}
\label{fig:dark-photon-limit-2026}
\end{figure}

The expected background for these two signal regions is estimated in a similar way as for the first analysis, leading to an almost background free search. The estimated background is dominated by neutrino interactions and is $0.03 \pm 0.01$ ($0.05 \pm 0.03$) events for the track segment (one track) signal regions. After unblinding zero events were observed in either signal region and the results were used to set the exclusion limit shown in Fig.~\ref{fig:dark-photon-limit-2026}, excluding dark photons with mass up to 200~MeV. 

\subsubsection{Neutral final states}
\label{sec:ALPs}
FASER can probe several types of ALP models~\cite{FASER:2018eoc} and is particularly sensitive to scenarios in which the ALP couples to the SU(2)$_L$ gauge bosons. In this model, the coupling to SU(2)$_L$ gauge bosons is present before electroweak symmetry breaking (EWSB)~\cite{izaguirre2016new, gori2020koto, Kling_2020}. After EWSB, the ALP couples to both photons and the weak gauge bosons. This model is labeled ALP$_W$, and its phenomenology is determined by the ALP mass ($m_a$) and coupling ($g_a$). 

\begin{figure}[ht]
\centering
\includegraphics[width=0.99\textwidth]{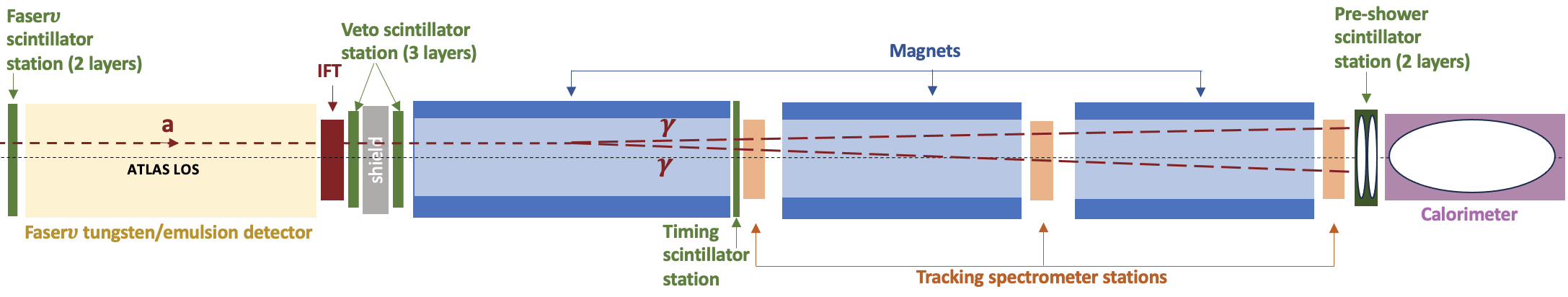}
\caption{A sketch of an ALP (a) decaying into two photons inside FASER. (This figure is taken from Ref.~\cite{FASER:2024bbl}.)
}
\label{fig:ALP-sketch}
\end{figure}

The search for ALPs uses the data collected in 2022 and 2023, corresponding to an integrated luminosity of 57.7~fb$^{-1}$~\cite{FASER:2024bbl}. 
Since the signal does not produce charged particles, the analysis is agnostic to activity in the FASER tracker, and the effective decay volume is the full length of FASER from after the veto scintillators to before the preshower (corresponding to a decay volume length of 4~m, 2.7 times larger than that in the first $A'$ analysis discussed above). The main target of this search is the ALP$_W$ model discussed above, however the analysis is also interpreted in several other scenarios which lead to two or more photons in the final state, and are discussed later. 
A sketch of the signal topology is shown in Fig~\ref{fig:ALP-sketch}.
The event selection rejects any incoming charged particles with the same requirements on the signals from the veto scintillators in the $A'$ search, but additionally no signal in the larger Trigger/timing scintillators is required. A significant background arises from neutrino interactions in the calorimeter material, which can fake the signal topology. This is partially mitigated by requiring the signature in the simple scintillator-based preshower is consistent with an EM shower starting to develop. The requirement is on the ratio of the measured charge in the downstream preshower scintillator to the upstream one. For an EM shower the ratio should be greater than one as more charge is deposited downstream as the shower develops, whereas for a MIP the ratio should be close to one. This variable was studied and found to be effective for separating electrons from muons and pions in an SPS testbeam~\cite{testbeampaper}.
In order to further control the background from neutrinos interacting in the preshower or calorimeter, it was necessary to have a tight requirement on the calorimeter energy of $E>$1.5~\tev .~\footnote{The upgraded preshower discussed in Section~\ref{sec:upgrades}, will allow the neutrino background to be controlled with a much looser calorimeter energy selection, opening up sensitivity to additional signal parameter space.} 
Despite this tight selection, the analysis is sensitive to a large part of the signal parameter space, with typical efficiencies for ALPs decaying in FASER of 75\%.

\begin{figure}[ht]
\centering
\includegraphics[width=0.49\textwidth]{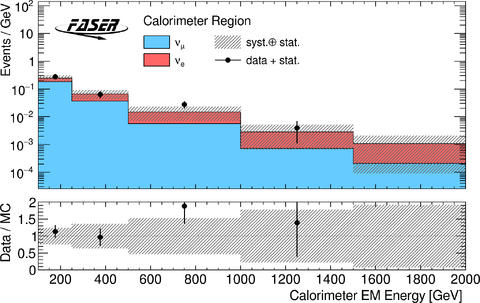}
\includegraphics[width=0.49\textwidth]{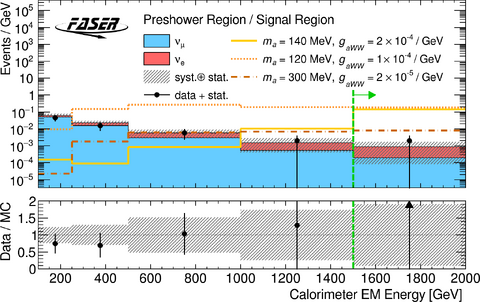}
\caption{(left) Data control region with the preshower-ratio selection inverted for validating the ALP background estimate, showing the data and neutrino background estimate as a function of the calorimeter energy. (right) The lower bins form a data control region (with all other selections the same as for the signal region)  for validating the ALP background estimate, whereas the highest energy bin is the analysis signal region. Also shown are three example ALP$_W$ signal models. 
(These figures are taken from Ref.~\cite{FASER:2024bbl}.)
}
\label{fig:ALPs-regions}
\end{figure}

Although several background processes were studied the only background which is non-negligible is from neutrinos interacting in the detector. This is estimated with large MC simulation samples, with the neutrino production and interaction modeling following the recipe described in Section~\ref{sec:neutrinoPhys}. The predicted background is 0.44 $\pm$ 0.05 (stat.) $\pm$ 0.39 (syst.) events. The background estimate is validated in data control regions with reverted selections on the calorimeter energy or preshower-ratio, as shown in Fig.~\ref{fig:ALPs-regions} (note the highest energy bin in the right hand plot is showing the signal region).

Similarly to the $A'$ search the systematic uncertainties on the expected signal yield are dominated by theoretical modeling uncertainties ($\mathcal{O}(60\%)$), whereas the experimental uncertainties are much smaller with the largest related to the calorimeter energy scale ($\mathcal{O}(15\%)$) and the preshower-ratio selection ($\mathcal{O}(8\%)$).

\begin{figure}[ht]
\centering
\includegraphics[width=0.6\textwidth]{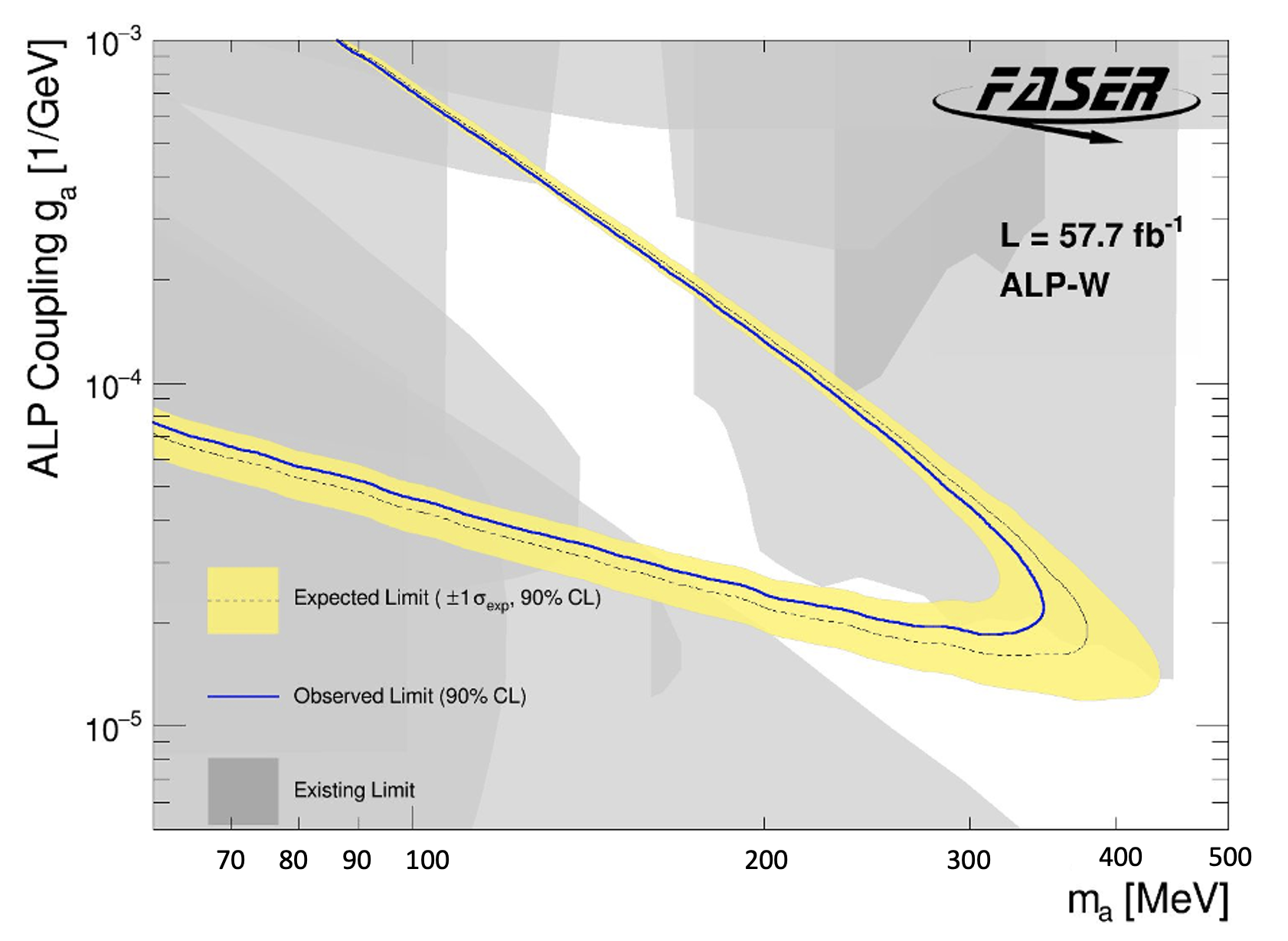}
\includegraphics[width=0.6\textwidth]{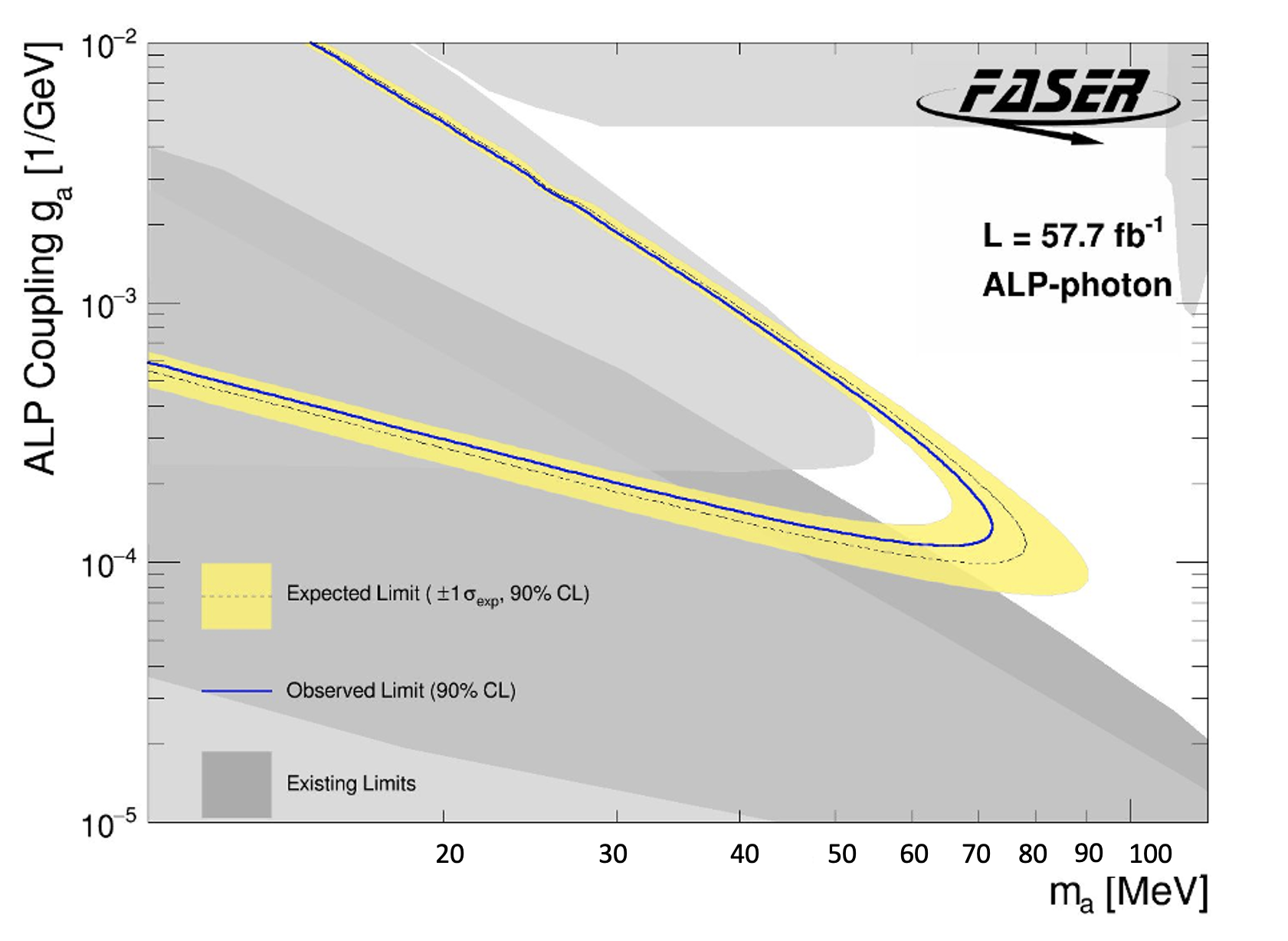}
\caption{FASER exclusion for two ALP models. (left) The exclusion for the ALP-W model; (right) The exclusion for the ALP-photon model. 
(These figures are taken from Ref.~\cite{FASER:2024bbl}.)
}
\label{fig:ALPs-exlcusion}
\end{figure}

After validating the background estimates the signal region was unblinded and one event was observed as can be seen in Fig.~\ref{fig:ALPs-regions} (right), consistent with the background estimate. The data were used to produce exclusion curves in the ALP signal parameter space using the same statistical methods as for the $A'$ search. Fig.~\ref{fig:ALPs-exlcusion} (left) shows the corresponding exclusion curve at 90\% CL for the ALP-W model.

The result was interpreted in two other ALP models with two photon final states, where the ALP only couples to photons~\cite{Beacham:2019nyx}, and the ALP only couples to gluons~\cite{Aloni:2018vki}. In addition, three other non-ALP models, with two or more photons in the final state, were considered: an Up-philic scalar model, a U(1)B vector boson model and a Type-I “fermiophobic” two Higgs doublet model, with new sensitivity achieved in all of these models.

The model with the ALP coupling solely to photons is based on an interesting production mechanism originally discussed in Ref.~\cite{Feng:2018noy}. The ALP is produced via the Primakoff process, when high energy photons produced in the LHC collisions in ATLAS and traveling along the LOS, interact with the material in the TAN (a copper absorber, 140~m from the IP, designed to protect the super conducting LHC magnets from the collision debris in the forward direction). ALPs produced in this way typically travel along the LOS with high energy. However, compared to the other BSM searches in FASER the ALP must travel only 340~m to decay inside FASER, rather than the 480~m needed for particles produced at the IP. The exclusion contour for this model is shown in Fig.~\ref{fig:ALPs-exlcusion} (right). 

\newpage
\subsection{Neutrino results}

FASER has released several neutrino results using analysis of both the electronic detector, and the FASER$\nu$ emulsion detector. The electronic detector can observe muon neutrino CC interactions in the FASER$\nu$ tungsten via a muon-appearance signature, with the momentum and charge of the muon being measured in the spectrometer, allowing to separate $\nu$ and $\overline{\nu}$. Whereas the emulsion detector can observe neutral vertices with associated leptons to identify $\nu_e$, $\nu_\mu$ and eventually $\nu_\tau$ CC interactions as well as neutral current (NC) interactions, but without separating $\nu$ and $\overline{\nu}$. In the future it is expected that FASER will be able to combine these two approaches and be able to measure the charge of muons from $\nu_\mu$ (and $\nu_\tau$ with $\tau \to \mu$ decays) vertices observed in the emulsion, but this has not been done yet.   For all approaches the incoming neutrino flux is predicted as described in Section~\ref{sec:neutrinoPhys}. 

\subsubsection{Pilot run emulsion detector}
\begin{figure}[ht]
\centering
\includegraphics[width=0.35\textwidth]{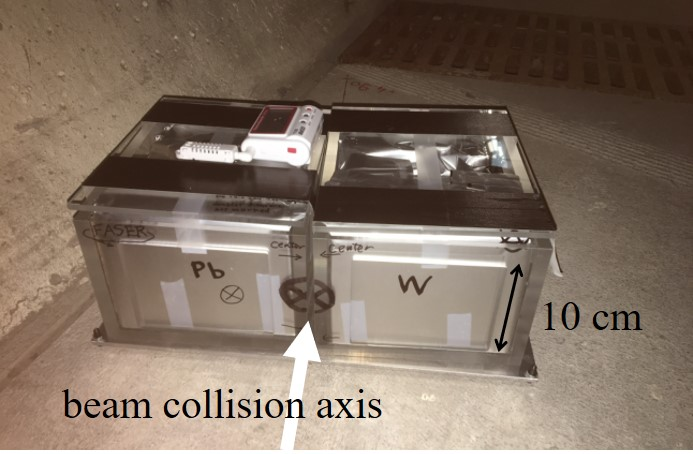}
\includegraphics[width=0.63\textwidth]{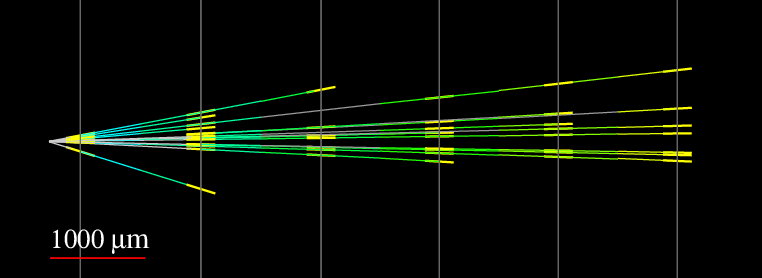}
\caption{ (left) A picture of the pilot emulsion detector installed in TI12 in 2018. (right) An event display showing a candidate neutrino interaction vertex from the pilot emulsion detector analysis. (These figures are taken from Ref.~\cite{FASER:2021mtu}.) 
}
\label{fig:pilotDetector}
\end{figure}

As part of the initial feasibility studies for FASER, a small 29~kg tungsten/lead emulsion detector, shown in Fig.~\ref{fig:pilotDetector} (left), was installed into the FASER location for a few weeks in 2018 LHC running and was exposed to a luminosity of 12.2~fb$^{-1}$. This detector was used to measure the muon background rates, but the data was also analysed to search for neutrino candidates~\cite{FASER:2021mtu}. For the neutrino analysis a target mass of 11~kg was considered. Given the small size of the detector no lepton identification was possible, which increases the background from neutral hadron interactions. To allow sufficient background rejection a Boosted Decision Tree (BDT) was used to separate the signal from the background. A fit to the BDT classifier yielded 6.1 signal and 11.9 background vertices, with the background-only hypothesis being disfavored with a statistical significance of 2.7$\sigma$. The most signal-like vertex is shown in Fig.~\ref{fig:pilotDetector} (right).

\subsubsection{Neutrino results using the electronic detector}
{\bf First observation of collider neutrinos with the electronic detector}

\begin{figure*}[ht]
\centering
\includegraphics[width=0.99\textwidth]{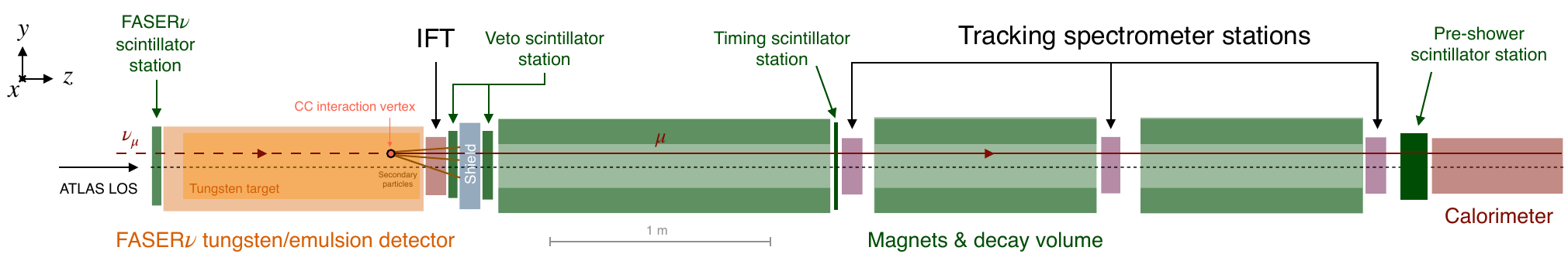}
\caption{Sketch of the electronic neutrino analysis, taken from Ref.~\cite{FASER:2023zcr}.
}
\label{fig:elec-neutrino-EVD}
\end{figure*}

\begin{figure}[ht]
\centering
\includegraphics[width=0.55\textwidth]{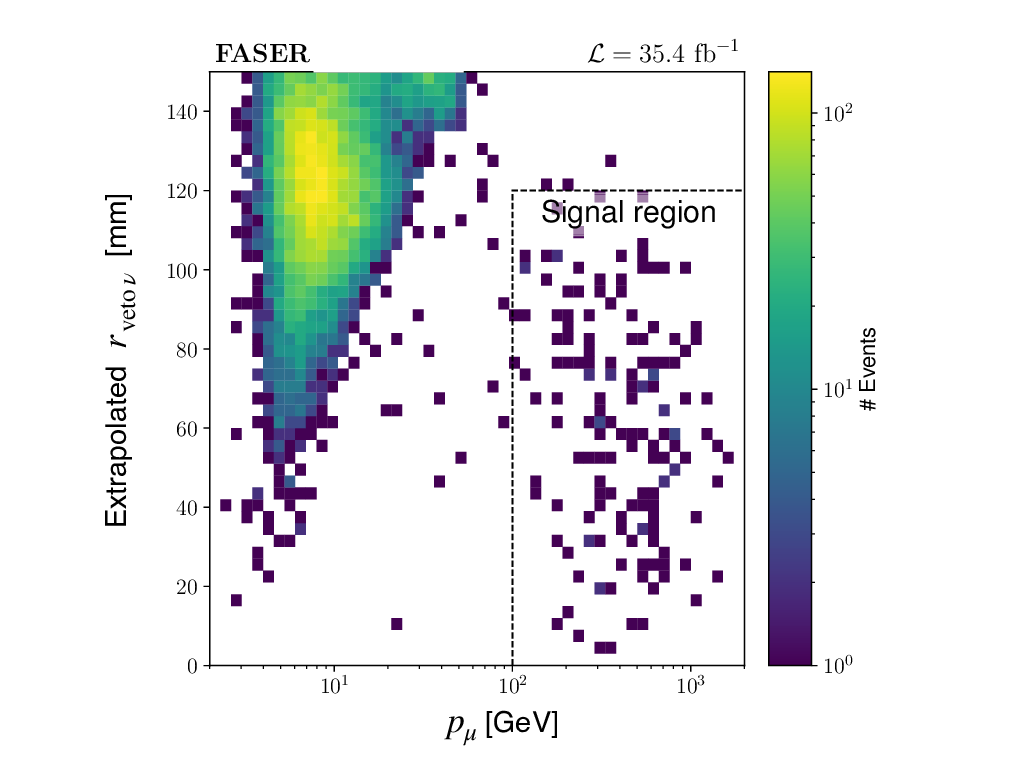}
\includegraphics[width=0.44\textwidth,height=0.37\textwidth]{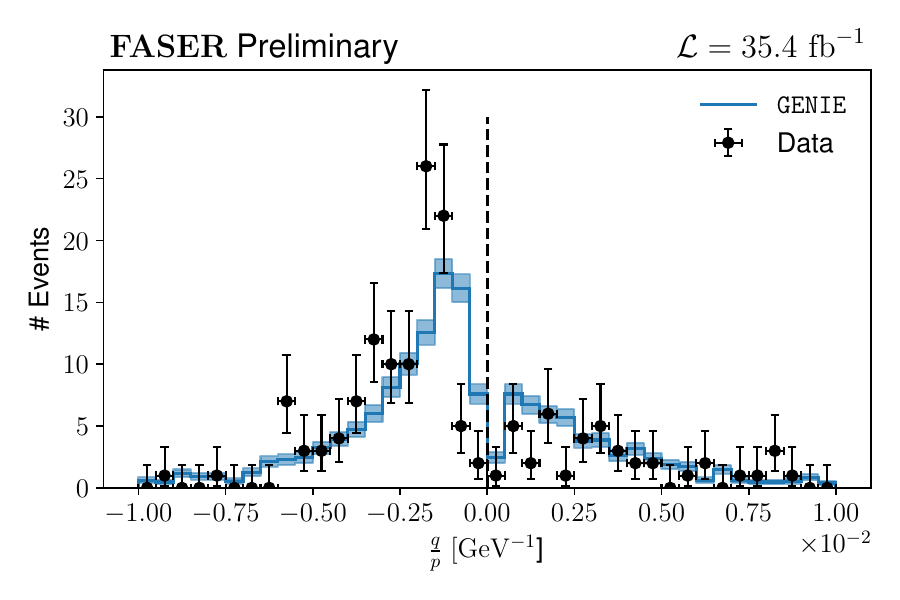}
\caption{(left) The muon momentum ($p_\mu$) versus the transverse distance from the extrapolated muon track to the centre of the veto scintillators (r$_{\rm{veto}\nu}$) showing the clear separation between the signal and background events. (right) The $q/p$ distribution for the selected events compared to the neutrino simulation (GENIE). (These figures are taken from Ref.~\cite{FASER:2023zcr}.)
}
\label{fig:elec-nu-1}
\end{figure}

FASER observed neutrinos produced at a particle collider for the first time in March 2023~\cite{FASER:2023zcr}, using the dataset recorded in 2022, corresponding to 35.4~fb$^{-1}$. The analysis was designed to observe a robust signal for $\nu_\mu$ CC interactions but not to correct the observation for detector effects. As sketched in Fig.~\ref{fig:elec-neutrino-EVD}, events were selected by requiring no signal in the two veto scintillators in front of the target, and a reconstructed track in the spectrometer with momentum greater than 100~GeV. To mitigate background from incoming muons missing the veto scintillators, the track when extrapolated to the front veto system was required to be within 120~mm from the detector center. With this selection 151 $\pm$ 41 signal events were expected with a negligible background. On unblinding the data 150 events were observed, constituting the first observation of neutrino interactions at a collider with a large statistical significance. Fig~\ref{fig:elec-nu-1} left shows the observed data in the plane of the track momentum and the extrapolated track distance from the detector center (r$_{\rm{veto}\nu}$), this shows a very clear signal at high momentum and with r$_{\rm{veto}\nu}$ less than 120~mm, well separated from the background. Fig~\ref{fig:elec-nu-1} right shows, the measured charge divided by momentum ($q/p$) for the selected events, also compared to the expectation from GENIE simulation, which shows that both neutrinos and anti-neutrinos are observed with about the expected ratio, and that the sample includes neutrinos with high energies above several hundred GeV.  

\newpage

{\bf First $\nu_\mu$ differential measurements with the electronic detector}
\label{sec:e-neutrino}

In December 2024, FASER released the first measurement of the muon neutrino interaction cross section in bins of neutrino energy and separated for neutrino and antineutrino~\cite{FASER:2024ref}. As for the first observation of neutrinos at the LHC, this analysis focused on muon neutrino CC interactions in the FASER$\nu$ target with the electronic detector components, and uses the measured momentum and charge of the reconstructed muon candidate to infer if the interaction was from a neutrino or antineutrino and its energy. The measurement was carried out using the 2022 and 2023 dataset corresponding to about 66~fb$^{-1}$ of data. Events are selected in a similar way to the previous analysis (discussed above) with some small modifications to improve the signal efficiency and to reduce the systematic uncertainties. 

The signal is defined as $\nu_\mu$ CC neutrino interactions in a fiducial volume corresponding to a 200~mm diameter cylinder in the target, centered on the electronic detector axis, and with neutrino energy greater than 100~GeV.  The analysis selects 362 data events, corresponding to $338.1 \pm 21.0$ observed $\nu_\mu$ CC candidates in the defined fiducial volume, with $298.4 \pm 42.6$ events expected. The background is dominated by non-$\nu_\mu$ CC neutrino interactions, and $\nu_\mu$ CC neutrino interactions outside the fiducial volume. Correcting for the selection efficiency yields a total of 1242.7 $\pm$ 137.1 CC muon neutrino interactions inside the fiducial volume.

\begin{figure}[ht]
\centering
\includegraphics[width=0.49\textwidth]{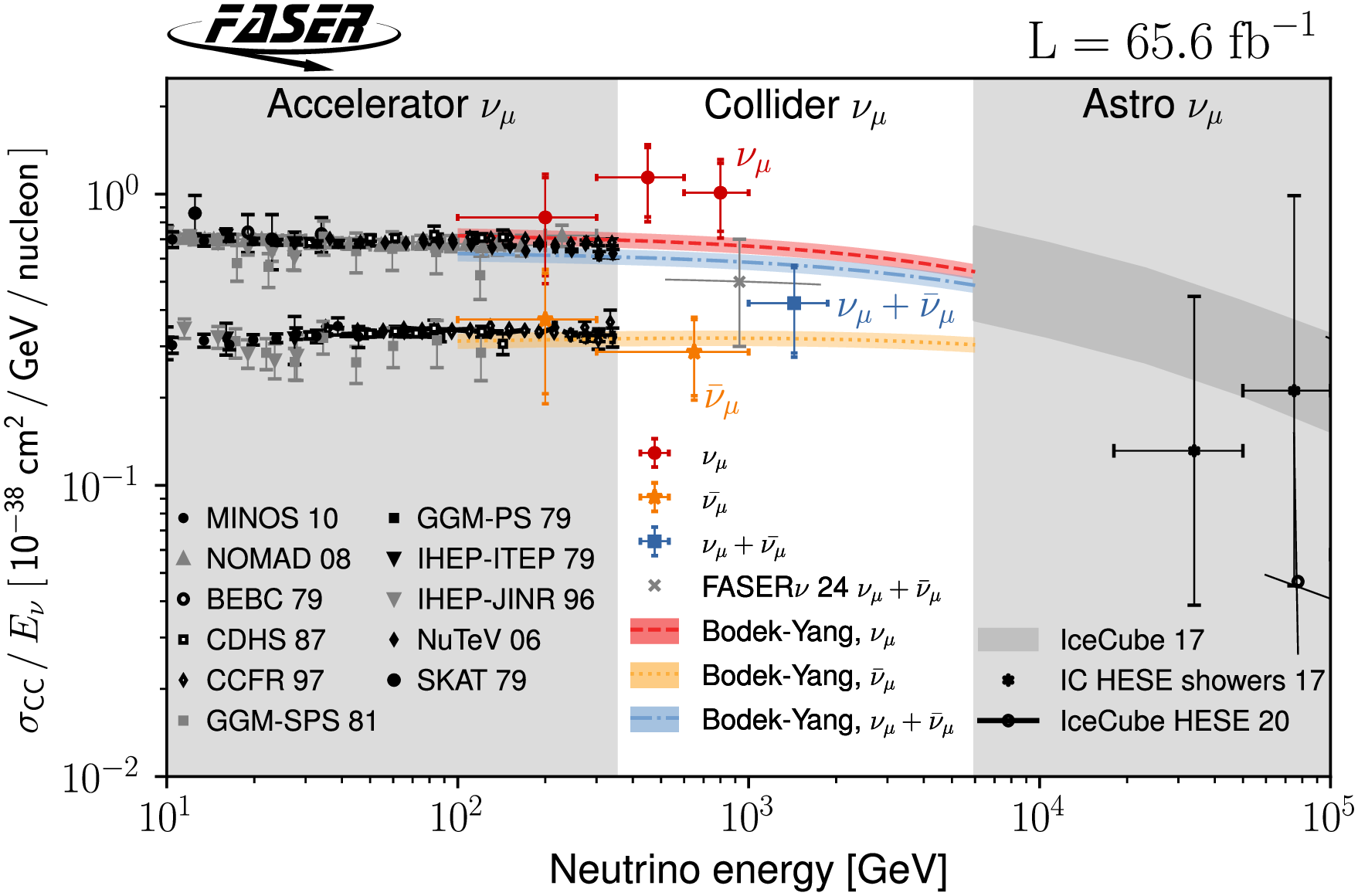}
\includegraphics[width=0.49\textwidth]{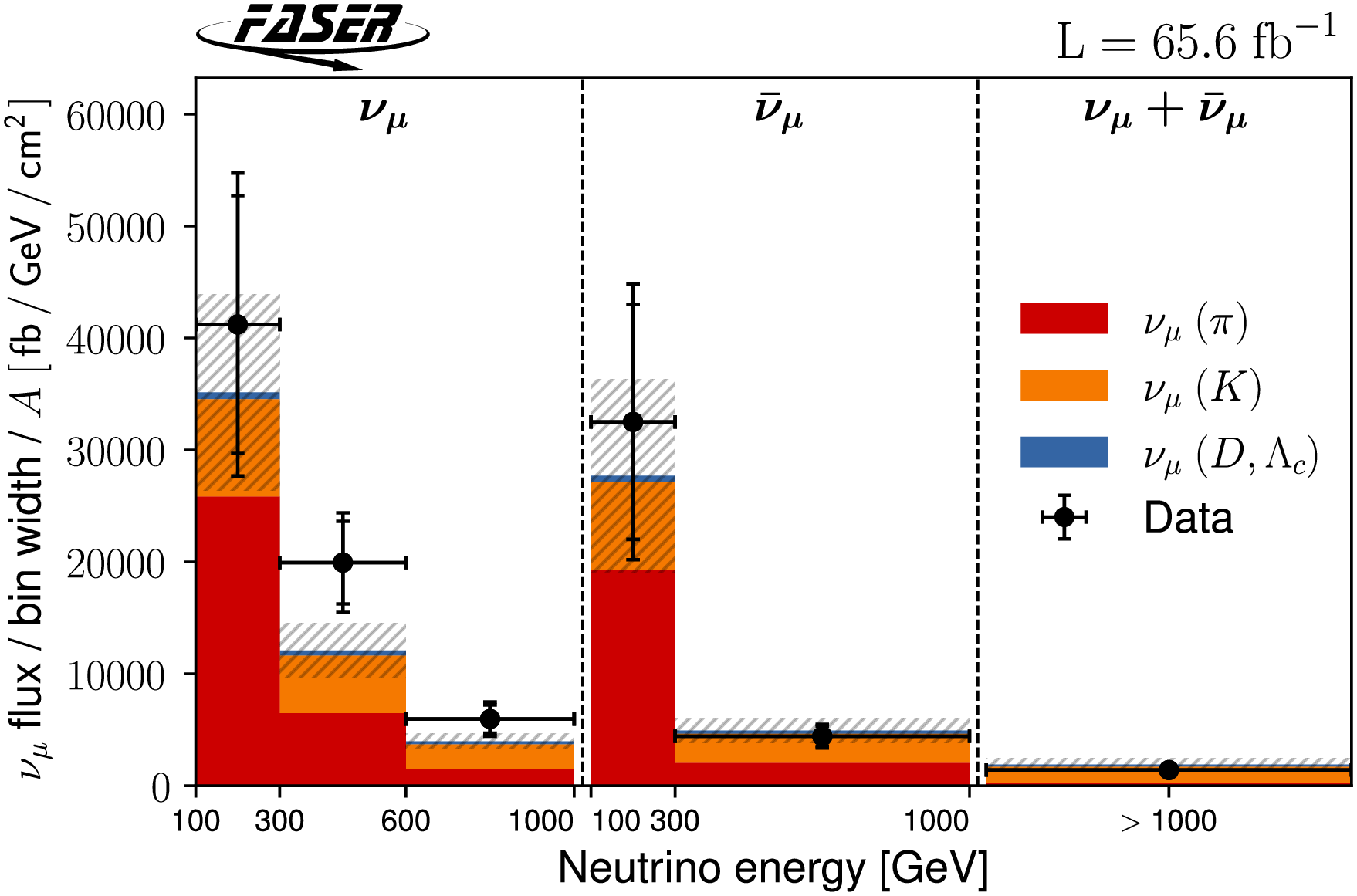}
\caption{(left) The measured neutrino and antineutrino cross section as a function of the neutrino energy. Coloured points show the FASER measurements, and the coloured lines the expectation. Also shown are measurements from beam dump experiments at lower energy, and from IceCube at very high energy. (right) The FASER measurement of the muon (anti)neutrino flux, with the expectation from pion, kaon and charm decays shown as the coloured histograms. (These figures are taken from Ref.~\cite{FASER:2024ref}.)
}
\label{fig:elect-neutrino2}
\end{figure}

For signal events passing the event selection, the muon retains on average 80\% of the neutrino energy. The {\it calibrated} muon momentum $p'_\mu$=$p_\mu$/0.8 is therefore used to determine the number of selected events as a function of neutrino energy, as follows. The selected events are divided into six bins of the ratio of charge to calibrated momentum ($q/p'_\mu$) with bin edges:
\begin{align}
\left[ -\frac{1}{100}, -\frac{1}{300}, -\frac{1}{600}, -\frac{1}{1000}, \frac{1}{1000}, \frac{1}{300}, \frac{1}{100} \right] \, \ \rm{/GeV} \, . \nonumber
\end{align} 
This gives three momentum bins for negatively charged muons (which originate from neutrino interactions), two momentum bins for positively charged muons (from antineutrino interactions), and one high momentum bin including muons of either charge with $p'_\mu >$ 1~TeV, since the charge identification performance degrades at high momentum. The binning is defined to have a similar number of expected events in each bin. The selected events are unfolded to give the number of (anti)neutrino events in bins of neutrino energy with the same bin ranges. The results are interpreted in two ways, firstly the expected neutrino flux (from simulation, as described in Section~\ref{sec:physicscase}) is used to derive the (anti)neutrino cross section as a function of energy, which is shown in Figure~\ref{fig:elect-neutrino2} (left). Secondly, the theoretical cross section is assumed and the neutrino flux is derived from the data, as shown in Fig.~\ref{fig:elect-neutrino2} (right). The measured flux is then used to put a first constraint on forward charged pion and kaon production at the LHC. The dominant uncertainties affecting the cross section measurement arise from the event statistics and systematic uncertainties related to the neutrino flux, and the modeling of the neutrino interactions, whereas experimental systematic uncertainties are sub-dominant.

\begin{figure}[ht]
\centering
\includegraphics[width=0.45\textwidth]{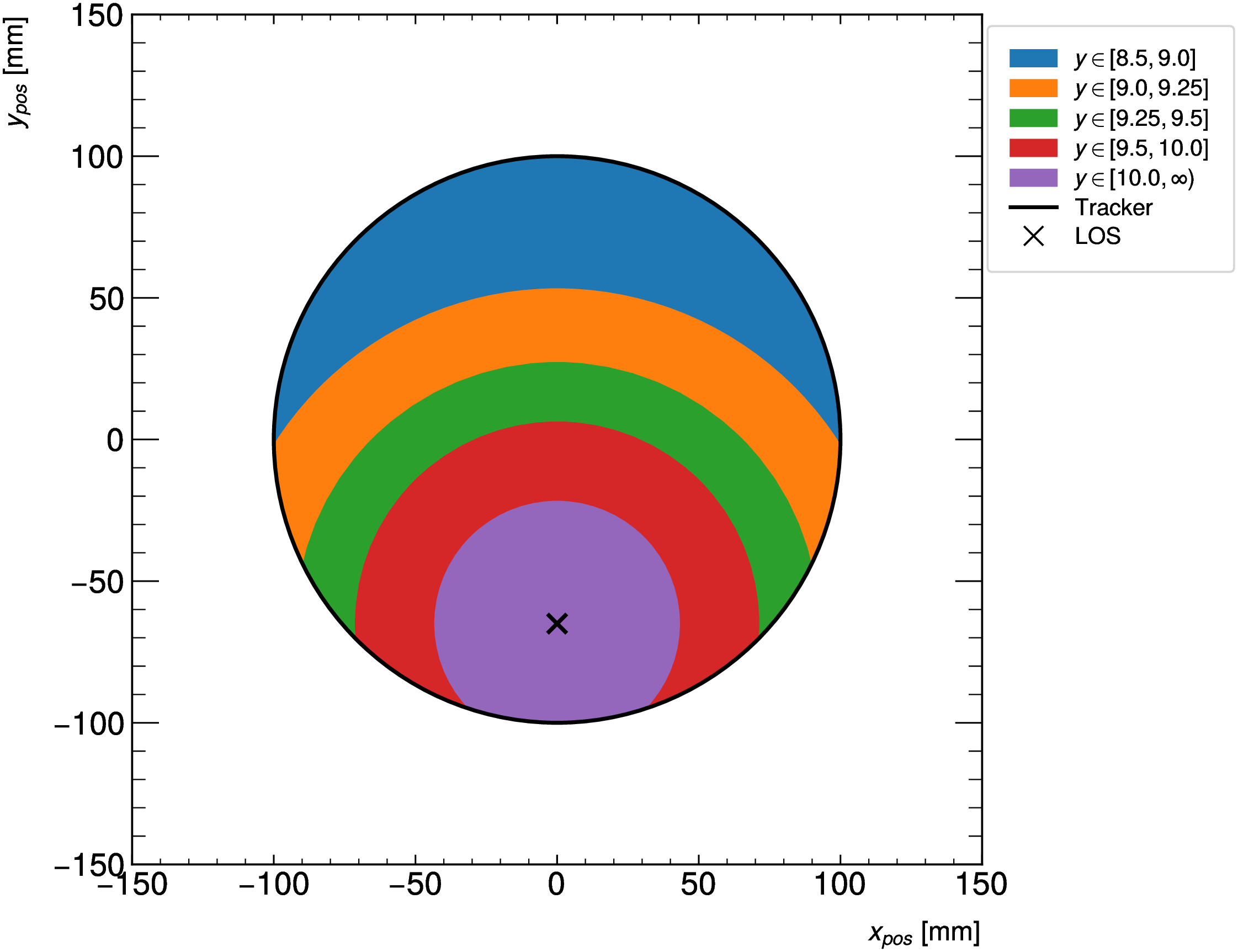}
\includegraphics[width=0.49\textwidth]{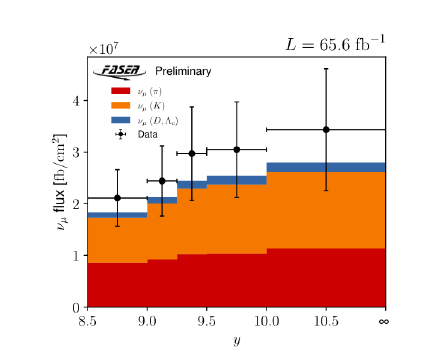}
\caption{(left) The different annular bins used to measure the neutrino flux as a function of rapidity ($y$). (right) The measured neutrino flux as a function of rapidity ($y$). (These figures are taken from Ref.~\cite{FASER:2927629}.)
}
\label{fig:neut-rapidity}
\end{figure}

A followup of this analysis was carried out using the same dataset and with the same event selection and background estimation, but where the results were presented as a function of neutrino rapidity~\cite{FASER:2927629}. The data were binned based on the transverse distance of the reconstructed muon track from the LOS when extrapolated longitudinally to the center of the FASER$\nu$ detector, and then unfolded into five bins of the neutrino rapidity ($y$). Fig.~\ref{fig:neut-rapidity} (left) shows the different annular regions around the LOS and in the spectrometer acceptance, that correspond to the different rapidity bins. Fig.~\ref{fig:neut-rapidity} (right) then shows the measured neutrino flux as a function of the rapidity 

{\bf Latest $\nu_\mu$ measurements with the electronic detector}

\begin{figure}[ht]
\centering
\includegraphics[width=0.9\textwidth]{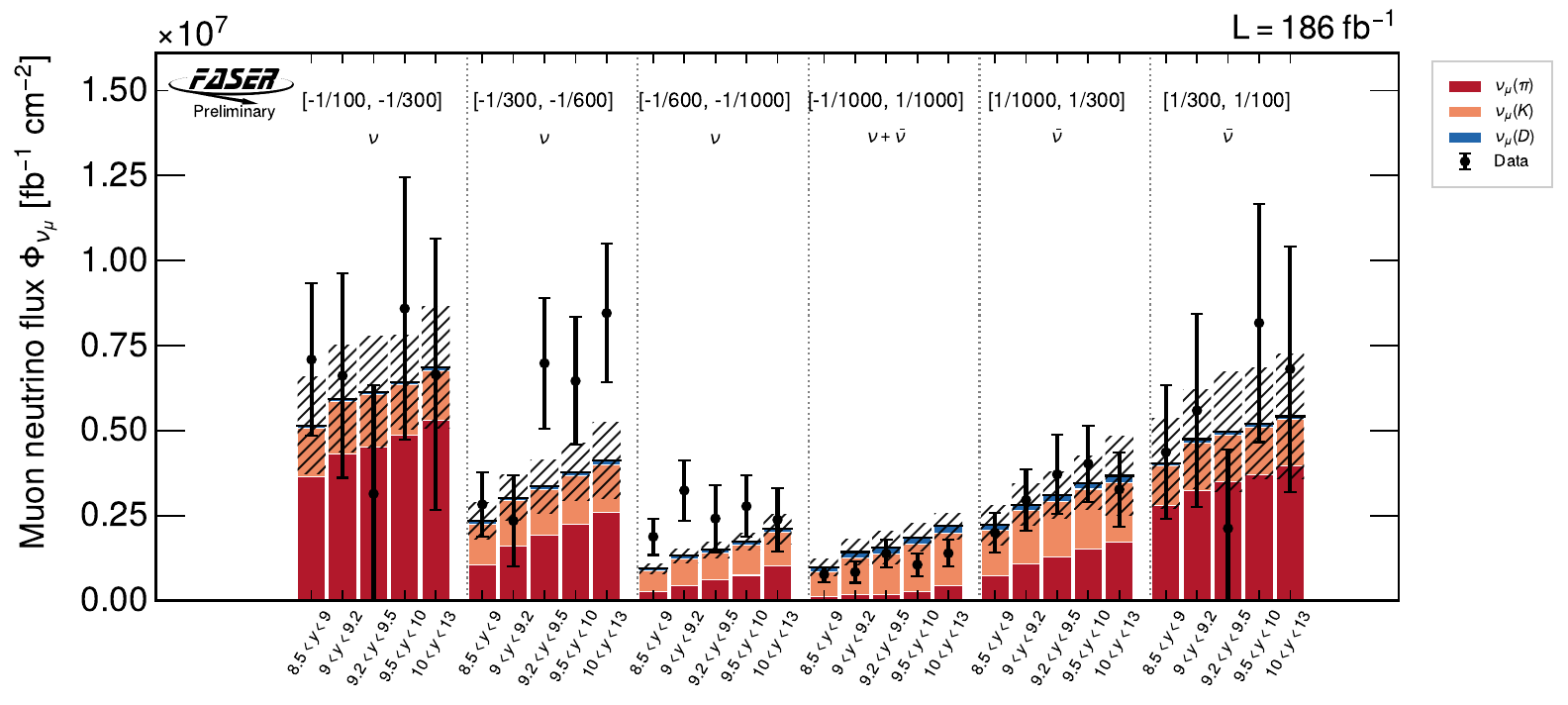}
\caption{The measured neutrino flux in bins of energy and rapidity ($y$). (These figures are taken from Ref.~\cite{FASER:elecNeut2026}.)
}
\label{fig:eneut_2d-binning}
\end{figure}

In March 2026 an updated analysis was released~\cite{FASER:elecNeut2026} in which the measured $\nu_\mu$ neutrinos and antineutrinos were unfolded into a 2D binning of energy and rapidity. The analysis followed the same strategy as described above but included the 2024 data to use a total dataset of 186~fb$^{-1}$. After subtracting the expected backgrounds, $766.8 \pm 29.6$
 interaction events are observed. The unfolded measurement is presented in Fig.~\ref{fig:eneut_2d-binning}.

{\bf Observation of $\nu_e$ with the electronic detector}

In March 2026 the first search for $\nu_e$ CC interactions with the electronic detector was released~\cite{FASER:ALPtrino2026}. This analysis was motivated by the fact that the calorimeter control region from the ALPs search (discussed in Section~\ref{sec:ALPs} and shown in Fig.~\ref{fig:ALPs-regions}) was dominated by $\nu_e$ CC interactions for calorimeter energy above about 500~GeV. The analysis selects events with large calorimeter energy and no signal in any of the veto scintillators, with the data binned in four bins of calorimeter energy from 250~GeV. The background is dominated by muon neutrino interactions in the calorimeter material, since the $\nu_\mu$ interactions typically do not produce a high energy electromagnetic shower the background is reduced by selecting events with large calorimeter energy. However, given that the incoming flux of $\nu_\mu$ is almost an order of magnitude larger than the signal $\nu_e$ flux, care needs to be taken to remove the $\nu_\mu$ background. This is done by using the measured $\nu_\mu$ measurement described in Section~\ref{sec:e-neutrino}, where the measured events are unfolded into the calorimeter energy bins used in this analysis. This approach removes the large uncertainty on the neutrino flux in the background estimation. 

\begin{figure}[ht]
\centering
\includegraphics[width=0.47\textwidth]{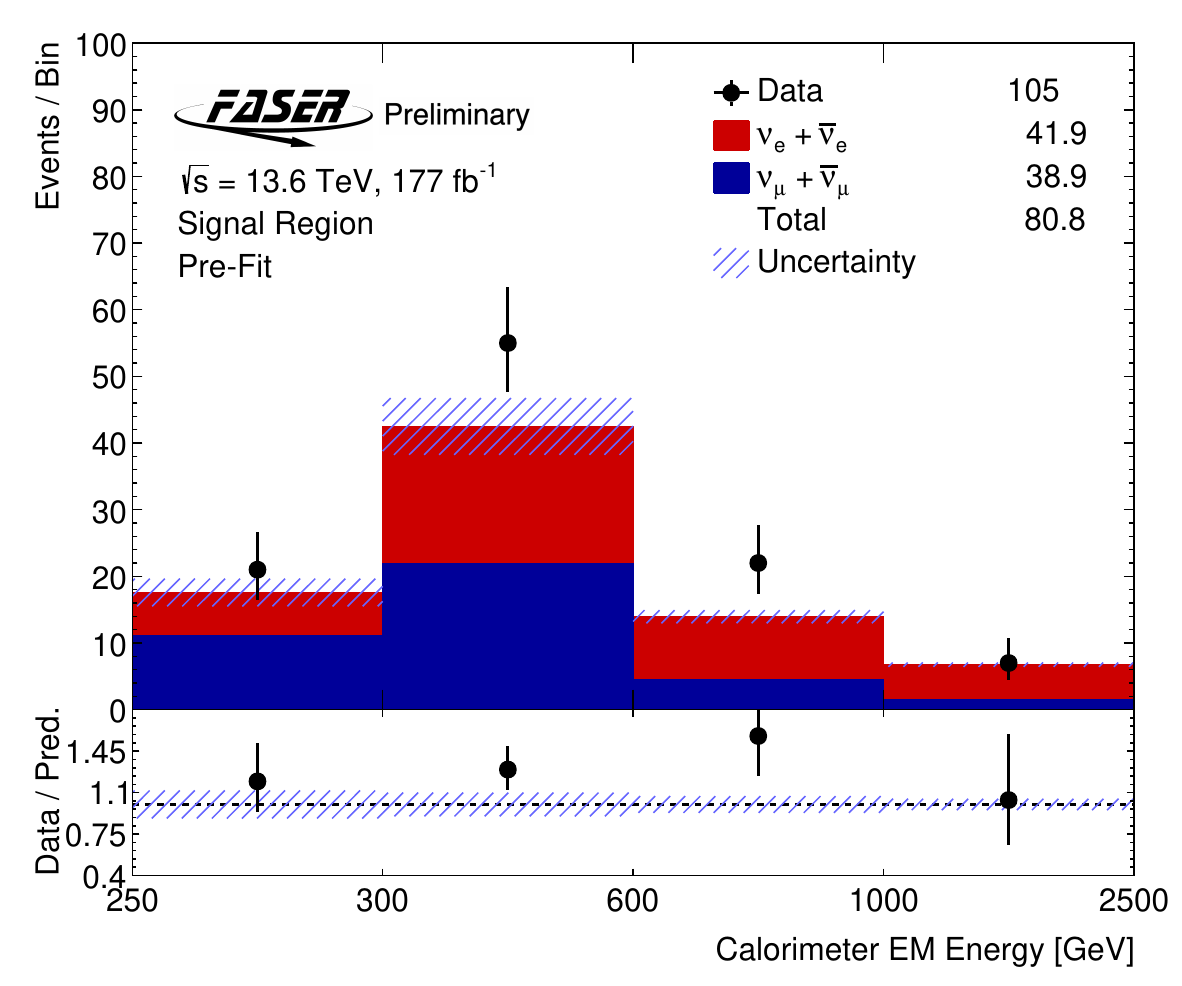}
\includegraphics[width=0.47\textwidth]{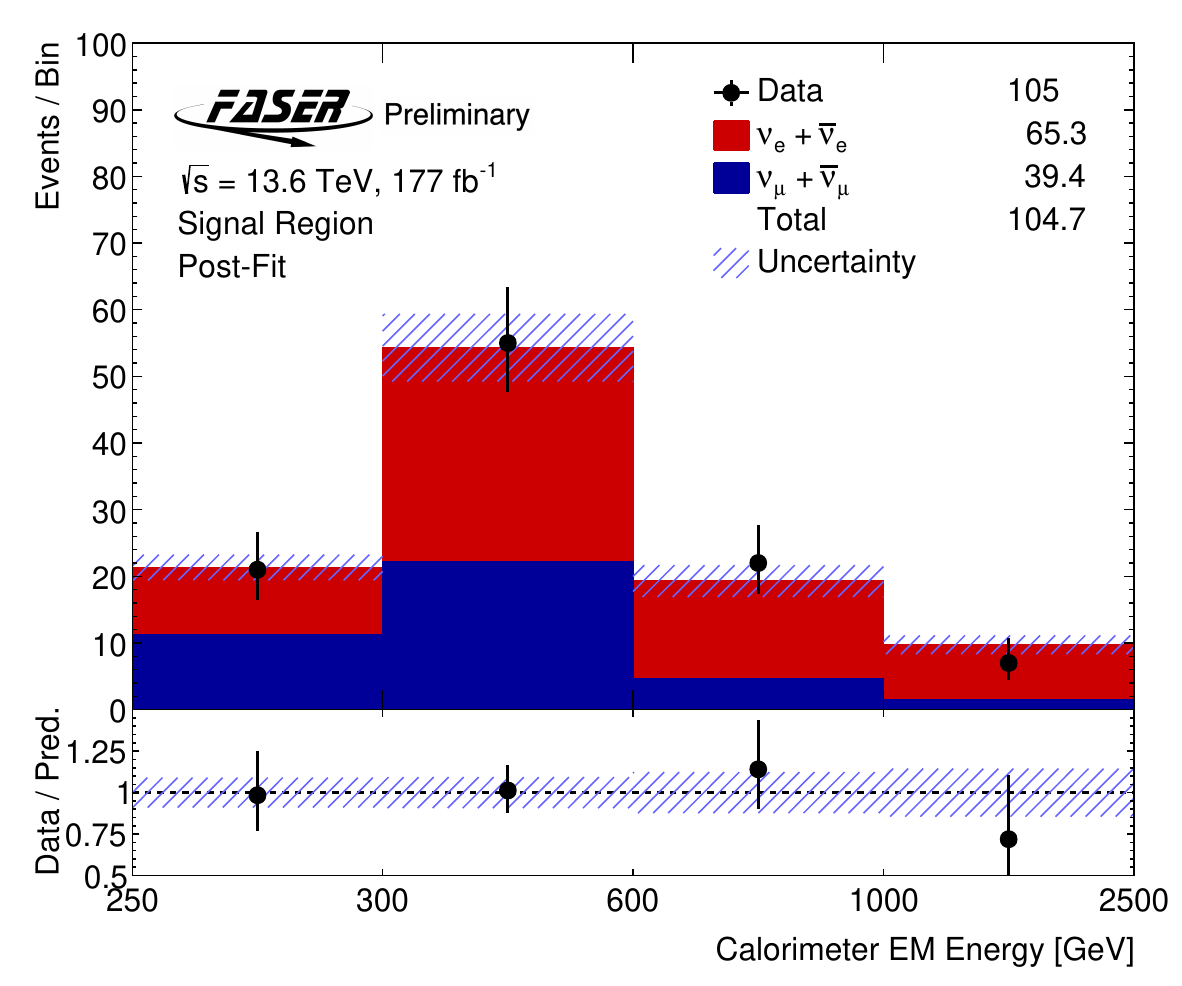}
 \includegraphics[width=0.47\textwidth]{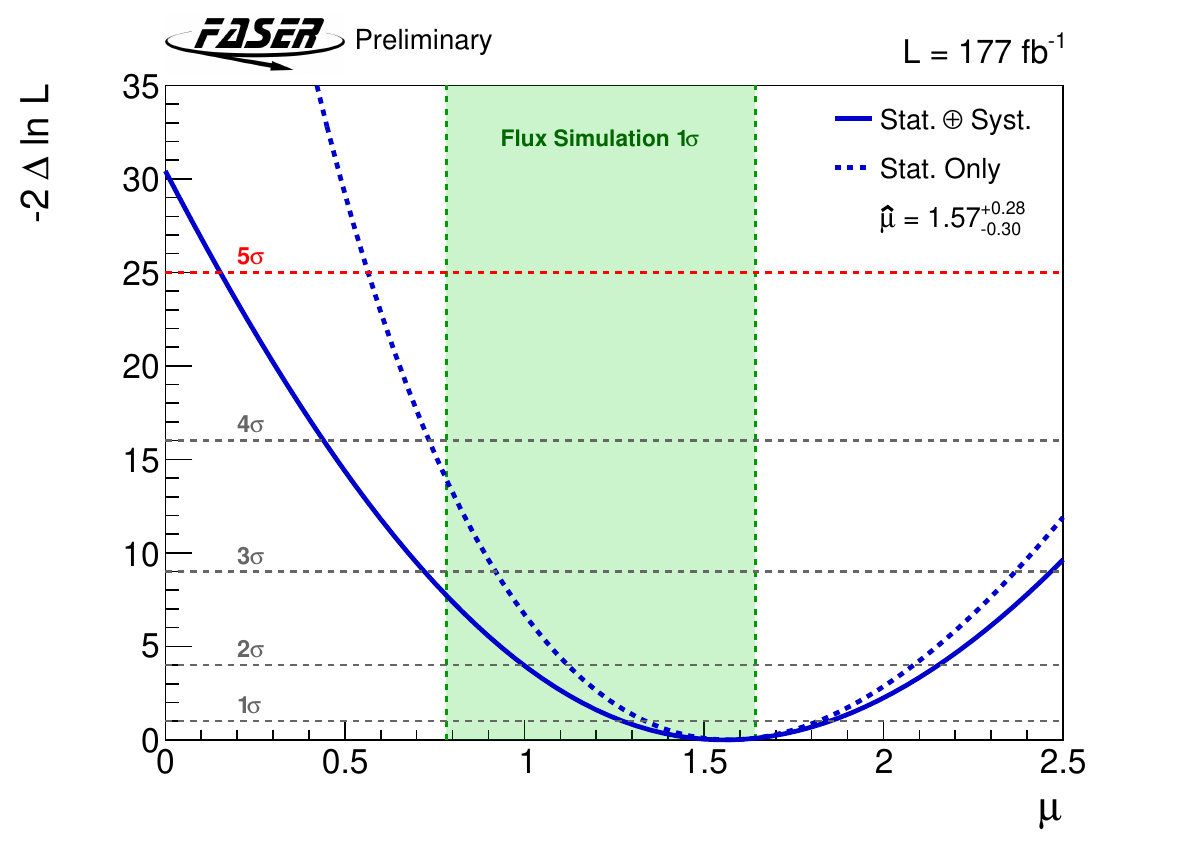}
\caption{(top left) The number of events observed in the calorimeter energy bins, compared to the expectation from $
\nu_\mu$ and $\nu_e$. (top right) The same with the $\nu_e$ normalization fitted to the data. 
(bottom) The likelihood scan for the $\nu_e$ normalization parameter $\mu$.
(These figures are taken from Ref.~\cite{FASER:ALPtrino2026}.)
}
\label{fig:ALPtrino}
\end{figure}

As shown in Fig.~\ref{fig:ALPtrino}, a clear excess of events above the $\nu_\mu$ background-only expectation is observed. This corresponds to $65 \pm 12$ events, consistent with the expected electron-neutrino signal of $42 \pm 27$. The background-only hypothesis is
rejected with a significance of 5.5 standard deviations, as shown the bottom plot in Fig.~\ref{fig:ALPtrino}. This constitutes the first observation of collider $\nu_e$ interactions with an electronic detector.

\clearpage

\subsubsection{Results with FASER\texorpdfstring{$\nu$}{nu}}

{\bf First FASER$\nu$ results}

Charged current neutrino interaction vertices are searched for in the first FASER$\nu$ emulsion box exposed to 9.5 fb$^{-1}$ of data in 2022. The first result~\cite{FASER:2024hoe}, used a fraction of the emulsion that had been analyzed at that time, corresponding to a target mass of 130~kg. Vertices are required to have at least five charged particle tracks associated to the vertex, with no upstream track therefore forming a neutral vertex. In order to reduce the background from neutral hadron interactions high-energy vertices are selected by requirements on the track angles in the longitudinal plane. An electron or muon candidate with energy/momentum greater than 200 GeV is also required to be associated to the vertex. Finally, a back-to-back topology between the lepton and hadron system in the transverse plane is required. 

The main background arises from neutral-hadron interactions, where the neutral hadrons are produced in muon initiated photo-nuclear processes in the rock in front of FASER, or in the FASER$\nu$ detector. It is strongly suppressed by the vertex selections, especially the requirement of a high-energy electron or muon associated to the vertex. The neutral hadron background passing the selection is estimated using high-statistics Monte Carlo simulation samples.  
The estimate is validated by comparing the number of vertices without a high-energy selection between data and simulation, leading to a 50\% uncertainty on the background estimate. For the $\nu_\mu$ analysis, an additional background comes from NC neutrino interactions, which can mimic the signal and is estimated with simulation. The total background estimate is $0.025 \pm 0.013$ vertices for the $\nu_e$ selection and $0.22 \pm 0.08$ vertices for $\nu_\mu$ selection.  

\begin{figure}[hbt!]
    \centering
    \includegraphics[width=0.98\textwidth]{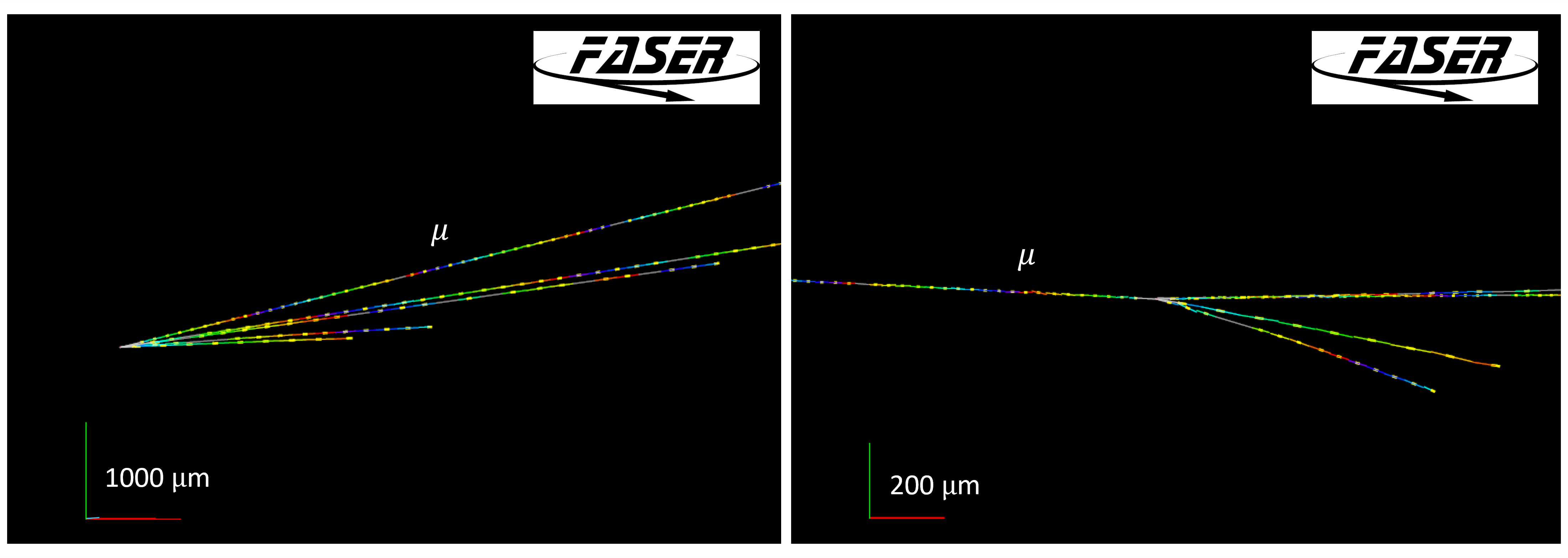}
    \includegraphics[width=0.98\textwidth]{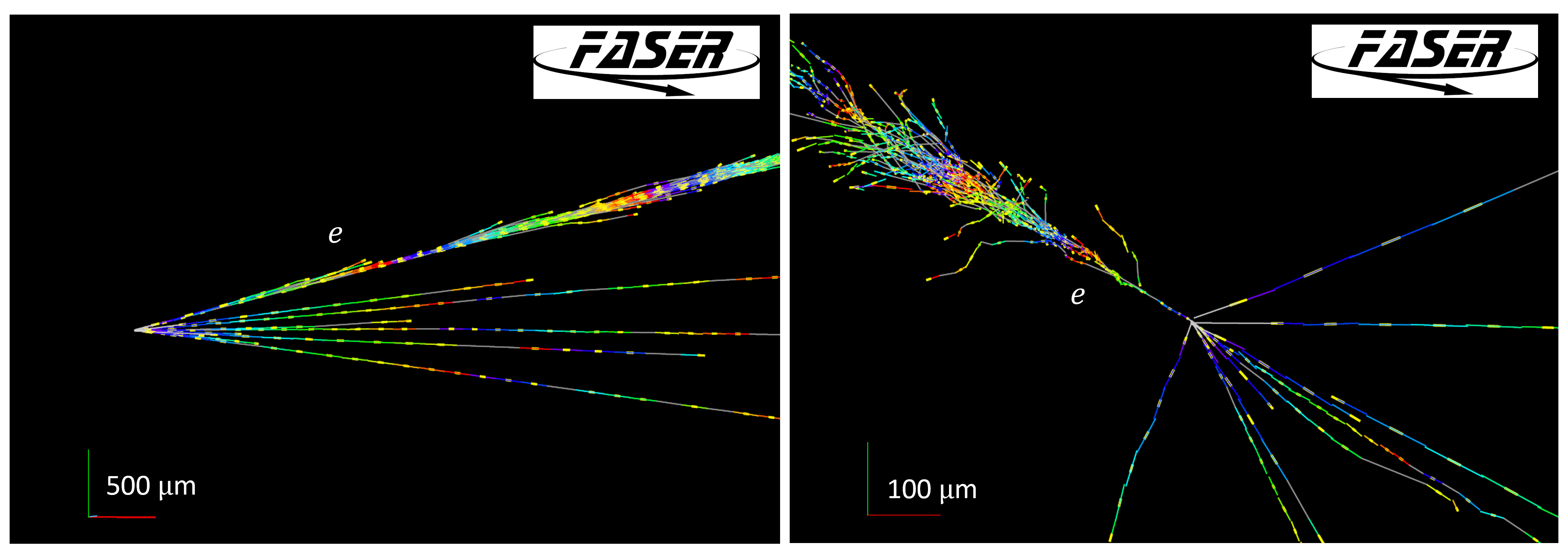}
    \caption{Example candidate neutrino interaction vertices in FASER$\nu$. Top shows a muon neutrino candidate, bottom an electron neutrino candidate. Left shows a side view, whereas right shows the transverse view (along the incoming neutrino direction). (Figures are taken from Ref.~\cite{FASER:2024hoe}.)}
    \label{fig:nu-cands}
\end{figure}

\begin{figure}[hbt!]
    \centering
    \includegraphics[width=0.35\textwidth]{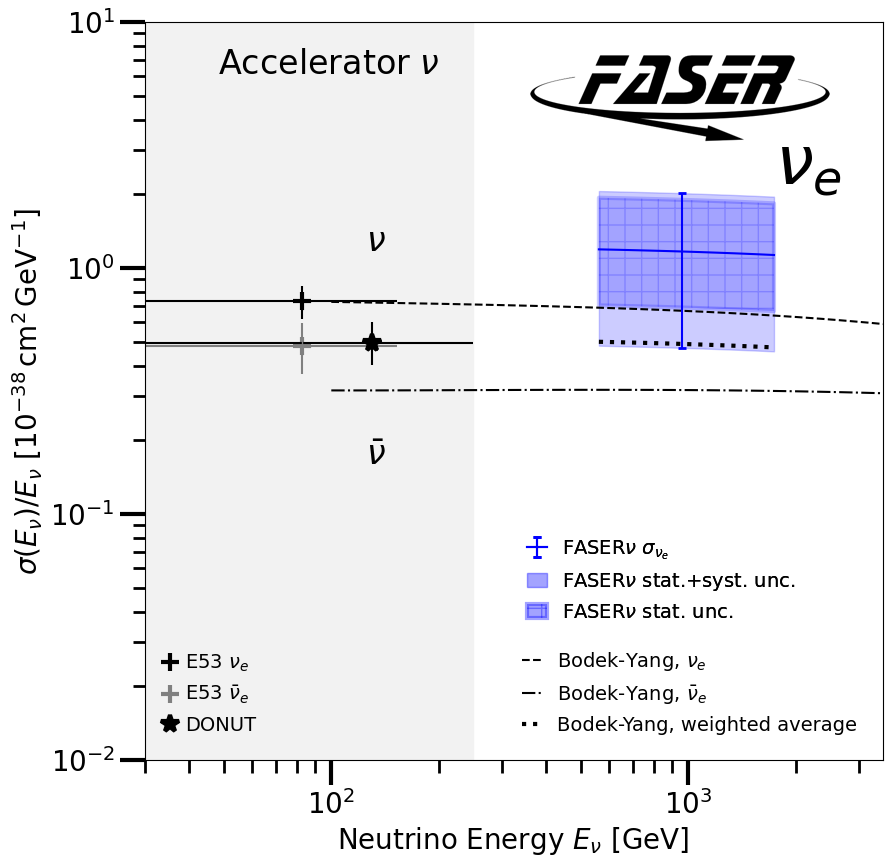}
    \includegraphics[width=0.62\textwidth]{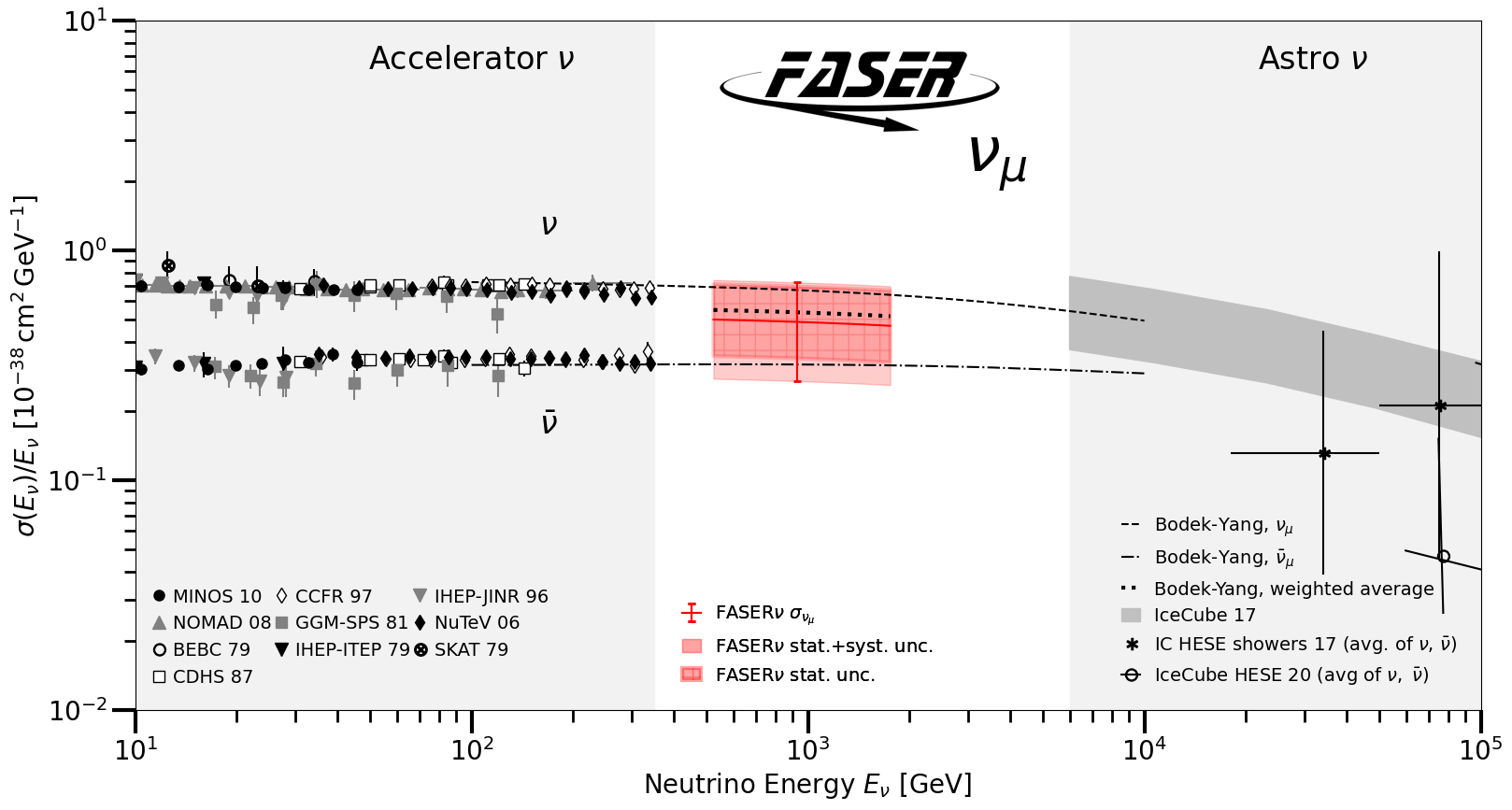}
    \caption{The neutrino interaction cross section measured by FASER$\nu$ for electron neutrinos (left) and muon neutrinos (right). (These figures are taken from Ref.~\cite{FASER:2024hoe}.)}
    \label{fig:fasernu-xsec}
\end{figure}

Five (eight) electron (muon) neutrino vertex candidates are observed in the data, which are used to measure the neutrino interaction cross section. Examples of candidate vertices are shown in Fig.~\ref{fig:nu-cands}. The cross section is calculated by scaling the theoretical cross section by the ratio of the observed to expected neutrino yields. It is shown in Fig.~\ref{fig:fasernu-xsec} as a function of the neutrino energy, where the energy range shown corresponds to the 68\% region for simulated neutrino interactions to pass the analysis selection. Note, since FASER$\nu$ can not distinguish between neutrino and anti-neutrino the measured cross section is the weighted average of the two.

The main uncertainties on the cross section measurements arise from data statistics, the uncertainty on the incoming neutrino flux (especially for the $\nu_e$ channel), and the uncertainty on the selection efficiency related to the neutrino interaction modeling.

The analysis was repeated, analyzing a larger volume of the same FASER$\nu$ box (exposed to 9.5~fb$^{-1}$), corresponding to a target mass of 315~kg (2.5 $\times$ larger than in the previous analysis). 
Table~\ref{tab:fasernuconf-summary_cand} shows the observed number of neutrino vertex candidates, as well as the expected background and signal.
The observations are consistent with the expectation within the large uncertainties. 

\begin{table}
\begin{center}
\begin{tabular}{lll}
\hline
\ & $\nu_e$ CC& $\nu_{\mu}$ CC\\
\hline
Expected signal      & 2.8--7.2 & 16.2--28.7 \\
Expected background  & 0.06$^{+0.04}_{-0.02}$ & 0.54$^{+0.22}_{-0.17}$ \\
Observed events      & 5        & 20 \\
\hline 
\end{tabular}
\caption{Summary of the expected signal, the expected background and the number of observed neutrino vertex candidates from Ref.~\cite{Ariga:2927714}.}
\label{tab:fasernuconf-summary_cand}
\end{center}
\end{table}

{\bf Latest FASER$\nu$ results}

\begin{figure}[h]
\centering
\includegraphics[width=0.90\linewidth]{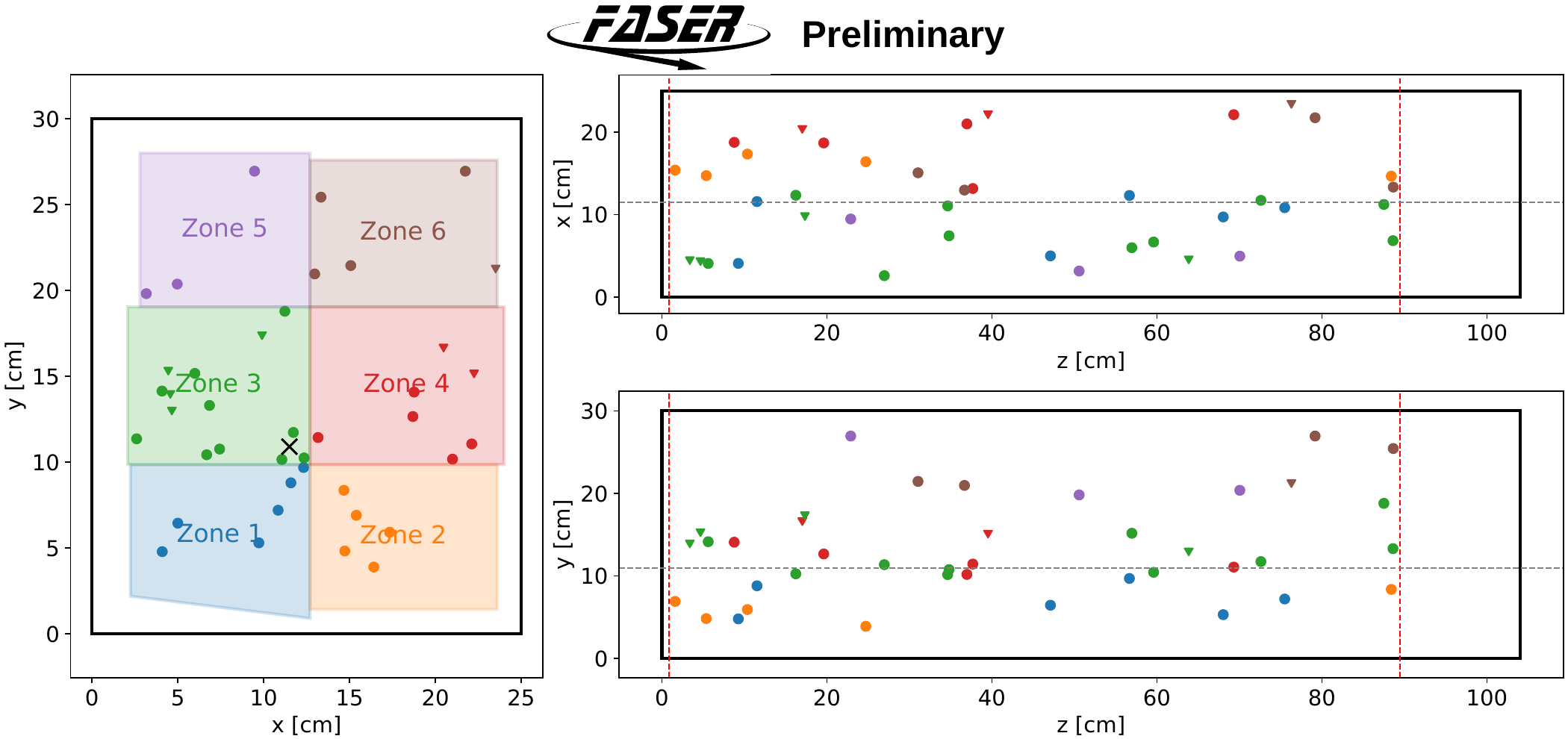}
\caption{The location of the reconstructed neutrino interaction vertices in the FASER$\nu$ detector. (left) in the transverse plane (the different zones are used in the reconstruction procedure). (right) in the longitudinal direction). (These figures are taken from Ref.~\cite{FASER:FASERnu2026}.)
}
\label{fig:FASERnu26_vertex-posn}
\end{figure}

The analysis was updated in March 2026~\cite{FASER:FASERnu2026}, in which the target mass was increased to 680~kg (2.2 $\times$ larger than for the previous analysis). For the $\nu_e$ selection, seven events are observed with an expected background of $0.1 \pm 0.1$, and an expected signal of $7.7^{+6.3}_{-2.5}$.
For the $\nu_\mu$ selection 33 events are observed with an expected background of $1.2 \pm 0.5$, and an expected signal of $40.0^{+14.5}_{-9.7}$. 
The observations are consistent with the SM expectations in both channels. The location of the reconstructed neutrino vertices in the transverse and longitudinal planes are shows in Fig.~\ref{fig:FASERnu26_vertex-posn}.

\begin{figure}[h]
\centering
\includegraphics[width=0.750\linewidth]{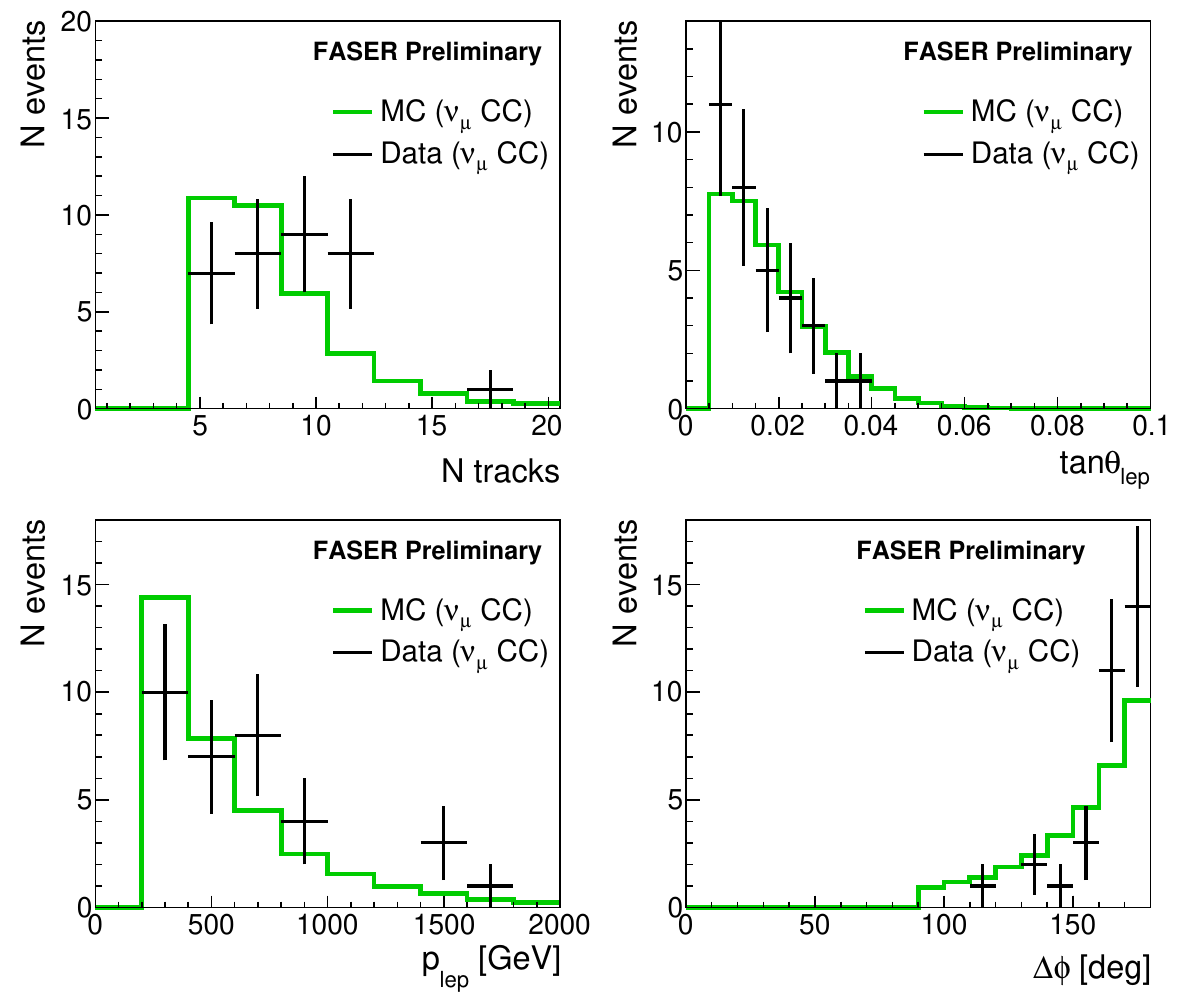}
\caption{Data (black points)  and MC simulation (green histogram) distributions of the track multiplicity (N tracks), lepton angle ($\tan \theta_{\text{lep}}$), lepton momentum ($p_{\text{lep}}$), and $\Delta\phi$ for the $\nu_{\mu}$ CC signal that passed the selection criteria. The MC simulation distributions are normalized to the number of observed events. (These figures are taken from Ref.~\cite{FASER:FASERnu2026}.)
}
\label{fig:plots_numuCC_ev}
\end{figure}

The larger number of $\nu_\mu$ candidates allows a comparison of the vertex properties with the expectation from simulation as shown in Fig.~\ref{fig:plots_numuCC_ev}. This shows, for example, the expected back-to-back topology for a CC neutrino interaction between the lepton and the hadronic system in the transverse plane ($\Delta \phi$). The other distributions show reasonable agreement between the simulation and data.

\begin{figure}[h]
\centering
\includegraphics[width=0.45\linewidth]{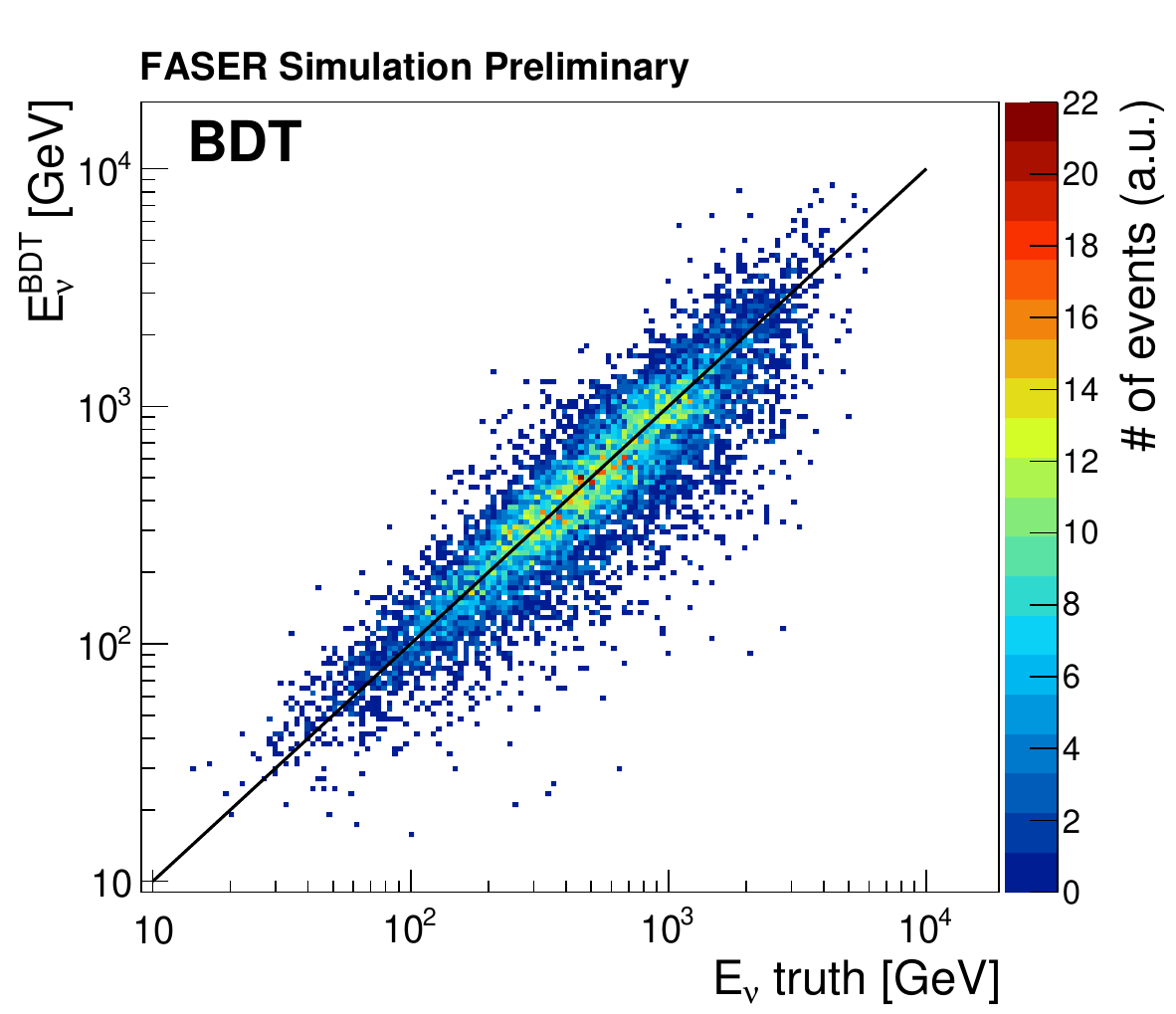}
\includegraphics[width=0.45\linewidth]{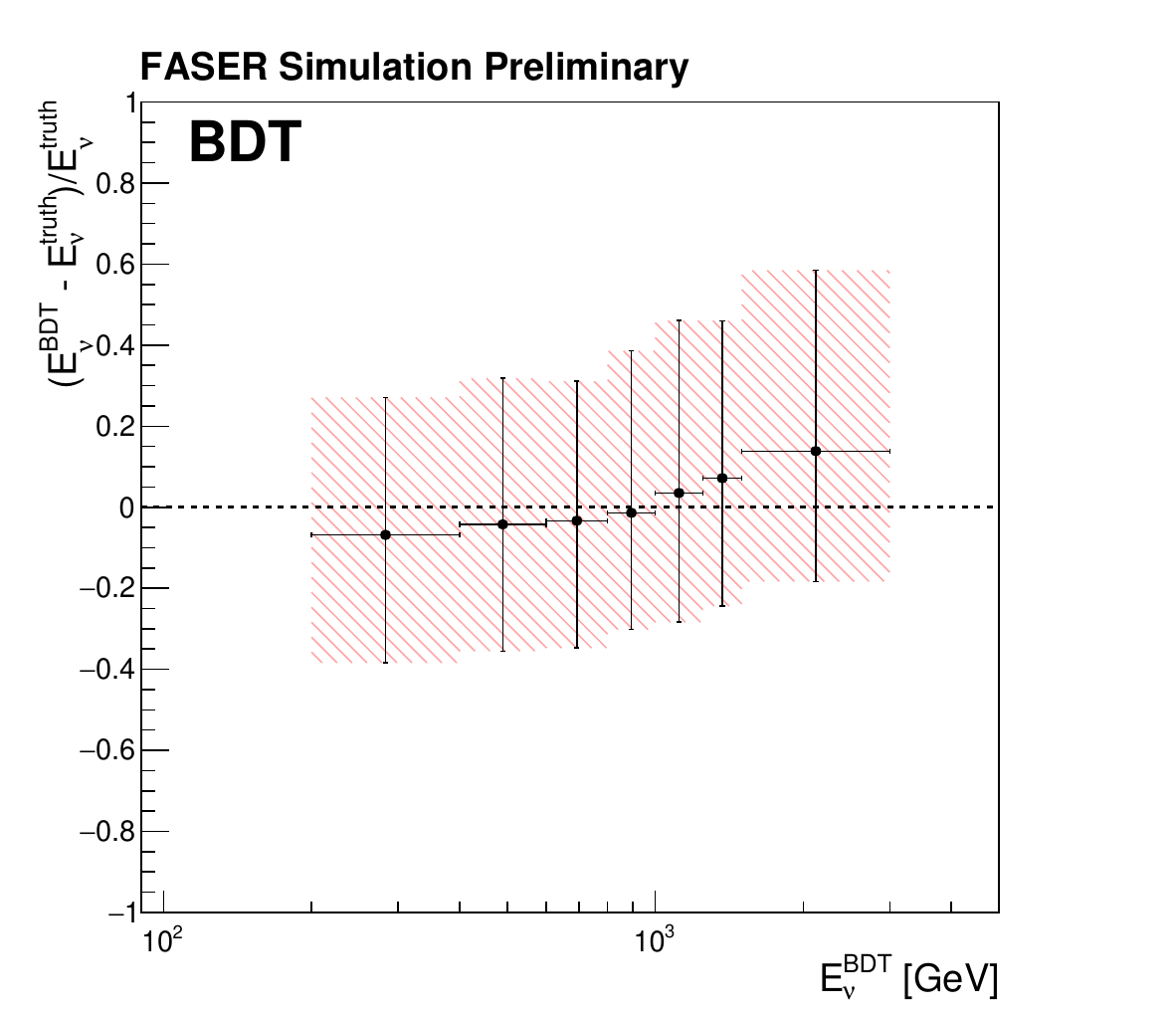}
\caption{(left) The reconstructed neutrino energy versus the true energy from simulation. (right) The reconstructed neutrino energy resolution versus energy. 
(These figures are taken from Ref.~\cite{FASER:FASERnu2026}.)
}
\label{fig:FASERnu-Eres}
\end{figure}

\begin{figure}[h]
\centering
\includegraphics[width=0.45\linewidth]{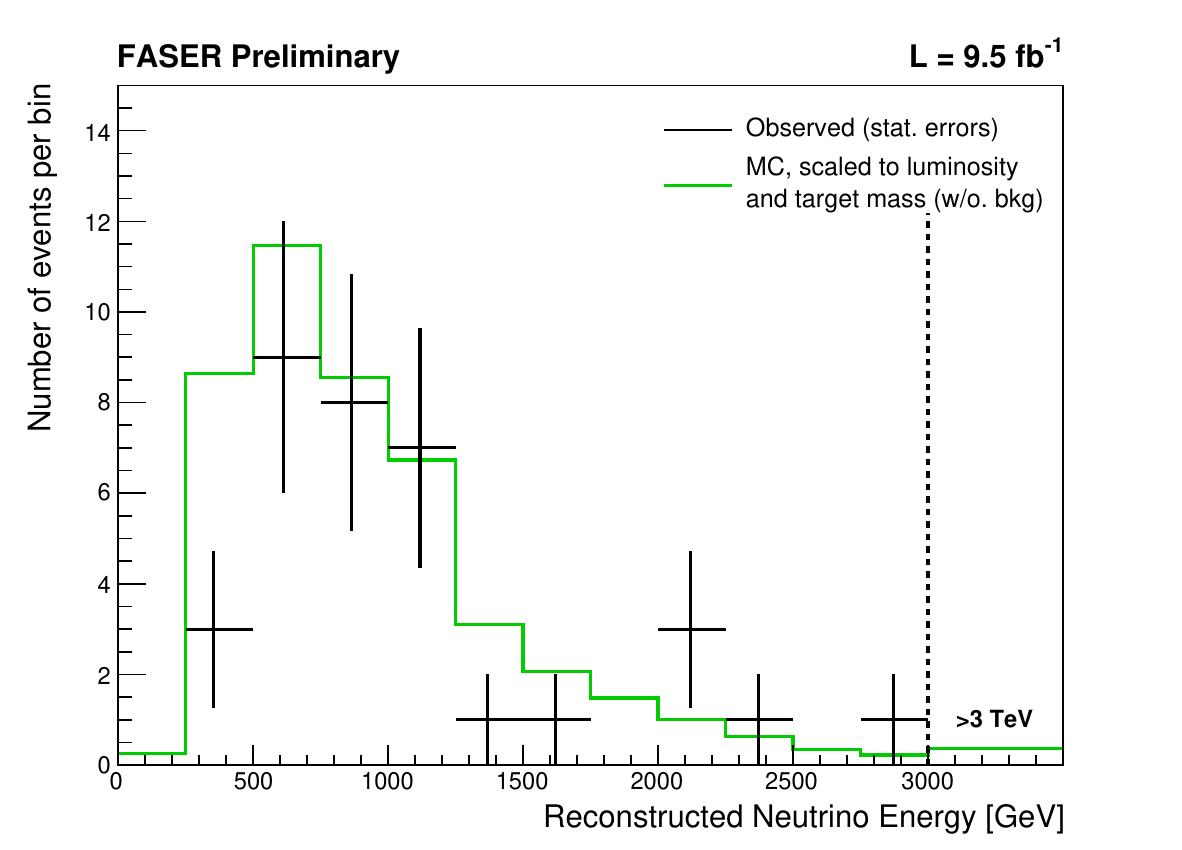}
\caption{The energy distribution for selected FASER$\nu$ $\nu_\mu$ CC  interactions compared to simulation. 
(This figure is taken from Ref.~\cite{FASER:FASERnu2026}.)
}
\label{fig:FASERnu-Eneut}
\end{figure}

\begin{figure}[h]
\centering
\includegraphics[width=0.59\linewidth]{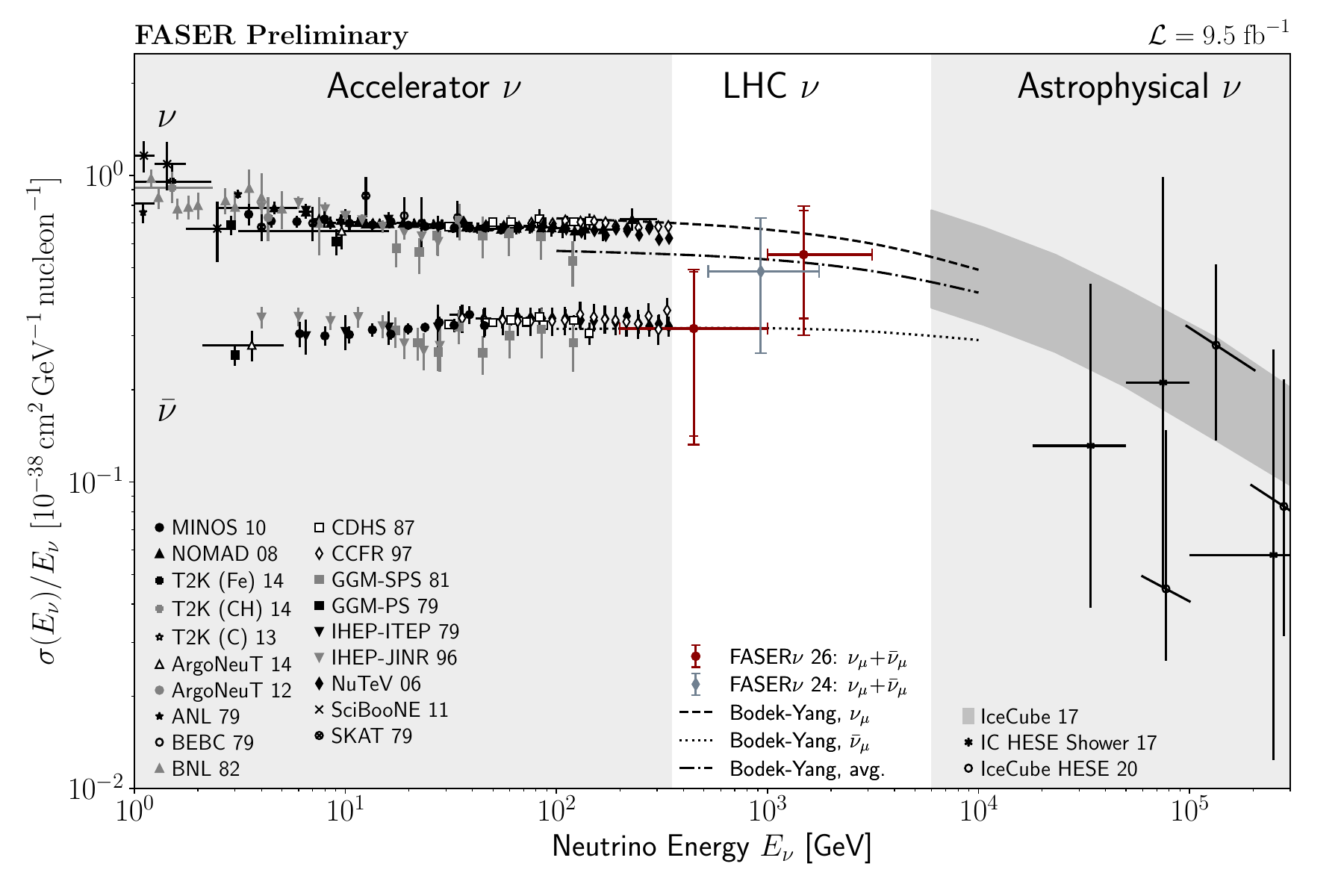}
\includegraphics[width=0.39\linewidth]{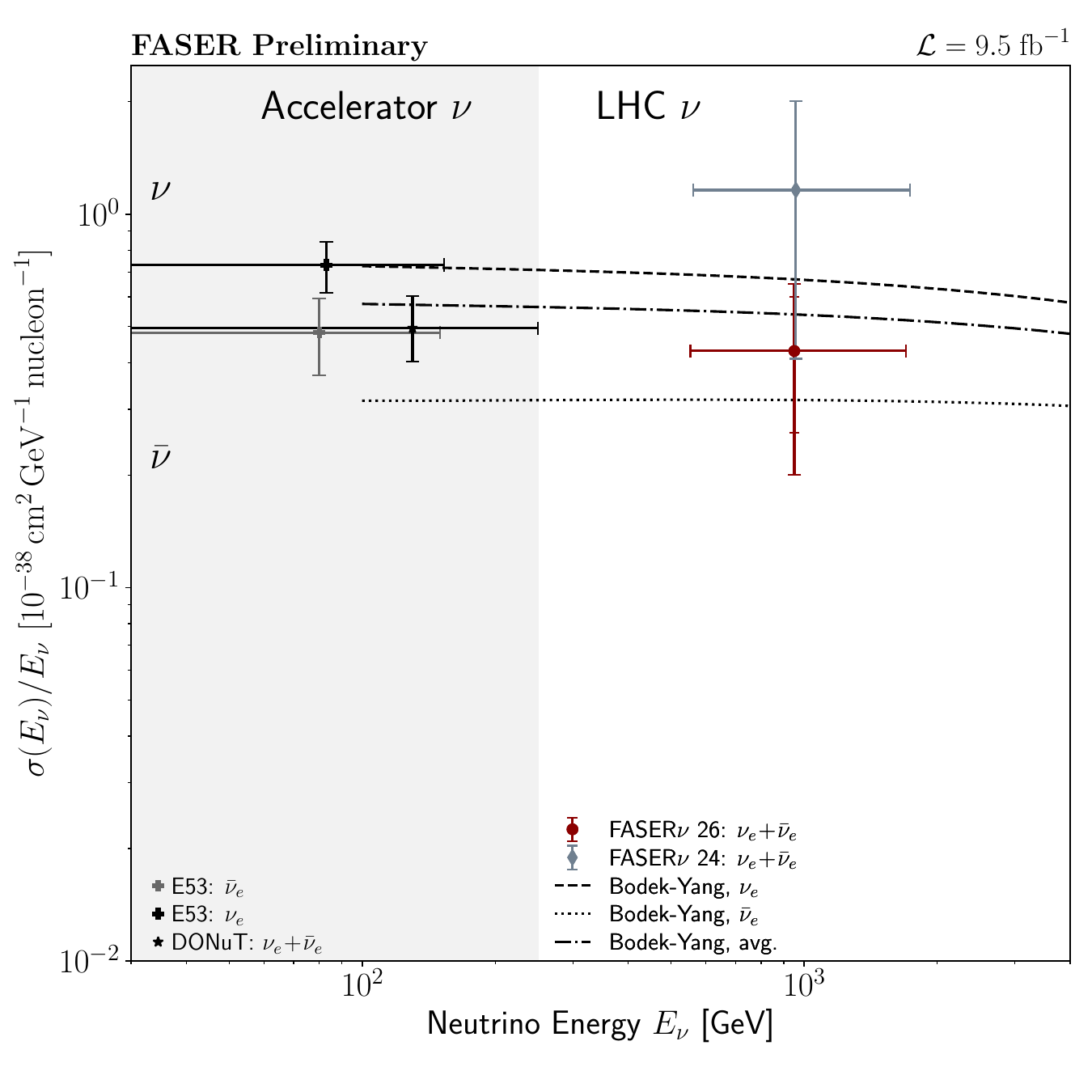}
\caption{The measured neutrino interaction cross section (averaged between $\nu$ and $\overline{\nu}$) for the most FASER$\nu$ analysis. (left) For $\nu_\mu$ in two bins of energy. (right) For $\nu_e$ with no energy binning.  
(These figures are taken from Ref.~\cite{FASER:FASERnu2026}.)
}
\label{fig:FASERnu-xsec}
\end{figure}

For the $\nu_\mu$ channel the energy of the neutrino candidates is estimated using a BDT algorithm. The BDT uses as input the following variables: the daughter muon momentum $P_\mu$, the inverse of the daughter muon angle tangent 1/tan $\theta_\mu$, and the sum of charged hadron momenta $\Sigma P_{chg.had}$. The muon and charged hadron momenta are reconstructed as described in Section~\ref{sec:performance} and in more detail in Ref.~\cite{FASER:2026hzm}. The BDT is trained on simulated neutrino samples, and shows a good linearity, with a resolution in the range of 30 - 40\%, as shown in Fig.~\ref{fig:FASERnu-Eres}. The neutrino energy distribution for the selected $\nu_\mu$ vertices and the expectation from simulation are shown in Fig.~\ref{fig:FASERnu-Eneut}, where the simulation describes the data well. 
The selected $\nu_\mu$ events are unfolded to give the interaction cross section in two bins of neutrino energy. The result is shown in Fig.~\ref{fig:FASERnu-xsec} (top). The measured $\nu_e$ cross section is shown in Fig.~\ref{fig:FASERnu-xsec} (bottom) with no energy binning. 

In addition, work has started to search for neutrino interaction events producing a charm hadron~\cite{FASER:charm2026}. This allows to probe the strange content of the proton, and is also important for validating short-lived particle reconstruction, a pre-requisite for searching for $\nu_\tau$ events.

\clearpage

\subsubsection{Implication of FASER neutrino results on forward hadron production}

\begin{figure}[ht]
\centering
\includegraphics[width=0.35\textwidth]{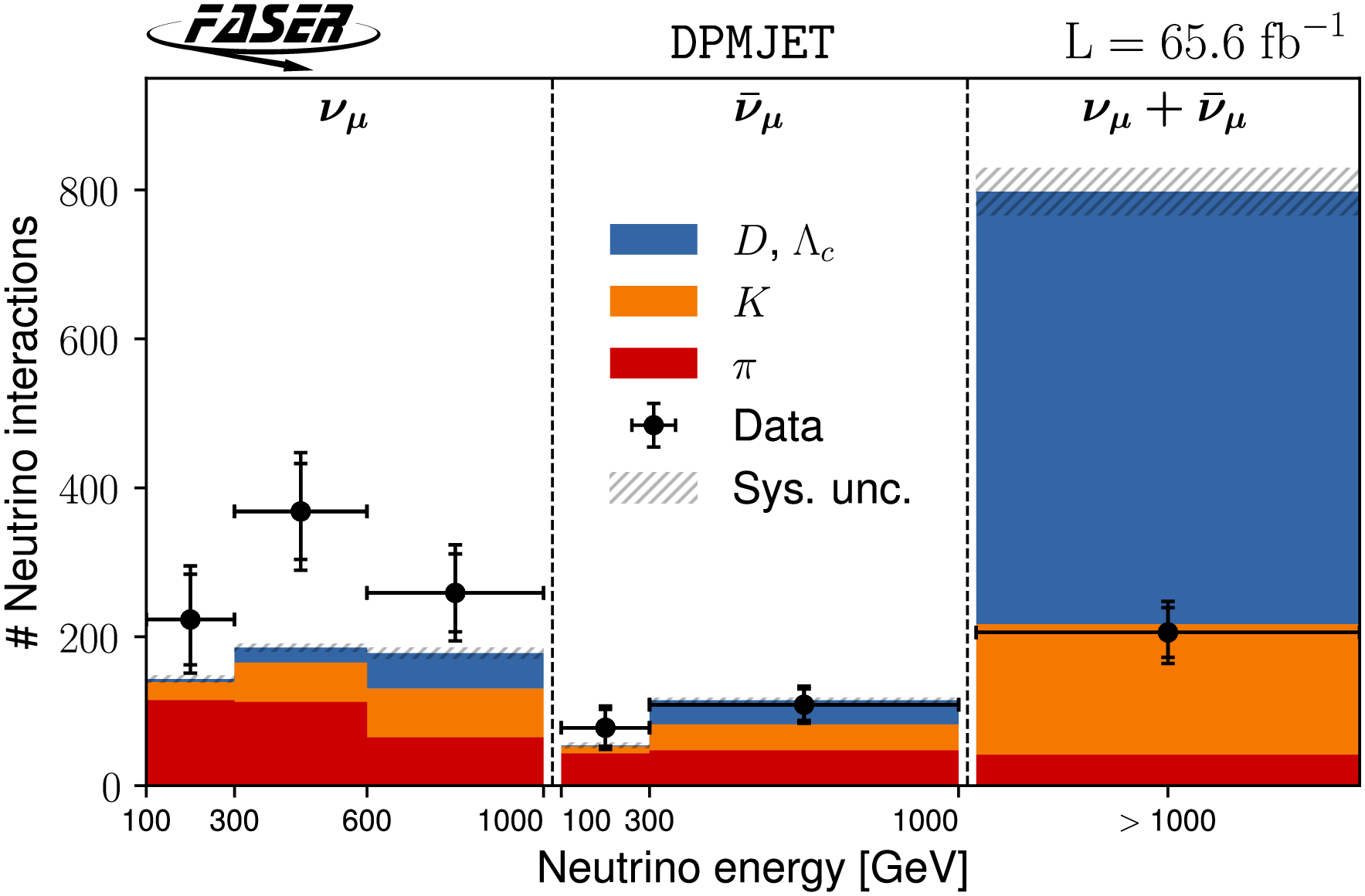}
\includegraphics[width=0.64\textwidth]{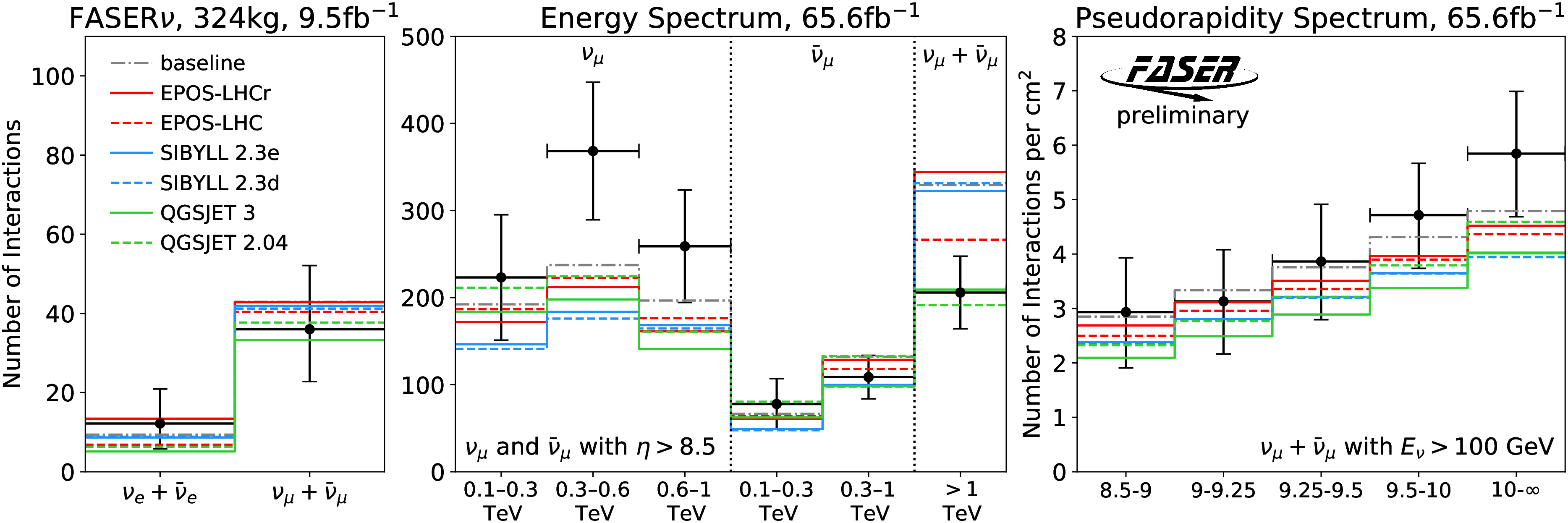}
\caption{(left) A comparison between the FASER data and the DPMJET generator~\cite{Fedynitch:2015kcn}. (right) A comparison between FASER measurements and the EPOS~\cite{Pierog:2013ria}, SIBYLL~\cite{Riehn:2019jet} and QGSJET~\cite{Ostapchenko:2010vb} generators. (Taken from Ref.~\cite{FASER:Astro}.)
}
\label{fig:forward-hadron}
\end{figure}

The FASER neutrino results are already able to provide useful constraints on different forward hadron production models. Fig.~\ref{fig:forward-hadron} (left) shows a comparison between the FASER measurement of the neutrino flux~\cite{FASER:2024ref}, where the flux has been predicted using the DPMJET generator~\cite{Fedynitch:2015kcn}. This generator is not able to describe the data for high energy $\nu_\mu$ production. Fig.~\ref{fig:forward-hadron} (right) shows a comparison of several FASER neutrino measurements (not including the updates from 2026) with the expectation from different forward hadron production models~\cite{FASER:Astro}. None of the models is able to reproduce the data across all bins. Future FASER measurements, with larger datasets will allow different hadronic interaction models to be more precisely probed.  

    \clearpage
    \newpage    
	\section{Future}\label{sec:future}
    
Since FASER started operating, the LHC schedule was updated to continue Run 3 until mid-2026, when Long Shut Down 3 (LS3) will start with an expected duration of 3 years. FASER will continue to operate in 2026, with an additional dataset of around 30~fb$^{-1}$ expected.
In 2023, FASER was approved to continue operating in Run 4~\cite{Boyd:2882503,approval4-RBminutes} when an additional 780~fb$^{-1}$ of data is expected to be delivered. The collaboration is actively investigating possible detector upgrades to improve the performance during Run 4 operations, especially since it may not be feasible to run with the FASER$\nu$ emulsion with the higher luminosity expected in Run 4. Different upgrade options are discussed in Ref.~\cite{FASER:2025myb}. 

For 2026 running two prototypes of possible upgrade detectors will be installed behind FASER in order to provide input to the upgrade decision making process. 
The first of these is the FASERCal prototype, consisting of 20 layers of 40 $\times$ 40 cm$^2$ made up of 1~cm$^3$ scintillating cubes which are read-out by wavelength shifting fibers in three dimensions. 
This detector concept is based on the SuperFGD~\cite{T2K:2019bbb}, a key component of the upgraded near detector in the T2K neutrino experiment. 
The other prototype is the AHCAL~\cite{Shi:2025epp} which is a 5~tonne steel / scintillator-pad detector, which was designed as a hadronic calorimeter prototype for the CEPC Higgs factory. Simulation studies show that this detector (with 4 $\times$ 4~cm$^2$ sized scintillator pads) could have good performance for neutrino measurements in Run 4. The detector was shipped from China and commissioned at CERN in late-2025.

In the longer term FASER collaborators, along with others, have proposed the Forward Physics Facility (FPF), a new dedicated facility to maximize the physics in the far forward region of the LHC collisions in the HL-LHC era. The proposed facility would house four complementary experiments for neutrino physics and BSM searches. The experiments would collect $\mathcal{O}(10^6)$ neutrino interactions and have unprecedented sensitivity for several BSM scenarios. A letter of intent~\cite{FPF:2025bor} was submitted to the LHCC in September 2025.

    \clearpage
    
    \newpage    
	\section{Summary}\label{sec:summary}
    
The FASER experiment was designed as a cheap and small experiment to search for light weakly-coupled new particles and to study neutrinos produced at a collider for the first time. The detector was designed, constructed, installed and commissioned in record time during LHC Long Shutdown 2, and was ready to take first physics data at  the start of LHC Run 3 in 2022. The collaboration were able to quickly analyse the data and released first results for the winter conferences in March 2023, including the first ever observation of neutrinos produced at a collider. Since then several new results have been released including the first observation of collider electron neutrinos, the first measurements of the neutrino interaction cross section in the TeV energy range, and searches for dark photons and ALPs constraining unexplored regions of signal parameter-space motivated by dark matter. The steady release of physics results has been based on the smooth detector operations, with the detector collecting more than 97\% of the delivered luminosity, and the excellent detector performance with all sub-detector systems performing as, or better than, the design requirements. The FASER Collaboration is investigating several options for detector upgrades which could be implemented in LHC Long Shutdown 3 to enhance the physics programme for Run 4, and maximise the physics output from the HL-LHC.
    \clearpage
	
	\newpage
	\section*{Acknowledgements}
	It is a great pleasure to thank all members of the FASER Collaboration for their contributions to the work described in this paper. Especial thanks to Jonathan L. Feng, Brian Petersen, Felix Kling, Aki Ariga, Tomoko Ariga, Eric Torrence and Tomohiro Inada for useful input on the manuscript. \\   

	\bibliography{references}

	\newpage
	\appendix
	\renewcommand*{\thesection}{\Alph{section}}
	

\end{document}